\documentclass[12pt, oneside]{article} 

\usepackage{amsmath}  				
\usepackage{geometry}
\usepackage{amssymb}
\usepackage{mathtools}
\usepackage[T1]{fontenc}
\usepackage[utf8]{inputenc}
\usepackage[parfill]{parskip}    		
\usepackage{graphicx}
\usepackage{mathtools}

\allowdisplaybreaks
\usepackage[justification=centering]{caption}
\usepackage{mathtools}
\numberwithin{equation}{section}

\usepackage{xfrac}
\usepackage{amssymb}
\usepackage[table]{xcolor}
\usepackage{booktabs}
\usepackage{mdwtab}
\usepackage{breqn}
\usepackage{lmodern}
\def\a{\alpha}

\def\e{\epsilon}

\def\g{\gamma}

\def\k{\kappa}
\def\l{\lambda}

\def\o{\omega}
\def\p{\pi}

\def\s{\sigma}
\def\t{\tau}

\def\G{\Gamma}

\def\cL{{\cal L}}
\usepackage{float}
\usepackage{xcolor}
\usepackage{lscape}
\usepackage{subfig}
\usepackage{amsthm}
\usepackage{feynmp-auto}
\usepackage{graphicx}
\usepackage{caption}

\def \nn {\nonumber}
\def \be  {\begin{equation}}
\def \ee  {\end{equation}}
\def \bea  {\begin{eqnarray}}
\def \eea  {\end{eqnarray}}

\begin{document}

\begin{flushright}
YITP-SB-2023-28
\end{flushright}

\begin{center}
\Large \bf Local infrared safety in\\  time-ordered perturbation theory
\end{center}

\bigskip

\centerline{George Sterman and Aniruddha Venkata}

\begin{center}
C.N.\ Yang Institute for Theoretical Physics and Department of Physics and Astronomy\\
Stony Brook University,
Stony Brook, NY 11794-3840 USA
\end{center}

\centerline{\today}


\date{}					

\abstract{We develop a general expression for weighted cross sections in leptonic annihilation to hadrons based on time-ordered perturbation theory (TOPT).
The analytic behavior of the resulting integrals over spatial momenta can be analyzed in the language of Landau equations and infrared (IR) power counting.  For any infrared-safe
weight, the cancellation of infrared divergences is implemented locally at the integrand level, and in principle can be evaluated numerically in four dimensions.   
We go on to show that it is possible to eliminate unphysical singularities that appear in time-ordered perturbation theory for arbitrary amplitudes. This is done by
reorganizing TOPT into an equivalent form that combines classes of time orderings into a ``partially time-ordered perturbation theory".  
 Applying the formalism to leptonic annihilation, we show how to derive diagrammatic expressions with only physical unitarity cuts.}

\tableofcontents

\section{Introduction}

In the light of prospects for increasingly high-statistics data from the Large Hadron Collider and proposed facilities, the need for precision in perturbative calculations of collider cross sections is widely recognized \cite{Heinrich:2020ybq,Caola:2022ayt,Boughezal:2022cbl}.   A recurring theme in these discussions has been the possibility of carrying out calculations  directly in four dimensions \cite{TorresBobadilla:2020ekr,Anastasiou:2020sdt,Anastasiou:2022eym}.   For physical observables like cross sections, this involves canceling infrared (IR) singularities at the level of integrands, resulting in expressions amenable to numerical calculation \cite{Hernandez-Pinto:2015ysa,Sborlini:2016gbr,Sborlini:2016hat,Aguilera-Verdugo:2019kbz,Capatti:2019ypt,Capatti:2019edf,Aguilera-Verdugo:2020set,Capatti:2020ytd}.   
The potential for such an approach is implicit in formal proofs of the infrared safety of jet and related weighted cross sections \cite{Sterman:1978bj,Sterman:1979uw}, based on light-cone ordered perturbation theory, and in proofs of factorization based on time-ordered perturbation theory (TOPT)  \cite{Bodwin:1984hc,Collins:1985ue,Sterman:1993hfp,Sterman:1995fz}.   In each case, cancellations of infrared divergences locally in momentum space become manifest after integration over loop energies or  light-cone variables.  The same fundamental observation is found in Refs.\  \cite{Hernandez-Pinto:2015ysa} -
\cite{Capatti:2020ytd}, which carry out energy integrals using the technique of  Loop Tree Duality \cite{TorresBobadilla:2020ekr,Catani:2008xa}.   In this paper, inspired by these results  and by the general analyses of perturbation theory in Refs.\ \cite{TorresBobadilla:2021ivx,Sborlini:2021owe,Borinsky:2022msp,Capatti:2022mly}, we return to time-ordered perturbation theory, to provide a complementary perspective on these important results.   

We will reconsider infrared  (IR) safe cross sections for hadronic states produced in lepton pair annihilation from this point of view.   In this case, unitarity implies that cross sections can be computed as sums over the final states that correspond to cuts of vacuum polarization diagrams.  The contribution of any fixed cut, of course, includes infrared divergences, which cancel in the sum over cuts.    The result of our analysis is a manifestly power-counting finite expression for such a cross section, given as a sum over time-ordered, cut vacuum polarization diagrams.  All terms coming from a given diagram to this sum are evaluated as one three-dimensional integration per loop of the original uncut diagram, without change of variables as we  sum over the cuts (states) of the diagram.   

In the course of our discussion in TOPT, we will encounter apparent, unphysical contributions to the cross section, associated with the vanishing of energy denominators corresponding to cuts that divide the diagram into more than two parts.  We'll see how these unphysical contributions cancel in the sum over time orders.   This observation opens the door to a reformulation of TOPT, in which unphysical energy denominators are altogether absent.  Such a form, in fact, has already been derived in Ref.\ \cite{Capatti:2022mly}, from a distinct but related point of view.   We will present an alternative derivation, based on what we will call partially time-ordered perturbation theory (PTOPT).  These results, like those in Ref.\ \cite{Capatti:2022mly}, are quite general, and provide a reformulation for arbitrary amplitudes, and to light-cone- as well as time-ordered perturbation theory.

 In Sec.\ \ref{sec:TOPT} we will review some properties of time-ordered perturbation theory, and we will observe that within a single time-ordered diagram, many contributions to the imaginary part are unphysical, involving the disconnected production of finite-energy particles out of the vacuum. 
We will show, however, that upon summing over time orders, these contributions cancel among themselves at the amplitude level.  In Sec.\ \ref{sec:local-irs} we construct a locally IR finite formula for weighted cross sections in TOPT for leptonic annihilation to hadrons.  Each such expression is closely related to the imaginary part of the vacuum polarization amplitude for the current that couples to quantum chromodynamics in the Standard Model.    We will derive the analogs of Landau equations for this class of cross sections, along with power counting at the ``pinch surfaces" where these equations are satisfied.  

In subsequent sections, we show how to reorganize TOPT to eliminate these unphysical singularities manifestly, by developing partially time-ordered perturbation theory, with results similar to those of Ref.\ \cite{Capatti:2022mly}.   These partial sums over time orders can be carried out algorithmically, and in effect reduce the number of terms necessary to compute amplitudes and cross sections.   To this end, in Sec.\ \ref{Poset and caus}, we will recall a mathematical structure that underlies each time order: the partially ordered set (poset), and show by examples how partial ordering can be used to eliminate unphysical cuts. 

Section \ref{Formula} is dedicated to combining the full TOPT contributions associated with each poset, implementing the constraints associated with the poset. In doing so, we will  sum over all time orders consistent with a poset. Finally,  in Sec.\ \ref{sec:weighted-cs}, we will return to weighted cross sections in leptonic annihilation, and adapt  our formulas for the weighted cross section case.   We conclude with a short summary and comments on future directions.

\section{Time-ordered perturbation theory and leptonic annihilation}
\label{sec:TOPT}

Time-ordered perturbation theory provides a systematic method for integrating all loop energies in an arbitrary diagram \cite{Sterman:1993hfp,Sterman:1995fz}.  For the treatment of electroweak annihilation, the diagrams of interest are two-point correlators of electroweak currents.   More generally, we can consider any amplitude $A\left (q'_j,q_i \right)$, where $q_i$ represents a set of incoming momenta, and $q_j'$ a set of outgoing momenta, any or all of which may be off-shell.  In time-ordered perturbation theory, the expression for such an amplitude is of the form
\begin{equation}
A\left (q'_j,q_i \right)\ =\sum_{G\in {G_A}}\; \int d\cL_G \; \sum_{\tau_G} \; \mathbb{N}_{\tau_G} \; \prod_{i=1}^{N_G}\frac{1}{2\omega_i} \prod_{s=1}^{V_G-1}  \frac{i}{E_{s}-\sum_{j\in  s} \omega_{j }+i\epsilon},
\label{eq:topt-general}
\end{equation}
where $G_A$ represents the set of graphs in Lorentz-invariant perturbation theory that contribute to the amplitude and  $\tau_G$ represents the time orders of the $V_G$ vertices of each graph $G$, with $N_G$ lines.   Each ordering, $\tau_G$ of the vertices specifies a set of states, labeled $s=1 \dots V_G-1$, whose total energy is the sum of the on-shell energies of all the particles in that state, defined to flow from earlier to later times.  
 The denominators in Eq.\ (\ref{eq:topt-general}) are  ``energy deficits", for each state $s$, the difference between the sum of the on-shell energies $\omega_i=\sqrt{\vec p_i^2+m_i^2}$ for lines of mass $m_i$ in state $s$ and the external energy, $E_s$, that has flowed into the diagram before state $s$.   The external net energy of each state depends on the specific ordering $\tau_G$.
 
In Eq.\ (\ref{eq:topt-general}), $\int d\cL_G$ represents the spatial loop momentum space of the graph $G$, which is the same for every time order.   Each line momentum $\vec p_j$ is a linear function of loop and external momenta, which in this way determine the energies $\omega_j$.
The factor $\mathbb{N}_{\tau_G}$ represents the perturbative numerator, consisting of overall constants and spin-dependent momentum factors, with the energies of every line evaluated on-shell, that is with energies $\omega_i$ defined as above.    The signs of these energies are always positive for flow from the earlier to later vertex connected by the line in question.   In this way, the numerator factor depends on the time order.

\subsection{Cross sections in TOPT}

We consider annihilation cross sections for leptons to hadrons, at lowest order in electroweak couplings.  In this case, we can represent our cross sections as
\bea
\sigma(Q)\ &\equiv& \ \sum_{N}\, (2\pi)^4\delta^4 \left (q - P_N\right)\, \big | \langle 0 \big | \, {\cal J}(0)\, |N\rangle |^2\, ,
\label{eq:sigma-def-1}
\eea
where the  $\cal J$ represents an  electroweak current, implicitly contracted  with tree-level leptonic tensors and vector boson propagators, which we suppress, and where $q^2=Q^2$.   The latter will play no role in these arguments, and we will consistently absorb them into the currents.  As a total cross section, $\sigma(Q)$ is related by the optical theorem to a forward-scattering amplitude, here a two-point correlation function
of the currents ${\cal J}$,
\bea
\sigma(Q)\ &=& \ {\rm Im}\, \Pi(Q) \, ,
\nn\\[2mm]
\Pi(Q) \ &=&\ i\, \int d^4x\, e^{-iq\cdot x} \langle 0| T[ {\cal J}(0) {\cal J}(x)] | 0 \rangle\, .
\label{eq:optical}
\eea
In this paper, we will use time-ordered perturbation theory to derive expressions for sets of weighted cross sections that generalize Eq.\ (\ref{eq:sigma-def-1}),
\bea
\Sigma[f,Q]\ &\equiv& \  \sum_{N}\, (2\pi)^4\delta^4 \left (q - P_N\right)\, f(N)\,  \langle 0 | \, {\cal J}(0)\, |N\rangle |^2\, ,
\label{eq:sigma-def}
\eea
with $f(N)$ a function of the kinematic variables that describe state $N$, and with currents normalized as for the total cross section, Eq.\ (\ref{eq:sigma-def-1}). Our goal will be to derive a set of expressions that exhibit manifestly the cancellation of infrared divergences in such quantities \cite{Sterman:1978bj,Sterman:1979uw,Komiske:2020qhg}.   First, however, we will discuss a bit more how the general relations (\ref{eq:sigma-def-1}) and (\ref{eq:optical}) are realized in TOPT.

In the notation of Eq.\ (\ref{eq:topt-general}), for the two-point correlation function of Eq.\ (\ref{eq:optical}),  a single momentum $q$ flows into and out of the diagram, and we may choose to work in its rest frame, $q^\mu=(Q,\vec 0)$.   In this case, our vacuum polarization diagram $\Pi(Q)$ is given by
\begin{eqnarray}
\Pi(Q)\ &=&\sum_{G\in G_\Pi}\; \int d\cL_G \;\prod_{i=1}^{N_G}\frac{1}{2\omega_i}\;  \sum_{\tau_G}\; \mathbb{N}_{\tau_G} \; \pi_{\tau_G}(Q,\cL_G)\, ,
\nn\\[2mm]
\pi_{\tau_G}(Q,\cL_G)\ &=&\  \prod_{s=1}^{V_G-1}  \frac{i}{Q\lambda_s-\sum_{j\in  s} \omega_{j }+i\epsilon}\, ,
\label{eq:forwardscattering}
\end{eqnarray}
where the sum is now over the set of diagrams $G_\Pi$ that contribute to the two-point correlation function that mediates the leptonic annihilation process.   We have suppressed vector indices of the currents
associated with the decay of the mediating vector bosons.
In this expression, the state-dependent factor $\lambda_s$ enforces the condition that the net external energy in state $s$ is positive only when state $s$ occurs after momentum $q$ flows into the diagram, here and below at vertex $i$, and before the same energy flows out, here and below at vertex $o$,
\bea
\lambda_s\ &=&\ 1\, , \quad  o>s\ge i\, ,
\nn\\[2mm]
\lambda_s\ &=&\ \hspace{-2mm} -1\, , \quad  i>s\ge o
\nn\\[2mm]
&=&\ 0\, , \quad {\rm otherwise}\, .
\label{eq:lambda-s-def}
\eea
In Eq.\ (\ref{eq:forwardscattering}), the set of states $s$ depends implicitly on the time order, $\tau_G$.   Finally, we notice that energy denominators are negative semi-definite for massless particles and negative definite for massive particles except when $o>s\ge i$.   

Each covariant diagram with $V_G$ vertices provides $V_G!$ terms, the sum over $\tau_G$ in Eq.\ (\ref{eq:forwardscattering}), each with $V_G-1$ denominators.   
The vanishing of a subset of these denominators results in a branch cut in the function $\Pi(Q)$. Each such branch cut corresponds to an on-shell intermediate state that is at or above threshold.   
These are the physical, or unitarity cuts of the underlying diagram. In time-ordered perturbation theory,
such states separate a vacuum polarization diagram $G$ into a connected amplitude and complex-conjugate amplitude,
where external momentum flows into the amplitude and out of the complex conjugate.   Clearly, such cuts are possible only in orderings for which 
the vertex $i$ is earlier than vertex $o$.   By unitarity in the form of the optical theorem, the sum of these singularities gives the total cross section for
leptonic annihilation or equivalently the decay width of the relevant off-shell electroweak boson.
We shall see below, however, that even when $o>i$ in the uncut diagram, many cuts of $\pi_{\tau_G}$ in Eq.\ (\ref{eq:forwardscattering}) are actually unphysical, corresponding to the production of particles out of the vacuum.    One of the aims of our discussion is to 
reorganize the sum of time orders to show how these unphysical singularities always cancel.  In Secs.\ \ref{Poset and caus} and \ref{Formula}, we will show how to reorganize the time-ordered series, reducing the number of terms and
eliminating unphysical singularities without having to rely on this cancellation.   

Setting aside for a moment the cancellation of unphysical cuts, let us write the imaginary part of the amplitude $\Pi(Q)$, Eq.\ (\ref{eq:forwardscattering}), as a sum over its intermediate states, and generalize the resulting sum over final states to more general weighted cross sections.   
 To do so, we will need first to  write the TOPT expression for $\Pi(Q)$ as a sum over fixed states, given by cuts $C$ of an arbitrary diagram.   Here each $C$ is a particular state $s$ in Eq.\ (\ref{eq:forwardscattering})
 that is set on shell.
 Summing over $C$, we know from the optical theorem that we reconstruct the imaginary part of the forward scattering graph in Eq.\ (\ref{eq:forwardscattering}),
\begin{eqnarray}
{\rm Im}\, \Pi(Q)\ &=&\  
\sum_G\, \sum_{C}\, \sum_{\tau_{L[G/C]}}\, \sum_{\tau_{R[G/C]}}\, \pi^{(C)}_{\tau_{L[G/C]}\, \cup\, \tau_{R[G/C]}}(Q)
\nn\\[2mm]
\pi^{(C)}_{\tau_{L[G/C]}\, \cup\, \tau_{R[G/C]}}(Q)\ &=&\ \int d\cL_G \;  \mathbb{N}_{\tau_G} \; \prod_{i=1}^{N}\frac{1}{2\omega_i}\;
 \prod_{s=C+1}^{V_G-1}  \frac{i}{Q\lambda_s-\sum_{j\in  s} \omega_{j }-i\epsilon}
 \nn\\[2mm]
&\ & \hspace{5mm}\times\  (2\pi)\, \delta \left(Q-\sum_{j\in  C}\omega_{j}\right)
 \prod_{s=1}^{C-1}  \frac{i}{Q\lambda_s-\sum_{j\in  s} \omega_{j }+i\epsilon} 
\, ,
\nn\\
\label{eq:Im-T}
\label{eq:forwardscattering-cuts}
\end{eqnarray}
where $\tau_{L[G/C]}$ represents the time orders of the graph $G$ on the left of the cut, $C$, and $\tau_{R[G/C]}$ represents the time orders of the graph $G$ on the right of cut $C$. 
The union of these orders, $\tau_{L[G/C]} \cup \tau_{R[G/C]}$, uniquely specifies an ordering,  $\tau_G$ of the full diagram, with all vertices in the conjugate after all vertices in the amplitude\footnote{In Eq.\ (\ref{eq:forwardscattering-cuts}) and below, factors associated with later times in the uncut diagram are written to the left, following time ordering conventions.  In all figures, however, we will choose our final states to the right of initial states.  Thus, $\tau_{L[G/C]}$ refers to time orders on the left of the cut diagram.}.  For a given underlying diagram, $G$, orderings $\tau_{L[G/C]}$ and $\tau_{R[G/C]}$ specify diagrammatic contributions to the amplitude and complex conjugate amplitude, respectively, to produce final state $C$ by the action of current ${\cal J}$.   As above, the function $\mathbb{N}_{\tau_G}$ absorbs overall factors associated with vertices, and the overall factor of $i$ in Eq.\ (\ref{eq:optical}), all of which we do not exhibit.   We note, however, that each vertex is associated with a factor of $i$, and that when all denominators are real and nonzero, the diagram is real.   Note that to satisfy the delta function we must have $\lambda_C=+1$, so that the external momentum must flow into the diagram in the amplitude ($s<C$) and out from the complex conjugate ($s>C$).   Then $\lambda_s=0$ or $1$, and not $-1$, in both the amplitude and complex conjugate.
   
A complete discussion of the proof of Eq.\ (\ref{eq:forwardscattering-cuts}), including symmetry factors, is given in Ref.\ \cite{Sterman:1993hfp}, but here we observe that it involves the relationship between time orders of the full diagram and of amplitudes separated by its cuts,
\bea
\sum_{\tau_{G}}\sum_{C}\ =\ \sum_{C}\sum_{\tau_{L[G/C]}}\sum_{\tau_{R[G/C]}}\, ,
\label{eq:C-tau-sums}
\eea 
which the summations satisfy by definition.  A subtle feature of the sums is that the diagrams $L[G/C]$ and $R[G/C]$ specified by these orderings need not be connected on the left and right of the cut.  As a result, Eq.\ (\ref{eq:forwardscattering-cuts}) includes unphysical terms that are not found on the right-hand side of Eq.\ (\ref{eq:sigma-def-1}), which defines the cross section in terms of connected amplitudes.    In the next sub-section, we show how such contributions occur, and that they always cancel in the sum over time orders, as they must by the optical theorem.   Notice further that many of the final states $C$ on both sides of this relation do not contribute to the imaginary part, because they correspond to states for which $\lambda_s$ equals $0$ or $-1$ in Eq.\ (\ref{eq:lambda-s-def}).  The full cross section, of course, involves a leptonic tensor, and depends on the specific electroweak currents at the vertices we have labeled $i$ and $o$.    As noted above, for simplicity, we have suppressed these familiar factors, which play no direct role in our discussion, and refer to expressions like Eq.\ (\ref{eq:forwardscattering-cuts}) as cross sections.

We now consider a weighted lepton annihilation cross section where the sum over final states $C$ is weighted by a set of functions $f_C(p_i)$, which depend only on particle momenta appearing in state $C$. 
The relevant TOPT energy factors, denominators, and energy delta function for this weighted cross section can be organized as
\begin{eqnarray}
\Sigma[f,Q]\ &=&\   \sum_G\sum_{C}  \sum_{\tau_{L[G/C]}}\sum_{\tau_{R[G/C]}}\, \int d\cL_G \; \; \mathbb{N}_{\tau_G} \; \prod_{i=1}^{N_G}\frac{1}{2\omega_i}\, \sigma^{(C)}_{\tau_{L[G/C]} \cup \tau_{R[G/C]}}[f,\cL_G]\, ,
\nn\\[2mm]
\sigma^{(C)}_{\tau_G}[f,\cL_G,Q] &=&  \prod_{s=C+1}^{V_G-1}  \frac{i}{Q\lambda_s-\sum_{j\in  s} \omega_{j }-i\epsilon}\  f_{C}(\vec{q}_1\ldots \vec{q}_{k_C})
\nn \\[2mm]
&\  & \hspace{5mm}  \times\ (2\pi )\, \delta \left(Q-\sum_{j\in  C}\omega_{j}\right) \prod_{s=1}^{C-1}  \frac{i}{Q\lambda_s-\sum_{j\in  s} \omega_{j }+i\epsilon}\, ,
\label{eq:denominator-fs}
\end{eqnarray}
where we understand that $\tau_G=\tau_{L[G/C]} \cup \tau_{R[G/C]}$.
Here $f_C(\{\vec q_i\})$ is a weight function, depending in general on the spatial momenta, $\{\vec q_i\}=\{\vec{q}_1\ldots \vec{q}_{k_C}\}$ and masses of particles in the state $C$. This expression differs from the corresponding factor in the imaginary part of the forward scattering graph only by the weight function $f_C$, and setting $f_C=1$, gives back Eq. (\ref{eq:forwardscattering-cuts}).   At this point, we observe that the sum over time orders in Eq.\ (\ref{eq:denominator-fs}) can be thought of as generated from connected diagrams only, so that the sum does not include the disconnected diagrams encountered in Eq.\ (\ref{eq:forwardscattering-cuts}).   Because, as we observed above and will show in the next section, these contributions cancel among themselves, the sums are nevertheless effectively identical.

In what follows, we will study infrared safe weight functions, for which
\bea
f_{C}(\vec{q}_1,\ldots \vec q_i \dots \vec q_{j-1}, \xi \vec q_i, \vec q_{j+1}, \dots \vec{q}_{k_C})\ 
&=&\
\nn\\ [2mm]
&\ & \hspace{-25mm}
f_{C/j }(\vec{q}_1,\ldots (1+\xi)\vec q_i, \dots \vec q_{j-1},  \vec q_{j+1}, \dots  \vec{q}_{k_C})\, ,
\nn\\
\label{eq:irs}
\eea
for any real $\xi>-1$,
where $C/j$ denotes a state  with $k_C-1$ particles.   Such weights are unchanged by the emission of zero-momentum particles ($\xi=0$) or by the emission or recombination
of massless collinear particles.   This is a familiar condition, of course, which ensures the cancellation of soft and collinear (collectively, infrared) singularities.   Requirements on the manner in which the weight functions approach the equalities of Eq.\ (\ref{eq:irs}) are discussed in Refs.\ \cite{Sterman:1979uw,Komiske:2020qhg}.

Our  goal below is to exhibit an expression for such a weighted cross section in which the cancellations implied by Eq.\ (\ref{eq:irs}) can be made explicit, so that all cancellations take place in four dimensions, without infrared regularization.   First, however, we resolve the apparent difference between the sums over time-ordered diagrams in the unitarity condition, Eq.\ (\ref{eq:forwardscattering-cuts}), and in the corresponding expression for weighted cross sections, Eq.\ (\ref{eq:denominator-fs}).    

\subsection{Unphysical cuts in TOPT}
\label{sec:unphys}

Let us now turn our attention to unphysical cuts in time-ordered perturbation theory.  By unphysical cuts, we refer to states that contribute to 
 the forward-scattering diagram but which, when they go on-shell (that is, when the corresponding denominator vanishes), do not separate the
 diagram into connected diagrams to the right and left.   Such cuts of the diagram do not appear in the sum over final states in the optical theorem, Eq.\ (\ref{eq:optical}).
 What we will show is that the sum over all time-ordered diagrams that share any specific unphysical cut vanishes.  This result appears to be the analog in TOPT of the
 cancellation of so-called spurious singularities in loop-tree duality \cite{Capatti:2020ytd}.

\begin{figure}[h]
\begin{center}
  \includegraphics[width=5cm]{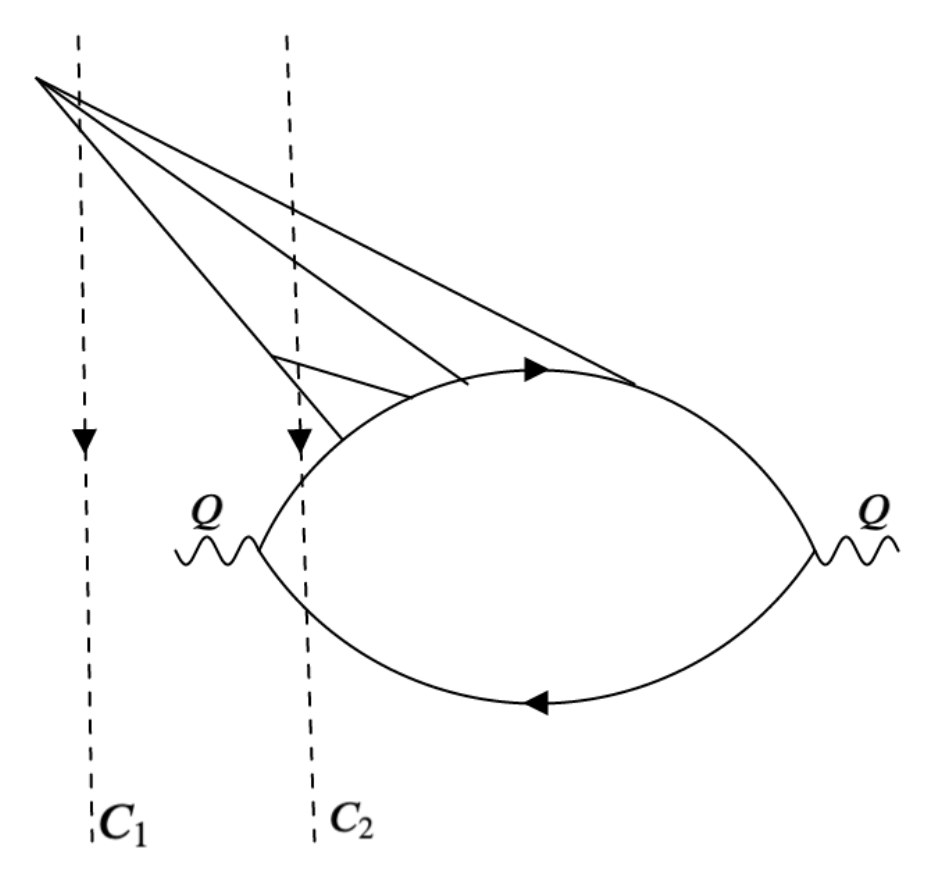}
  \caption{An example of a graph with unphysical cuts of two types. Here, no external momentum is ordered earlier than state $C_1$,  while momentum $Q$ enters prior to state $C_2$. The state $C_1$ is a vacuum cut and the state $C_2$ is a pseudo-physical cut.}
  \label{TOPT-up-example}
    \end{center}
\end{figure}

To establish some notation, we say the vertices are in the set $V=\{i,b_1,b_2,b_3,\ldots b_n,o\}$, and $i$, $o$ are the vertices where the external momentum $q$ flows in and out respectively. We take $(b_{P_1}, b_{P_2}, \ldots , b_{P_{n+2}})$ to represent an arbitrary time order labeled by the permutation $P$. 
The way in which unphysicsl cuts appear in TOPT is illustrated by the low-order example
in Fig.\ \ref{TOPT-up-example}.    In TOPT, intermediate states may include particles created from the vacuum, with no available external energy.   If such a state
precedes the vertex at which momentum flows into the diagram, as in state $C_1$ of the figure, the corresponding energy denominator is negative semi-definite, vanishing only if all particles  are massless, and then only on the set of zero measure where all particles carry zero spatial momentum. All denominators in states such as
this are of the form $\frac{i}{-\sum_j \omega_j+i\epsilon}$.  We shall refer to such states as ``vacuum states".

State $C_2$ in Fig.\ \ref{TOPT-up-example} describes the amplitude for the same production of particles from the vacuum,  which, however,
appears mixed with particles that share the energy flowing into the diagram.  
Denominators like those of state $C_2$ can be of the form $\frac{i}{Q-\sum_j\omega_j+i\epsilon}$.
The resulting mixed denominators can vanish in this configuration
and can contribute to the imaginary part of the specific time-ordered diagram. We refer to these states as ``pseudo-physical".

 We  now prove that the cross section evaluated on a pseudo-physical cut is identically zero. To do so, we start with the expression for the denominators, $s$, of the amplitude $a_{L[G/C]}\left(Q,\cL_{L[G/C]}\right)$, $C\ge s$ for any final state $C$ of diagram $G$, Eq.\ (\ref{eq:forwardscattering-cuts}) summed over its time orders,
\bea
2\pi \delta \left(Q-\sum_{j\in C}{\omega}_{j}\right)\, a_{L[G/C]}\left(Q,\cL_{L[G/C]}\right)
&=&\  2\pi \delta  \left(Q-\sum_{j\in C}{\omega}_{j}\right)
\nn\\[2mm]
&\ & \hspace{-20mm} \times \  \sum_{\tau_{L[G/C]}}\prod_{s=1}^{C}  \frac{i}{Q\lambda_s -\sum\limits_{j\in  s} \omega_{j }+i\epsilon}  \, ,
\label{eq:a-L-topt}
\eea
where we exhibit as well the energy-conserving delta function of final state $C$.  (Note that $C$ denotes both the final state and the integer label on the vertex that immediately precedes the final state.)  At this point, diagram $a_L$ may or may not be connected.   What we will show is that if it is not connected, the sum over its time orderings vanishes when state $C$ is on-shell.

A  useful technique in our analysis, here and below, will be  a representation of the denominators in terms of integrals over times,
\begin{equation}
2\pi \delta \left (Q - \sum_{j\in C}\omega_j \right)\, a_{L[G/C]}\,   =\, \sum_{P(L[G/C])}\prod_{\alpha_P=1}^{C}\int_{-\infty}^{t_{\alpha_P+1}}dt_{\alpha_P} \ 
 e^ {-i \left (\delta_{\alpha_P, i} Q \;
 + \; \eta^{(\alpha_P)}_j (\omega_j-i\epsilon) \right )\, t_{\alpha}}\, .
\label{eq:integral-form}
\end{equation}
Here, we have written the sum over time orders as the sum over all ordered permutations, $P(L[G/C])$, of the set of $C$ time variables $\{ t_{\alpha} \}$ associated with the vertices $\alpha$ of diagram $L[G/C]$.   For each order, $P$, we define $\eta^{(\alpha_P)}_{j}$ as the incidence matrix of the vertex 
$\alpha_P$, defined in a given time order by
\bea
\eta^{(\alpha_P)}_{j} \ &=& 1\, , \quad {\rm line}\ j\ {\rm enters}\ {\rm vertex}\ \a_P\, ,
\nn\\[2mm]
\eta^{(\alpha_P)}_{j} \ &=& -1\, , \  {\rm line}\ j\ {\rm leaves}\ {\rm vertex}\ \a_P\, ,
\nn\\[2mm]
\eta^{(\alpha_P)}_{j} \ &=& 0\, , \quad {\rm otherwise}\, .
\label{eq:incidence-def}
\eea
All lines are emitted and then absorbed, except for lines that appear in (final) state $C$.  As a result, we have in Eq.\ (\ref{eq:integral-form}), for any state $s\le C$,
\bea
\sum_{\alpha_P\le s} \eta^{(\alpha_P)}_j\ &=&\ 0\, ,\ j\, \rlap{/}{\in}\, s\, ,
\nn\\[2mm]
\sum_{\alpha_P\le s} \eta^{(\alpha_P)}_j\ &=&\ -1\, ,\ j \in s\, .
\label{eq:sum-eta-alpha}
\eea
The term $\delta_{\alpha_P, i} Q $ in Eq.\ (\ref{eq:integral-form}) contributes the energy flowing into the diagram at vertex $i$, which must always be in the diagram specified by the time order $\tau_{L[G/C]}$, for the final state $C$ to be on-shell.   Using this result and Eq.\ (\ref{eq:sum-eta-alpha}), we readily see that each $\alpha_P$ time integral produces the TOPT denominator in Eq.\ (\ref{eq:a-L-topt}) for the state immediately following it, times a phase whose exponent is proportional to the following time, $t_{\a_P+1}$.
We define $t_{C+1}=\infty$, for the final integral in each order, which produces the delta function that enforces the condition that the energy flowing in, $Q$, equals the energy flowing out ($\sum_{j\in C}\omega_j$).

For every ordering $P(L)$ in Eq.\ (\ref{eq:integral-form}), the energy-conservation delta function is generated by the final time integral, which is unbounded on both sides: $\int_{-\infty}^{\infty} dt_{C}$.   The vertex associated with this integral depends, of course, on the specific time order.  At this point, it will be useful to rewrite (\ref{eq:integral-form}) by introducing an auxiliary maximum time, $t_{max}$ that is the same for all orders.   We do this by using the  following identity for each order $P(L)$,
\bea
 \int_{-\infty}^{\infty} dt_{C}\,  
 e^ {-i \left (\delta_{C, i} Q \;
 + \; \eta^{(C)}_j (\omega_j-i\epsilon) \right )\, t_C}
  \prod_{\alpha =1}^{C-1} \int_{-\infty}^{t_{\alpha +1}} dt_\alpha\, 
  e^ {-i \left (\delta_{\alpha, i} Q \;
 + \; \eta^{(\alpha)}_j (\omega_j-i\epsilon) \right )\, t_{\alpha}}
  &\ &
  \nn\\[2mm]
&\ & \hspace{-100mm}  =\ \left [-i(Q-\sum\limits_{j\in C}\tilde{\omega}_{j}) \right ]
\, \int_{-\infty}^{\infty} dt_{max}\, \int_{-\infty}^{t_{max}} dt_{C}\, e^ {-i \left (\delta_{C, i} Q \;
 + \; \eta^{(C)}_j (\omega_j-i\epsilon) \right )\, t_C}
\nn\\[2mm]
&\ & \hspace{-95mm} \times\ 
\prod_{\alpha =1}^{C-1} \int_{-\infty}^{t_{i+1}} dt_\alpha\, e^ {-i \left (\delta_{\alpha_P, i} Q \;
 + \; \eta^{(\alpha_P)}_j (\omega_j-i\epsilon) \right )\, t_{\alpha}}\, .
\label{eq:tmax-trick}
\eea
On the right-hand side of this equation, the integral over $t_C$ produces a denominator that cancels the overall factor of $-i(Q-\sum\limits_{j\in C}\tilde{\omega}_{j})$ in Eq.\ (\ref{eq:tmax-trick}), while the new, $t_{max}$, integral reproduces the energy conservation delta function, of the same argument.
We can implement this identity for all time the orders, $P$ in Eq.\ (\ref{eq:integral-form}).  

In the case when the amplitude breaks up into two (or more) disconnected pieces, $G_1$ and $G_2$, we may write it as the product of two factors, with $n_1$ and $n_2$ vertices each, $n_1+n_2=C$, with a sum over time orders, or permutations $P^{(1)}$ and $P^{(2)}$, of the independent  disconnected vertices, whose times can be integrated independently to $t_{max}$ in Eq.\ (\ref{eq:tmax-trick}).   For definiteness, we assume that vertex $i$, at which external energy flows in, attaches to the subdiagram $G_1$, with $n_1$ vertices.   For such a diagram, we then have in place of Eq.\ (\ref{eq:integral-form}), the equivalent form
\bea
a_{L[G/C]}\left(Q, \cL_{L[G/C]} \right) 2\pi \delta \left(Q-\sum_{j\in C} {\omega}_{j}\right)
&=& \ [-i(Q-\sum_{j\in C} {\omega}_{j})] \int_{-\infty}^{\infty} dt_{max} 
\nn\\[2mm]
&\ & \hspace{-20mm} \times\ \left( \sum_{P^{(1)}} \prod_{\alpha=1}^{n_1} \int_{-\infty}^{t_{\alpha+1}}dt_{\alpha} e^{-i(\delta_{P^{(1)}_\alpha, i} Q  + \eta^{(P^{(1)}_\alpha)}_j (\omega_j-i\epsilon))t_{\alpha}}\right)
\nn\\[2mm]
&\ & \hspace{-20mm} \times\ \left( \sum_{P^{(2)}} \prod_{\alpha=1}^{n_2} \int_{-\infty}^{\tilde{t}_{\alpha+1}}d\tilde{t}_{\alpha} e^{-i(  \eta^{(P^{(2)}_\alpha)}_j (\omega_j-i\epsilon))\tilde{t}_{\alpha}}\right),
\label{eq:disconnected}
\eea
where now we have defined  $t_{n_1+1}=\tilde{t}_{n_2+1}=t_{max}$.   Note that because of this independence, the number of terms in the sums of permutations is $n_1!n_2!$, down from $C!=(n_1+n_2)!$.   

The connected subdiagrams each possess the  properties of the sums over vertices in Eq.\ (\ref{eq:sum-eta-alpha}).
As a result, the final time integral $t_{max}$ inherits the same energy-conservation phase, and the integrals in (\ref{eq:disconnected}) can be done explicitly, giving for each choice of $P^{(1)}$ and $P^{(2)}$, the factor
\bea
-i\left( Q-\sum_{j\in C}{\omega}_{j} \right) 2\pi \delta \left(Q - \sum_{j\in C}{\omega}_{j}\right)
\times \left(\prod\limits_{s_2=1}^{n_2}\frac{i}{-\sum\limits_{j \in s_2} {\omega}_j+i\epsilon}\right)
 \left(\prod\limits_{s_1=1}^{n_1} \frac{i}{Q\lambda_{s,1}-\sum\limits_{j \in s_1} {\omega}_j+i\epsilon}\right)\, ,
 \nn\\
 \label{eq:two-components}
\eea
where $s_1$ and $s_2$ label states within the two disconnected factors, and where $\lambda_{s,1}$ is defined by analogy to Eq.\ (\ref{eq:lambda-s-def}), this time
for the diagram with $n_1$ vertices, through which the external energy $Q$ flows.   Recall that in the case of an amplitude, $\lambda_s=1$ or $0$ only.
Reorganized in this fashion, the time integrals in Eq.\ (\ref{eq:disconnected}) and the denominators that appear in Eq.\ (\ref{eq:two-components}) lack a denominator that cancels the overall factor $Q-\sum_{j\in C}\omega_j$.
For fixed time orders of the full diagram, such a denominator is always present, but it cancels in the sum over time orders, and the integrand vanishes almost everywhere in phase space because energy is conserved.

In summary, we have found that in every term, Eq.\ (\ref{eq:two-components}) that results from the sum over the relative time orders of the two disconnected diagrams while keeping their internal time orders fixed, the energy conservation delta function is multiplied by its own argument.
It is  easy to check that when the argument of the delta function vanishes, the cut denominators are generically  finite (except in a region with vanishing measure) while the quantity $\left(Q-\sum_{j\in C}\tilde{\omega}_{j}\right)$ multiplies the delta function, forcing the integral over the phase space to zero.

\section{Local Infrared safety}
\label{sec:local-irs}

Our interest here is in IR safe weighted cross sections, inclusive cross sections in which the IR singularities of their exclusive channels cancel among themselves. In this section, we construct a general expression that implements this cancellation locally in momentum space.   In principle, this can eliminate the need for infrared regularization. 

\subsection{Reorganized cross sections}

 When integrated over loop momenta as it stands, the arbitrary weighted cross section Eq.\ (\ref{eq:denominator-fs}) is a sum of infrared divergent terms in general, which, however, cancel in the sum over final states $C$ for each time ordering $\tau_G$.   To make this cancellation manifest, we use the distribution identity, 
\bea
2\pi \delta(x)=\frac{i}{x+i\epsilon}-\frac{i}{x-i\epsilon}
\label{eq:delta-identity}
\eea 
to rewrite the TOPT expression for a general weighted cross section, $\Sigma[f,Q]$, Eq.\ (\ref{eq:denominator-fs}).  Recalling the identity for sums over states in Eq.\ (\ref{eq:C-tau-sums}), we express $\Sigma[f,Q]$ as
\bea
\Sigma[f,Q] &=&   \sum_G\ \sum_{\tau_G}\,  \int d\cL_G  \; \mathbb{N}_{\tau_G} \;  \prod_{i=1}^{N_G}\frac{1}{2\omega_i}\;  \sum_{C} \, \sigma^{(C)}_{\tau_{L[G/C]} \cup \tau_{R[G/C]}}[f,\cL_G]\, .
\label{eq:Sigma-f-sum}
\eea
In this expression, we sum over all cuts of the vacuum diagram $G$ at fixed time-order $\tau_G$ and have used the independence of $\mathbb{N}_{\tau_G}$ from the choice of $C$.   
As noted above, the full set of cuts of any diagram will include in general unphysical cuts, that is, cuts that include
disconnected subdiagrams to the left and/or right of the cut.   In the previous section we have shown, however, that all such terms cancel once the sum over time orders is carried out.   In fact, in Sec.\ \ref{sec:weighted-cs} we will
show that we can re-express the cross section in a diagrammatic form in which all such cuts are absent.   For now,
however, we continue with this expression, in the knowledge that further cancellations will occur in a result that we will show is already infrared finite.

   We now apply the identity, Eq.\ (\ref{eq:delta-identity}) to the integrand of (\ref{eq:Sigma-f-sum}), using (\ref{eq:denominator-fs}) to get,
\bea
\sum_{C=1}^n\; \sigma^{(C)}_{\tau_{L[G/C]} \cup \tau_{R[G/C]}}[f,\cL_G,Q]\ &=&  \sum_C\ \prod_{s=C+1}^{n+1}  \frac{i}{Q\lambda_s-\sum_{j\in  s} \omega_{j }-i\epsilon}\  f_{C}(\vec{q}_1\ldots \vec{q}_{k_C})
\nn \\[2mm]
&\  & \hspace{0mm}  \times\, \left ( \frac{i}{Q\lambda_s-\sum_{j\in  C}\omega_{j} + i\epsilon} \, -\, \frac{i}{Q\lambda_s-\sum_{j\in  C}\omega_{j} - i\epsilon} \right) 
\nn\\[2mm]
&\ & \times \ \prod_{s=1}^{C-1}  \frac{i}{Q\lambda_s-\sum_{j\in  s} \omega_{j }+i\epsilon}\, .
\label{eq:denominator-fs-expand}
\eea
Then, simply collecting terms with  a fixed denominator structure yields,
\bea
\sum_C\; \sigma^{(C)}_{\tau_{L[G/C]} \cup \tau_{R[G/C]}}[f,\cL_G,Q]\  &=&\   \left(\prod_{s=1}^{n+1}  \frac{i}{Q\lambda_{s}-\sum_{j\in  s} \omega_{j }+i\epsilon}f_{n+1} \right.
\nn\\[2mm]
&\ & \left.\hspace{-30mm}  +\ \sum_{C=1}^{n} \prod_{s=C+1}^{n+1}  \frac{i}{Q\lambda_{s}-\sum_{j\in  s} \omega_{j }-i\epsilon} (f_{C}-f_{C+1}) 
\prod_{s=1}^{C}  \frac{i}{Q\lambda_{s}-\sum_{j\in  s} \omega_{j }+i\epsilon} \right.
\nn\\[2mm]
&\ & \left. \hspace{-20mm} -\ \prod_{s=1}^{n+1}  \frac{i}{Q\lambda_{s}-\sum_{j\in  s} \omega_{j }-i\epsilon} f_1 \right)\, .
\label{eq:reor}
\eea
In this equality, we have suppressed the arguments of the functions $f_C$, which are the momenta of individual particle lines, and hence linear combinations of the spatial loop momenta of diagram $G$.
We notice that if all weight functions $f_C$ were indeed equal  to unity, we would get back the imaginary part of the forward scattering graph. The first and last terms in this expression have the analytic structure of the vacuum polarization diagram and its complex conjugate and hence are individually IR finite \cite{Sterman:1978bj}.   Although Eq.\ (\ref{eq:reor}) is a simple reorganization of a standard expression, we will show that it provides a locally finite expression for the set of leptonic annihilation cross sections under consideration so long as the weight function satisfies the usual criteria for infrared safety in Eq.\ (\ref{eq:irs}).  We are not aware of such an expression in the previous literature.

In Eq.\ (\ref{eq:reor}), the sum of terms labeled $C=1\dots n$ are somewhat unusual, having no energy-conserving delta function.   Rather, they are given entirely by products of denominators with opposite $i\epsilon$ prescriptions.  We will refer to the product of denominators with $+i\epsilon$ as the ``generalized amplitude", and the product with $-i\epsilon$ as the ``generalized conjugate amplitude".   In the following subsection, we will study how infrared singularities can arise in TOPT generally, and in the product of generalized amplitudes and complex conjugates, and go on to verify that the sum in Eq.\ (\ref{eq:reor}) is infrared safe locally in momentum space without the need for infrared regularization.

\subsection{Analysis of pinch surfaces}
\label{sec:pinch surfaces}

The individual terms in Eq.\ (\ref{eq:reor}) have the same infrared singularities that are found in cross sections for fixed final states, which have explicit energy-conservation delta functions.   In that case, singularities can be identified  from solutions to the Landau equations for the amplitude and complex conjugate for each point in final-state phase space.   These solutions are satisfied on subspaces of momentum space sometimes referred to as pinch surfaces \cite{Sterman:1995fz}.

The analysis leading to pinch surfaces in amplitudes can be applied to the 
spatial integrals in Eq.\ (\ref{eq:Sigma-f-sum}) for each choice of order, $\tau_G$, and cut, $C$ in the form of Eq.\ (\ref{eq:reor}), to derive the Landau equations.   
We begin by combining state denominators via a Feynman parametrization, chosen so that the imaginary parts all add with the same sign at every point in parameter space.   This requires that we factor out a $(-1)$ for each denominator in the generalized conjugate amplitude, with a $-i\epsilon$,
\bea
&& \int d\cL_G \; \frac{\mathbb{N}_{\tau_G}}{\prod_{i=1}^{N} 2\omega_i } \left( f_C - f_{C+1} \right)\; 
\prod_{s=C+1}^{n+1}  \frac{i}{Q-\sum_{j\in  s} \omega_{j }-i\epsilon}\ 
\prod_{s=1}^{C}  \frac{i}{Q-\sum_{j\in  s} \omega_{j }+i\epsilon}
\nn\\[2mm]
&& \hspace{0mm}=\ \int  \frac{d\cL_G }{\prod_{i=1}^{N} 2\omega_i } \; \frac{ \mathbb{N}_{\tau_G}\, \left( f_C - f_{C+1} \right)\, (-1)^{n-C} \, [d\alpha_s]_{n+1}}{\left(  \sum_{s=1}^C \alpha_s \left( Q-\sum_{j\in  s} \omega_{j } \right)
- \sum_{s=C+1}^{n+1} \alpha_s \left( Q-\sum_{j\in  s} \omega_{j }  \right) +i\epsilon \right )^{n+1}}\, ,
\nn\\
\label{eq:Landau-1}
\eea
where, as usual,
\bea
[d\alpha_s]_{n+1}\ =\ n!\, \int_0^1 d\alpha_{n+1} \dots d\alpha_1\, \delta \left( 1 -\sum_{s=1}^{n+1} \alpha_s \right)\, .
\eea
We would like to show that for any infrared safe weight function $f$, Eq.\ (\ref{eq:Landau-1}) is finite without infrared regularization.   To do so, we must identify the origin
and strength of IR singularities in these TOPT expressions.

As in covariant perturbation theory, infrared singularities in Eq.\ (\ref{eq:Landau-1}) can arise whenever the loop integrals are pinched between coalescing singularities.
In this case, of course, we have three integrals per loop remaining.   Singularities arise from two sources, the product of particle energies $\omega_i$, and from
the full parameterized denominator.  We note that each energy factor, $\omega_i$ depends quadratically on the spatial loop momenta of particle $i$, and always produces
a pinch at the point $\omega_i=0$, for line $i$ massless.  

The identification of pinches from sets of on-shell denominators with lines of nonzero energy in Eq.\ (\ref{eq:Landau-1}) requires a
 time-ordered perturbation theory version of Landau equations, which follow the same pattern as for integrals in covariant perturbation theory.   In the
parameterized form, an off-shell denominator $D_s\ne 0$ must have $\alpha_s=0$.   Derivatives with respect to each loop momentum component of denominators with $D_s=0$ 
 must vanish,
 \bea
\frac{\partial}{\partial l^\mu}\, \left [ \sum_{s=1}^C \alpha_s \left(  \sum_{i \in  s} \omega_i \right)
- \sum_{s=C+1}^{n+1} \alpha_s \left(  \sum_{j\in  s}  \omega_j  \right ) \right ]\ =\ 0\, .
\label{eq:Landau-1a}
\eea
 Because the derivative of the energy $\omega_i$ of a line with respect to its momentum gives its velocity, $\vec \beta_i= \partial\omega_i / \partial \vec p_i$, the Landau equations
are given as linear sums in velocities.   For an arbitrary loop momentum $l$, we can thus write
\bea
 \sum_{s=1}^C \alpha_s \left(  \sum_{i \in  s} \eta_{l,i}\, \vec \beta_{i} \right)
- \sum_{s=C+1}^{n+1} \alpha_s \left(  \sum_{j\in  s} \eta_{l,j}\, \vec \beta_{j }  \right )\ =\ 0\, ,
\label{eq:Landau-2}
\eea
with the $\eta_{l,i,}=\pm 1,0$ incidence matrices,
\bea
\eta_{l.i} \ &=& \ \frac{\partial p_i^\mu}{\partial l^\mu} \quad {\rm any}\ \mu\, .
\eea
To be specific, we define the momenta $p_i^\mu$ to be in the direction of energy flow, so that $p_i^0=\omega(\vec p_i) \ge 0$.  Note that for the amplitude this is the direction toward the final state, $C$.   

The equations (\ref{eq:Landau-2}) can be satisfied for any loop that
appears in a subset of denominators $\{t_i\}$, $i=1\dots k$, with $t_1 \le t_C$ and $t_k \ge C+1$, because such terms include denominators with both ``$i\epsilon$" prescriptions.   A loop whose on-shell states appear only with $+i\epsilon$ or only with $-i\epsilon$ cannot give a solution to Eq.\ (\ref{eq:Landau-2}), {\it unless} all lines that carry this loop have collinear momenta.   We can think of such a loop as internal to a jet of collinear moving particles.   These solutions can be given a physical interpretation in the sense of Coleman and Norton \cite{Sterman:1995fz,Coleman:1965xm}, by identifying the Feynman parameters $\alpha_s$ as times so that their products with velocities are translations.    The Landau equation for any loop  then describes a sum of translations relative to the origin, in the direction of the velocities of lines carried by that loop.  

For loops that are internal to a given jet, their contributions to Eq.\ (\ref{eq:Landau-2}) cancel within each state, because $\eta_{li}=-\eta_{li'}$ if the line flows forward along line $i$ within the jet and back along line $i'$.   For loops that extend over states with $s<C$ to states with $s>C$, Eq.\ (\ref{eq:Landau-2}) requires that the contribution of states in these two categories cancel for the two jets individually.   Thus, for the jet along which the loop flows forward, we have
\bea
 \sum_{s=1}^C \alpha_s
- \sum_{s=C+1}^{n+1} \alpha_s\ =\ 0\, ,
\label{eq:Landau-3}
\eea
which has the interpretation that the jet flows forward away from the origin through a sequence of states up to state $C$, and then backward to the origin for the same amount of time in the same direction.
Thus, for such terms, the {\it only} momentum configurations that produce pinch singularities are those in which on-shell states differ only by the rearrangement of collinear momenta and the emission or absorption of lines with zero momentum.   These states are characterized by ``jets" of fixed total momentum, accompanied by arbitrary numbers of soft lines.   The succession of on-shell states with different numbers of jets always results in at least one internal loop of the amplitude or complex conjugate that flows between two jets.   For such a loop, the two jet velocities in Eq.\ (\ref{eq:Landau-2}) appear times parameters with only positive or only negative signs, which is inconsistent with a pinch.   The only exception is for loops that carry zero momentum between different jets.   Such ``soft" loops are unconstrained by the Landau equations, in both TOPT and covariant perturbation theory, although as noted above, there is a pinch whenever any line carries exactly zero momentum \cite{Collins:2020euz}.

We now turn to the role of off-shell states.
Let us refer to the full set of states $\{s_i\}$ for time order $\tau_G$ of vacuum polarization diagram $G$ as ${\cal S}=\{s_1\dots s_{n+1}\}$.   At a given pinch surface,  states in ${\cal S}$ are either on-shell, with $Q-\sum_{j\in s_i}\omega_j= 0$, or off-shell, $Q-\sum_{j\in s_i}\omega_j\ne 0$.  These simple considerations limit how off-shell states can appear at a pinch surface.   First of all, states adjacent to the external vertices ($i$ and $o$ above) at which momenta flow into and out of the forward-scattering diagram, can always be off-shell.  These states correspond to the ``hard part" of the scattering cross section.   For an arbitrary pinch surface, $\zeta$ there is thus an ``earliest" state, $s^{[\zeta]}_{\rm min}\ge s_1$, at which the relevant set of jets first appears, and a ``latest" state, $s^{[\zeta]}_{\max} \le s_{n+1}$, in the complex conjugate amplitude, where they last appear.   

We next examine when a subset of states can be off-shell at pinch surface $\zeta$.  Let us denote such an ordered subset as $\Gamma^{[\zeta]}=\{ \s^{[\zeta]}_i\}\subset {\cal S}$,  $\s^{[\zeta]}_i=\s^{[\zeta]}_{\rm min} \dots \le \s^{[\zeta]}_{\rm max}$, and assume that all of these  states are off-shell, that is, $Q-\sum_{j\in \s^{[\zeta]}_i}\omega_j\ne 0$, and consecutive.   There may be a number of these sets; first, let us consider the case with only a single such set, $\Gamma^{[\zeta]}$.
For any such set of off-shell states, $\G^{[\zeta]}$, we must have two sets of on-shell states:  on-shell states $\{ s^{[\zeta]}\}_{\Gamma<}$ with $s^{[\zeta]}_{\rm min} \le s^{[\zeta]} <\s^{[\zeta]}_{\rm min}$ that are before the off-shell states, and similarly  another set of on-shell states,   $\{s^{[\zeta]}\}_{ \Gamma >}$, all of whose elements are between $\s^{[\zeta]}_{\rm max}$ and $s^{[\zeta]}_{\rm max}$.   

Because states are ordered,  the off-shell states of $\Gamma^{[\zeta]}$ may be either entirely in the generalized amplitude or in the generalized conjugate amplitude, or may extend  between them both.  In all cases, however, one of the two sets of states, $\{s^{[\zeta]}\}_{\Gamma <,}$ and $\{s^{[\zeta]}\}_{\Gamma_>}$  will have only $+ i\epsilon$ or only $-i\epsilon$ in its denominators.   
We use these considerations to show that at the pinch surface, $\zeta$, no sets of lines that appear in the states of $\Gamma^{[\zeta]}$
can carry finite momentum between any pair of distinct jets of pinch surface $\zeta$.   

To see why, consider the case that $\Gamma^{[\zeta]}$ is in the amplitude (all $+i\epsilon$ in its denominators).  We suppose the two jets in question have velocities $\vec \beta_a$ and $\vec \beta_b$, in different directions.
Let us suppose that some set of lines carries finite momenta and connects these two jets.   We will then be able to find at least one loop consisting of lines within the two jets in all the on-shell states of $\{s^{[\zeta]}\}_{\Gamma<}$, closing the loop with the lines of finite momentum that appear in the states of $\Gamma^{[\zeta]}$ and lines that appear only in the off-shell states, $s\le s^{[\zeta]}_{\rm min}$.  The Landau equation for this loop would have no contributions from the states in $\Gamma^{[\zeta]}$ or from states with $s\le s^{[\zeta]}_{\rm min}$, because the Feynman parameters of off-shell states are all zero.  They do of course get contributions from the on-shell states, $\{s^{[\zeta]}\}_{\Gamma<}$. We can always define the loop in question to flow through one line of jet $a$ in the direction of the $a$-jet energy flow for each on-shell state.   The velocity of all such lines is the same as the jet velocity, $\vec \beta_a$.  The loop then flows back, against the direction of the $b$-jet, for which the corresponding particle velocities are all $\vec \beta_b$.    The resulting Landau equations, Eq.\ (\ref{eq:Landau-2}), are then 
\bea
\sum_{s\in \{s^{[\zeta]}\}_{\G<}} \alpha_s\, \left( \vec \beta_a- \vec \beta_b \right)\ =\ 0\, .
\label{eq:Landau-ab}
\eea
 Because $\vec \beta_a$ and $\vec \beta_b$ are in different directions, there can be no solution to Eq.\ (\ref{eq:Landau-ab}) for nonzero parameters $\alpha_s$.   As a result, $\zeta$ is not a pinch surface after all.   The generalization of this result to sets $\Gamma^{[\zeta]}$ in the complex conjugate amplitude or between the amplitude and conjugate is immediate.  Several sequences of off-shell states also follow the same pattern.

The only remaining possibility for an off-shell $\Gamma^{[\zeta]}$ is one for which some set of off-shell states is associated with  one or more {\it internal} loop momenta of a jet, which are  not in the direction of the jet.  Such a loop will also take a set of consecutive states $\s^{[\zeta]}$ off-shell.   We will therefore need to consider this possibility in our discussion of local finiteness.

\subsection{Logarithmic singularities and cancellation}

To identify which pinch surfaces result in actual infrared singularities in individual terms, we must recall the power counting analysis of covariant perturbation theory \cite{Sterman:1978bi,Collins:2011zzd}.   Rather than repeating the details of this analysis, we can rely on the basic result.   For a pinch surface of a cut vacuum polarization diagram, we identify the space of ``normal"  variables, which parameterize the space perpendicular to the pinch surface, and assign a dimensionless scaling variable (conventionally denoted by $\lambda$) so that each normal variable vanishes linearly as the scaling variable vanishes.   The fundamental starting point of this analysis, when applied to leptonic annihilation, which shows that singularities are at worst logarithmic, can be summarized as 
\bea
2L_J+4L_S+p_{\rm num} -N_J-2N_S\ \ge\ 0\, ,
\label{eq:pc-start}
\label{eq:covariant-pc}
\eea
where $L_J$ and $N_J$ are the total number of loops and lines in jet subdiagrams, respectively, while $L_S$ and $N_S$ are the soft loops and lines, and $p_{\rm num}$ is the scaling dimension of numerator factors at the pinch surface in question.   We will
see an example below.

We can apply Eq.\ (\ref{eq:covariant-pc}) to TOPT expressions like Eq.\ (\ref{eq:Landau-1}) in a straightforward fashion.
First, using the graphical Euler identity in the form
\bea
L_J+L_S\ =\ N_J+N_S - V + 1\, ,
\label{eq:Euler}
\eea
 we see immediately that Eq.\ (\ref{eq:covariant-pc}) may be written equivalently as
\bea
L_J+3L_S+p_{\rm num} -N_S - (V-1)\ \ge\ 0\, ,
\label{eq:to-pc}
\eea
where here $V$ is the total number of vertices in the on-shell TOPT diagram found by contracting all states that are off-shell at this pinch surface.  This is also the number of states that are pinched on-shell.  In this expression, of course, $L_J$ remains the total number of internal loops of all jets, and $L_S$ is the number of loops that carry zero momentum in all three remaining components.  For leptonic annihilation \cite{Sterman:1978bi}, the natural normal variables are all three remaining components of the $L_S$ soft loops, $\vec l_{\rm soft}\sim\lambda$, and for each jet loop momentum $\vec l_{\rm jet}$,  $l_{\perp,\, {\rm jet}}^2 \sim \lambda$.   The dependence of all denominators is then linear in $\lambda$.   Because the number of denominators is $V-1$, the inequality of Eq.\ (\ref{eq:to-pc}) ensures that divergences in TOPT are at worst logarithmic, just as in covariant perturbation theory. 

Notice that it is necessary to saturate the inequality (\ref{eq:to-pc}) for a specific time order and pinch surface to contribute to an infrared divergence.  This eliminates infrared divergences for time orders where a vertex that connects on-shell jet lines and/or soft lines appears between off-shell denominators, simply because choosing such an order sacrifices at least one on-shell state for an off-shell state.    For this reason, in identifying pinch surfaces, we may capture their leading infrared behavior by their reduced diagrams, found by shrinking any loop momenta internal to the jets but not in the jet direction to points.   The resulting time-ordered reduced diagrams automatically have the maximum number of on-shell denominators.   
The logarithmic nature of infrared divergences applies to each term in the original expression of the weighted cross section, Eq.\ (\ref{eq:denominator-fs}) as well as to its linear combinations in Eq.\ (\ref{eq:reor}) and (\ref{eq:Landau-1}) for an arbitrary weight function, $f_C$.  

In summary, a general weighted cross section encounters at worst logarithmic singularities in individual terms in the sum found by combining Eqs.\ (\ref{eq:Sigma-f-sum}) and (\ref{eq:reor}),
\bea
\Sigma[f,Q] &=&   \sum_G\ \sum_{\tau_G}\,  \int d\cL_G  \; \mathbb{N}_{\tau_G} \;  \prod_{i=1}^{N_G}\frac{1}{2\omega_i}
\nn\\
&\ &\ \times  \left(\prod_{s=1}^{n+1}  \frac{i}{Q\lambda_{s}-\sum_{j\in  s} \omega_{j }+i\epsilon}f_{n+1} \right.
\nn\\[2mm]
&\ & \left.\hspace{0mm}  +\ \sum_{C=1}^{n} \prod_{s=C+1}^{n+1}  \frac{i}{Q\lambda_{s}-\sum_{j\in  s} \omega_{j }-i\epsilon} (f_{C}-f_{C+1}) 
\prod_{s=1}^{C}  \frac{i}{Q\lambda_{s}-\sum_{j\in  s} \omega_{j }+i\epsilon} \right.
\nn\\[2mm]
&\ & \left. \hspace{0mm} -\ \prod_{s=1}^{n+1}  \frac{i}{Q\lambda_{s}-\sum_{j\in  s} \omega_{j }-i\epsilon} f_1 \right)\, .
\label{eq:Sigma-f-final}
\eea
In this form, we can use the properties of infrared safe weight functions $f_C$, Eq.\ (\ref{eq:irs}) to show that this expression is locally finite in loop momentum space if the sum over $C$ is carried out before integration.   

We first observe that
the first and third terms in parentheses are infrared finite because their denominators can produce no pinches, just as for the total cross section.
For the remaining terms in the sum, which involve denominators with both $i\epsilon$ signs, let us start by considering a ``leading" pinch surface, at which every state in Eq.\ (\ref{eq:Sigma-f-final}) is pinched on-shell.   Each pinch surface (leading or not) is at worst logarithmically divergent in the integral over normal variables.   Differences $f_C-f_{C+1}$ then need only vanish as any power of the normal variables for the $\cL_G$ integrals to be finite.  This is precisely the condition for IR safety found in Refs.\ \cite{Sterman:1979uw,Komiske:2020qhg}. 
For any infrared safe weight function, by Eq.\ (\ref{eq:irs}), the value of any $f_C$ is the same for every state that is pinched on shell.   This  ensures that the difference $f_C-f_{C+1}$ in Eq.\ (\ref{eq:Sigma-f-final}) vanishes as a power of the normal variables.   The integral in Eq.\ (\ref{eq:Sigma-f-final}) is then finite for all leading pinch surfaces.     
To extend this result to pinch surfaces with intermediate off-shell states, we simply note that for an off-shell state $\sigma^{[\zeta]}$, the corresponding denominator $D_{\sigma^{[\zeta]}}$ is real, and the terms proportional to $f_{\sigma^{[\zeta]}}$ cancel between conjugate terms in (\ref{eq:Sigma-f-final}).   The sum of all terms in Eq.\ (\ref{eq:Sigma-f-final}) is thus finite at an arbitrary pinch surface.   This is what we set out to demonstrate.

It is worth noting that a special class of event weights are those that count jet final states.   These weights take the value unity when the state satisfies some criterion, and zero elsewhere.  For IR safe jet algorithms, the weights will be identical for all states associated with a given pinch surface \cite{Sterman:1978bi}.   For such states, the differences $f_C-f_{C+1}$ are exactly zero for the full range of momentum space where the weight function is unity for both states, a range that includes all pinch surfaces.

\subsection{All-order example: energy in a ``cone"}

It is useful to give an explicit example that illustrates the local finiteness of Eq.\ (\ref{eq:reor}) for a specific infrared safe cross section at all orders.
To this end, we begin by introducing an abbreviated notation for energy denominators,
\bea 
D_s\ =\ E_{s}-\sum_{j\in  s} \omega_{j }\, ,
\label{eq:D-s-def}
\eea
where in our case, $E_s=Q\lambda_s$, with $\lambda=\pm1,0$.
In these terms, a contribution to the integrand for a general weighted cross section, Eq.\ (\ref{eq:reor}) becomes
\bea
\sum_{C=1}^{n+1}\, \sigma^{(C)}_{\tau_G}[f,\cL_G,Q] &=&\ \left(\prod_{s=1}^{n+1}  \frac{i}{D_s+i\epsilon}f_{n+1} 
+\ \sum_{C=1}^{n} \prod_{s=C+1}^{n+1}  \frac{i}{D_s-i\epsilon} (f_{C}-f_{C+1}) 
\prod_{s=1}^{C}  \frac{i}{D_s+i\epsilon} \right.
\nn\\[2mm]
&\ & \left. -\ \prod_{s=1}^{n+1}  \frac{i}{D_s-i\epsilon} f_1 \right)\, .
\label{eq:reor-D}
\eea
To illustrate how Eq.\ (\ref{eq:reor-D}) provides a locally finite sum,  we consider the example of a ``cone" weight, defined as the energy flowing into a cone fixed in space,
\bea
f_s^{(\Omega)}\ = \ \sum_{i\in_s \Omega} \omega_i\, .
\label{eq:cone-weight}
\eea
Here, $i \in_s \Omega$ means that particle $i$ flows into angular region $\Omega$ in state $s$.  For simplicity, we consider angular region $\Omega$ to
be fixed, so that this quantity is not a jet cross section in the usual sense.   Clearly, this weight for state $s$ is related to the denominator $D_s$ by
\bea
f_s^{(\Omega)}\ =\ Q\, -\, \sum_{i\rlap{/}\in_s \Omega} \omega_i\, -\, D_s \, ,
\label{eq:D-s-f-s} 
\eea
where $ \sum_{i\rlap{/}\in_s \Omega}$ sums the on-shell energies of all the particles of state $s$ that are outside cone $\Omega$.
This leads to a relation between weights in consecutive final states,
\bea
f_C^{(\Omega)} - f_{C+1}^{(\Omega)} \ &=&\  D_{C+1}\, -\, D_C\, +\, \sum_{i\rlap{/}\in_{C+1} \Omega} \omega_i\ -\  \sum_{i\rlap{/}\in_C \Omega} \omega_i
\nn\\[2mm]
&\equiv& D_{C+1}\, -\, D_C\, +\, \delta^{(\Omega)} \omega_C\, ,
\label{eq:f-s-to-deltaomega}
\eea
where in the first relation we use (\ref{eq:D-s-f-s}), and in the second we define the quantity
$\delta^{(\Omega)}\omega_C$.  As we see from its definition, $\delta^{(\Omega)} \omega_C$  counts the energy of particles that are radiated outside the cone in the transition from state $C$ to state $C+1$,  or that was outside the cone in state $C$, but whose energy is absorbed into the cone in the transition. We then have, from the general relation, Eq.\ (\ref{eq:reor-D}),
\bea
\sum_{C=1}^{n+1}\, \sigma^{(C)}_{\tau_G}[f^{(\Omega)},\cL_G,Q] 
&=&\ \left(\prod_{s=1}^{n+1}  \frac{i}{D_s+i\epsilon} \left( Q\, -\, \sum_{i\rlap{/}\in_{n+1} \Omega} \omega_i\, -\, D_{n+1} \right) \right.
\nn\\[2mm]
&\ & \left.  
+\ \sum_{C=1}^{n} \prod_{s=C+1}^{n+1}  \frac{i}{D_s-i\epsilon} \left ( D_{C+1}\, -\, D_C\, +\, \delta^{(\Omega)} \omega_C \right )  \prod_{s=1}^{C}  \frac{i}{D_s+i\epsilon} \right.
\nn\\[2mm]
&\ & \left. -\ \prod_{s=1}^{n+1}  \frac{i}{D_s-i\epsilon} \left( Q\, -\, \sum_{i\rlap{/}\in_{1} \Omega} \omega_i\, -\, D_{1} \right) \right )\, .
\label{eq:reor-f-to-D}
\eea
On the right, all terms $D_i$ cancel, using the distribution identity,
\bea
D_i\, \left( \frac{i}{D_i - i\epsilon}\, -\, \frac{i}{D_i + i\epsilon} \right) \ =\ D_i\, 2\pi\delta \left( D_i \right )\ =\ 0\, .
\eea
We thus have for the cone cross section an expression that depends on energies flowing into the cone in the first and last state, and on the energy transfer variables, $\delta\omega_C$,
\bea
\sum_{C=1}^{n+1}\, \sigma^{(C)}_{\tau_G}[f^{(\Omega)},\cL_G,Q] &=&\ \prod_{s=1}^{n+1}  \frac{i}{D_s+i\epsilon} \left( Q - \sum_{i\rlap{/}\in_{n+1} \Omega} \omega_i \right ) \ -\  \prod_{s=1}^{n+1} \frac{i}{D_s - i\epsilon} \left( Q - \sum_{j\rlap{/}\in_1 \Omega} \omega_j \right )
\nn\\
&\ & \hspace{5mm} +\ \sum_{C=1}^{n} \prod_{s=C+1}^{n+1}  \frac{i}{D_s-i\epsilon} \, \delta^{(\Omega)} \omega_C\, 
\prod_{s=1}^{C}  \frac{i}{D_s+i\epsilon}\, .
\label{eq:reor-cone-final}
\eea
For this expression, we recall that at an arbitrary pinch surface, all finite energy is carried in jets of exactly collinear particles.   For any such configuration, if state $C$ and state $C+1$ are both on-shell, $\delta^{(\Omega)} \omega_C=0$ because consecutive states have exactly the same energy flow.  That is, the quantities $\delta^{(\Omega)} \omega_C$ vanish for every such final state $C$.  The contributions of these terms are integrable.   

For nonzero values of any set of $\delta^{(\Omega)} \omega_C$ in Eq.\ (\ref{eq:reor-cone-final}), following the general argument of the previous subsection, we consider a set of states $\Gamma^{[\zeta]}$,  that are off-shell at an arbitrary pinch surface $\zeta$, due to loop momenta circulating in the subdiagram of a jet or between soft lines.  As above, on-shell states appear both before and after $\Gamma^{[\zeta]}$.  Since all jets must appear with the same total cone energy in each on-shell state at the pinch surface, all nonzero contributions to $\delta^{(\Omega)}\omega_C$ must cancel as we sum over off-shell states $\Gamma^{[\zeta]}$.  Thus, the product of off-shell denominators associated with states $\Gamma^{[\zeta]}$ is multiplied by a sum of energy transfers $\delta^{(\Omega)} \omega_C$ that cancel at the pinch surface.

It is worth noting that in the special case where $\Omega$ grows to the entire two-sphere, there is no radiation outside the cone and all the $\delta^{(\Omega)}\omega_C$ are zero,
 and we find 
 \bea
\sum_{C=1}^{n+1}\, \sigma^{(C)}_{\tau_G}[f^{(S_2)}] &=&\ Q\,   \left [ \prod_{s=1}^{n+1}  \frac{i}{D_s+i\epsilon}  \ -\  \prod_{s=1}^{n+1} \frac{i}{D_s - i\epsilon} \right ]\, .
\label{eq:reor-f-4pi}
\eea
This, as expected, is the energy times the imaginary part of the uncut TOPT diagram, in our notation, the total cross section.    

\subsection{Low-order example}

\begin{figure}[h]
\begin{center}
  \includegraphics[width=5cm]{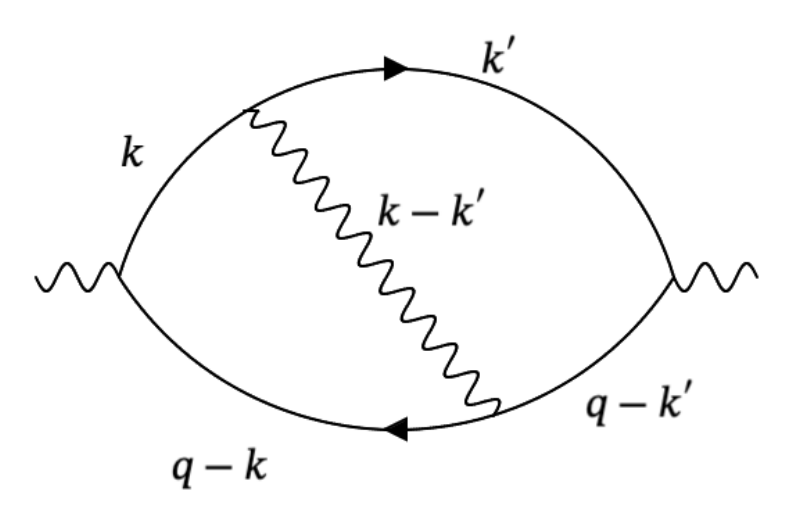} \quad \quad  \includegraphics[width=5cm]{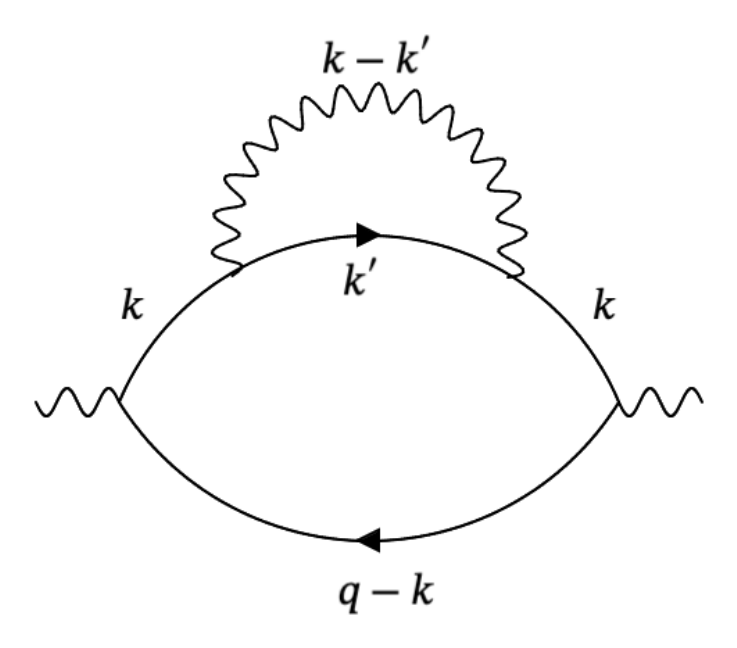}\\
  (1) \hspace{57mm} (2)
  \caption{Two loop vacuum polarization diagrams: (1) vector exchange, (2) fermion self-energy.}
  \label{fig:alphas-example}
    \end{center}
\end{figure}

To illustrate the formalism further with an explicit example, we consider the two next-to-lowest-order vacuum polarization diagrams in Fig.\ \ref{fig:alphas-example}.

For simplicity, we take the trace on the external vector current indices, and we do not consider color in our examples.  From the discussion above, the contributions of these diagrams to a general weighted cross section are given by
\bea
\Sigma_i[f,Q]
\ &=&\
ig^2\, \int \frac{d^3 \vec k}{(2\pi)^3}\, \frac{d^3 \vec k'}{(2\pi)^3} \frac { \mathbb{N}_i } { 2^4 \omega_k\, \omega_{q-k}\, \omega_{k'}\,  \omega_{k-k'} }\, \left ( \delta_{i1}\, \frac{1}{2\omega_{q-k'}} + \delta_{i2}\, \frac{1}{2\omega_k} \right )
\nn\\[2mm]
&\ &  \times \left[ \frac{f_{iA}}{ D^+_{iC} \, D^+_{iB} \, D^+_{iA} } \ -\ \frac{f_{iC}}{ D^-_{iC} \, D^-_{iB} \, D^-_{iA }} 
\ + \ \frac{f_{iB} - f_{iA}} { D^-_{iC} \, D^-_{iB} \, D^+_{iA} } \ +\ \frac{f_{iC} - f_{iB} }{ D^-_{iC} \, D^+_{iB} \, D^+_{iA} } \right ]\, ,
\nn\\
\label{eq:sigma_i}
\eea
where $i=1$ refers to the gluon exchange diagram and $i=2$ to the fermion self energy diagram.
\footnote{The case where the two-particle weights $f_A$ and $f_C$ and the three-particle state weight $f_B$ are defined to give
unity for a two-jet final state and zero otherwise, is of interest.   An example is the original cone jet cross section of Ref.\ \cite{Sterman:1977wj} at this order.
  Using the symmetry between the two-particle states, $A$ and $C$, the expression, Eq.\ (\ref{eq:sigma_i})  is readily seen to reduce to the total cross section minus the
three-jet cross section, as noted early on in Ref.\ \cite{Stevenson:1978td}, due to the exact cancellation of three- and two-particle momentum configurations in the two-jet region.
At this order, the resulting expression needs no infrared regularization.  }

In terms of the momentum assignments shown in the figure, energy denominators in Eq.\ (\ref{eq:sigma_i}) are given by
\bea
D^\pm_{1A}\ &=&\ D^\pm_{2A}\ =\ Q-2\omega_k \pm i\epsilon
\nn\\[2mm]
D^\pm_{1B}\ &=&\ D^\pm_{1B}\ =\ Q-\omega_k-\omega_{k'}-\omega_{k-k'} \pm i\epsilon\, ,
\nn\\[2mm]
D^\pm_{1C}\ &=&\ Q-2\omega_{k'} \pm i\epsilon\, , 
\nn\\[2mm]
D^\pm_{2C}\ &=&\ Q - 2\omega_{k} \pm i\epsilon\, .
\eea
The numerators simplify significantly in TOPT, where we must evaluate line momenta on-shell, $k^2=k'{}^2=0$,
and the $\mathbb{N}_i$ in Eq.\ (\ref{eq:sigma_i}) become
\bea
\mathbb{N}_1\ &=&\  -32 \left( k'\cdot (k-q) \right) \left (k\cdot (k'-q) \right) \, ,
\nn\\[2mm]
\mathbb{N}_2\ &=&\ -32 \, k \cdot k' \, q\cdot k \, .
\label{eq:N2}
\eea

To exhibit the pinch surfaces of these diagrams, we can choose spherical coordinates, with $\vec k$ in the $\hat z$ ($\theta_k=0$) direction, and the polar angle, $\theta'$ of $\vec k'$ measured relative to $\hat z$.   It is now natural to use the notation $k=|\vec k|=\omega_k$, and $k'=|\vec k'|=\omega_k'$,
but to leave $\omega_{k-k'}$ to represent $|\vec k - \vec k'|$, which is given by
\bea
\omega_{k-k'}\ &=& \sqrt{k^2+k'{}^2 - 2kk'\cos\theta'}
\nn\\[2mm]
&=& \sqrt{(k-k')^2 +2kk'(1-\cos\theta')}\, .
\eea
We assume that our weight functions can be expressed in terms of these variables only.

In Eq.\ (\ref{eq:sigma_i}), the first and second terms in square brackets are free of collinear pinches altogether, simply because their
energy denominators enter with the same sign for the $i\epsilon$s.   
We denote the remaining, potentially singular, terms as $\hat\Sigma_1$ and $\hat\Sigma_2$.   For $\hat\Sigma_1$, 
corresponding to gluon exchange, Fig.\ \ref{fig:alphas-example}(1), we have
\bea
\hat \Sigma_1[f,Q]
\ &=&\
i\, \frac{g^2}{8\pi^4}\, \int_0^\infty\, dk\, dk'\, \int_{-1}^1 d\cos\theta'\;
 \frac { k'\cdot (q-k)\, k\cdot(q-k') } { \omega_{k-k'} }
\nn\\[2mm]
&\ &  \times \left[   \frac{f_{1B}(k,k') - f_{1A}(k)} { (Q-2k'-i\epsilon) \, (Q- k-k'-\omega_{k-k'} - i\epsilon)\, (Q-2k +i\epsilon ) } \right.
 \nn\\[2mm]
&\ & \left.  +\ \frac{f_{1C}(k') - f_{1B}(k,k') }{(Q-2k'-i\epsilon) \, (Q- k-k'-\omega_{k-k'} + i\epsilon)\, (Q-2k +i\epsilon )}  \right ]\, ,
\nn\\
\label{eq:sigma1}
\eea
where we have indicated the momentum dependence of the weight functions.
For non-zero numerators, this integral has both collinear and soft pinches.  There is a collinear pinch at $k=Q/2,\, \theta'=0$, with
$|\vec k|\ > |\vec k'|$, between the second and third denominators of the first term in square brackets.   Similarly, a pinch appears at $k'=Q/2,\, \theta'=0$
with $|\vec k'|\  > |\vec k|$ between the first and second denominators of the second term in square brackets.  
These limits correspond to the gluon parallel to the quark (when $k=Q/2$) or the antiquark (when $k'=Q/2$). 
The soft pinch,
at which all three denominators vanish, appears in both
terms when $k$ and $k'$ both approach $Q/2$ with $\theta'=0$.   

All of these pinches lead to logarithmically divergent power counting, corresponding to two
denominators vanishing while two variables  ($k$ or $k'$ and $\theta'$) approach the endpoints of their integration regions. 
In all three cases, of course, the differences of the weight functions in the numerators
vanish, so long as they depend only on overall energy flow.   A simple example would be a weight function like the thrust,
for which $f_A=1$, while $f_B\sim 1-2k\cdot (k-k')/Q^2 \sim 1- 2kk'(1-\cos\theta')/Q^2$ near $\theta'=0$.
 
For $\hat\Sigma_2$, since the states $A$ and $C$ are identical for diagram 2, we can write
\bea
\hat \Sigma_2[f,Q]
\ &=&\
i\, \frac{g^2}{8\pi^4}\, \int_0^\infty\, dk\, dk'\, \int_{-1}^1 d\cos\theta'\;
 \frac { k k'{}^2 Q(1-\cos\theta') } { \omega_{k-k'} }
\nn\\[2mm]
&\ &  \times \left[   \frac{f_{2B}(k,k') - f_{2A}(k)} { (Q-2k-i\epsilon) \, (Q- k-k'-\omega_{k-k'} - i\epsilon)\, (Q-2k +i\epsilon ) } \right.
 \nn\\[2mm]
&\ & \left.  +\ \frac{f_{2A}(k) - f_{2B}(k,k') }{(Q-2k-i\epsilon) \, (Q- k-k'-\omega_{k-k'} + i\epsilon)\, (Q-2k +i\epsilon )}  \right ]\, .
\nn\\[2mm]
&=& \
i\, \frac{g^2}{8\pi^4}\, \int_0^\infty\, dk\, dk'\, \int_{-1}^1 d\cos\theta'\;
 \frac { k k'{}^2 Q(1-\cos\theta') } { \omega_{k-k'} }
\nn\\[2mm]
&\ &  \times \left[   \frac{f_{2B}(k,k') - f_{2A}(k) } { (\omega_{k-k'} - (k-k')  )^2 } (2\pi i) \delta \left( Q- k-k'-\omega_{k-k'} \right )\right]\, .
\label{eq:sigma2}
\eea
In the second relation, we have used the delta function to evaluate the squared denominator, $Q-2k$.  This denominator is negative semi-definite and vanishes only at the integration endpoint, $\cos\theta'\to 1$, where $k'$ is collinear to $k$.   Although it has an apparent double pole at $2k=Q$, the explicit numerator factor of $1-\cos\theta'$ reduces this to a single pole, and logarithmic power counting.   This relative suppression is a contribution ``$+1$" to the term $p_{\rm num}$ to infrared power counting in Eq.\ (\ref{eq:covariant-pc}), applied to the pinch surface where the self-energy consists of two collinear lines.    Here again, in Eq.\ (\ref{eq:sigma2}), the difference of weight functions in the numerator vanishes for any infrared safe weight and renders the integral finite.   This confirms that in the original form of the integration, the two- and three-particle singularities cancel.   
For this example, the suppression appears locally in loop momentum space, with the standard form of the self energy subdiagram.  At higher orders, however, the suppression requires in general integration over the internal loop momentum of the uncut self-energies  ($k'$ here), to realize their contributions to the suppression factor $p_{\rm num}$ in Eq.\ (\ref{eq:covariant-pc}).   For diagrams with more than a single self-energy on cut lines, we believe it will be natural to use alternative integrands for self-energies and their counterterms, which eliminate higher-order poles for the single-particle final state, as described for amplitudes in Refs.\ \cite{Anastasiou:2020sdt,Anastasiou:2022eym}.

\section{Posets in TOPT}
\label{Poset and caus} 

We now return to the question of pseudo-physical cuts in TOPT.   We recall that a pseudo-physical cut of a time-ordered graph disconnects the diagram into more than two connected parts.
In Sec \ref{sec:unphys}, we saw that the cross section evaluated on any pseudo-physical cut vanishes upon summing over time orders for the fixed cut.   In this section, we develop a poset formalism in order to show how pseudo-physical cuts can be avoided entirely.  We note that posets have also been useful in discussions of eikonal exponentiation \cite{Dukes:2013gea} and coordinate-space
amplitudes \cite{Erdogan:2017gyf}.

The treatment that follows has much in common with the recent discussion of ``flow-oriented" \cite{Borinsky:2022msp} and ``cross-free" representations of perturbation theory \cite{Capatti:2022mly}.  Here, we work directly from TOPT to find a number of related results, which, we believe, will provide intuition for applications, one of which we discuss in Sec.\ \ref{sec:weighted-cs}.  

 In this section, we will review some standard poset terminology, introduce our method, and show how it works in representative examples involving vacuum polarization diagrams. A general discussion will be provided in Sec.\ \ref{Formula}.

\subsection{Definitions}
\label{sec:def}

   The method we will use to reorganize TOPT 
  is based on the construction of  partially ordered sets (posets) on the vertices of the diagrams. 
  To do so, we impose a binary relationship among the vertices, which partitions the set of time orders into distinct posets.   A poset is a set together with a binary relationship.
  For any TOPT diagram, our binary relation can be defined from the incidence matrix introduced in the integral for an arbitrary  amplitude in Eq.\ (\ref{eq:integral-form}) and defined in Eq.\ (\ref{eq:incidence-def}).      We denote the incidence matrix by $\eta^{(b)}_{j}$, where the superscript $(b)$ represents a vertex and the subscript $j$ represents a line.  Entry $\eta^{(b)}_{j}$ is $+1$ if the line $j$ enters vertex $b$, $-1$ if the line $j$ exits vertex $b$ and, zero otherwise. Consider a TOPT diagram, given in the notation of Eq.\ (\ref{eq:integral-form}).
   If there is a line joining vertices $b_i,b_j$, and $t_{b_i}  \geq t_{b_j}$, for each line $k$ connecting both $b_i,b_j$, $\eta^{(b_i)}_k=1=-\eta^{(b_j)}_k$. On the other hand, if $t_{b_j} \geq t_{b_i}$ for each line $k$ between $b_i,b_j$, $\eta^{(b_j)}_k=1=-\eta^{(b_i)}_k$.

 We begin by defining 
an ordering between the vertices $\geq$, $V=\{b_1,b_2,b_3\ldots ,b_n\}$ of a Feynman graph $G$.  It satisfies:
\begin{itemize}
\item If  $b_i$ and $b_j$ are connected directly by one or more lines, then either $b_i \geq b_j$ or $b_j \geq b_i$, i.e., $b_i$ and $b_j$ are ordered by $\geq$.
The ordering $\ge$, abstracted from time ordering is transitive, $a\ge b,\ b\ge c\ \rightarrow a\ge c$.   It follows from the transitivity of the binary that if two vertices $b_i$ and $b_j$ are related by a sequence of increasing times, they are ordered by $\geq$.
\item If $b_i$ and $b_j$ are not connected by a line or a sequence of vertices that are increasing, then $b_i \not\geq b_j$ and $b_j \not\geq b_i$,  i.e., $b_i$ and $b_j$ are not ordered by $\geq$.   It will be useful to introduce the notation $b_i \sim b_j$, to mean $b_i \not\geq b_j$ and $b_j \not\geq b_i$.
\footnote{We shall assume that our TOPT diagrams do not include lowest-order tadpole subdiagrams, in which a single line emerges and is absorbed at a single vertex.  Such lines, which are removed by normal ordering, cannot be assigned a poset (or time) order.}
\end{itemize}

 We will say that $b_i > b_j$ if $b_i \geq b_j$ and  $b_i \neq b_j $ i.e., we require that $b_i, b_j$ are distinct in addition to being related. 
 
Having introduced the binary relation, we are ready to define the posets. 
A poset $D$ is the ordered pair $D= (V, \geq ) $, where the set $V$ is the set of vertices of a Feynman graph, taken together with the binary relationship $\geq$.

  It is clear that multiple time orders are compatible with a given poset, and in each time order, we can uniquely identify an underlying poset. Therefore, posets partition the set of all time orders.

 Our posets have the same information as the incidence matrix in the TOPT amplitudes, Eqs.\ (\ref{eq:integral-form}), (\ref{eq:sum-eta-alpha}).  It is clear that it is possible to read off a poset, given an incidence matrix.  Also notice that since the integrand in Eq.\ (\ref{eq:integral-form}) only depends on the incidence matrix, this is also fixed by the poset.  Therefore, posets are in one-to-one correspondence with incidence matrices.

 Let us make a few more definitions that relate to posets in order to develop this language more fully. An idea that we will find use for in what follows is that of minimal/maximal elements.  An element $b_i$ is said to be a minimal element of  a poset $D$ if  $\forall b_j \in D, b_j \geq b_i$ or $b_j \sim b_i$. A sequence of reducing vertices, starting from any vertex, always ends at a minimal element.

An element $b_i$ is said to be a maximal element of  a poset $D$ if  $\forall b_j \in D, b_i \geq b_j$ or $b_i \sim b_j$. A sequence of increasing vertices, starting from any vertex, always ends at a maximal element.

We will also find it useful to distinguish how two elements are related to each other. In particular if $a < b$, then either there exists a single line between the vertices $a$ and $b$ or there exists a sequence of increasing, connected vertices that starts at $a$ and ends  at $b$. It will be useful to reserve the symbol $\prec$ for the first scenario when $a$ and $b$ are directly connected through a line.  It is therefore natural to define a ``covering relation".
For elements $x,y \in V$, we say that $y$ covers $x$ if $y>x$ and there is no $z \in V$ such that $y>z>x$. We will denote it by $x \prec y$.
In a given graph, if $x \prec y$, then $x,y$ are vertices that are directly connected by one or more lines, while $y \geq x$ includes the possibility that $x,y$ are vertices that are connected by a sequence of lines. 

Having defined our posets, we will now go on to show how to use them to eliminate pseudo-physical cuts in some explicit examples of vacuum polarization diagrams.  A generalization to all processes at arbitrary loops will be given subsequently in Sec.\ \ref{Formula}.  

\subsection{Examples in vacuum polarization diagrams}\label{sec:examples}

Let us now look at some low order examples that will show
how the time integrals over minimal and maximal vertices can be carried out explicitly within a given poset. For definiteness, our examples will be drawn from the vacuum polarization diagrams of the lepton annihilation processes.  They illustrate features of quite general application and open the way to an analogous development for general amplitudes and Green functions in the next section.   

For vacuum polarization diagrams, we label vertices by permutations of the list, $V=(i,b_1,b_2, \dots, b_n,o)$, where the external momentum flows into the diagram at vertex $i$, and out at vertex $o$.   We can then divide all posets into four classes. 
\begin{itemize}
\item Posets with $o > i$  for which the external energy arrives at vertex $i$ before it leaves at vertex $o$.   
\item Posets with $i > o$  for which the external energy leaves at vertex $o$ before it arrives at vertex $i$
\item Posets with $o \sim i$, which split into two sub-classes by the imposition of an additional relationship in a time order: $t_o >t_ i$ and $t_i > t_o$. At the level of the poset, there is no definite order among the vertices $i(o)$, where the external energy enters(leaves). However, imposing the additional relationship $t_o>t_i$ restricts us to time orders within the poset where the external energy arrives at vertex $i$ before it leaves at vertex $o$.
\end{itemize}

 As we will see, there are unphysical cuts that one would like to eliminate in all four classes of posets. In what follows we will focus on time orders that satisfy $t_o >t_ i$. These are all time orders in posets with $o >i$, and in posets with $o \sim i$, with time orders $t_o>t_i$. These time orders carry at least one physical cut but contain both physical and unphysical cuts in general. There are no denominator singularities in time orders with $t_i > t_o$, and the integrand is completely real (and negative semi-definite).

Using the notation introduced in Eq.\ (\ref{eq:forwardscattering}), we consider here diagrammatic integrands, including numerator factors, written as sums over time orders $\tau_G$ of vacuum polarization integrands for an arbitrary diagram, $G$,
\bea
\pi_G[Q,\cL_G]\ &=& \sum_{\tau_G}\; \mathbb{N}_{\tau_G} \; \pi_{\tau_G}(Q,\cL_G)\, .
\label{eq:poset-denominator}
\eea
We observe that although the notation $\mathbb{N}_{\tau_G}$ suggests that the numerator factor  depends on the time order, it is the same for all time orders within a poset $D$. The numerator factor of a graph in covariant perturbation theory is derived from the Feynman rules and the numerator factor in TOPT is obtained by replacing the energy of a line $j$ by  the on-shell value of the energy $\pm \omega_j$, where the sign is determined by the direction of the line alone.  Therefore, the numerator factor is the same for all time orders within a given poset. We will henceforth use the notation $\mathbb{N}_{D}$ to emphasize the dependence of the numerator factor. 

Given that posets partition time orders into non-overlapping subsets, we may write $\pi_G$ in Eq. (\ref{eq:poset-denominator}) as
\bea
 \pi_G(Q,\cL_G)\ &=& \sum_{\tau_G}\; \mathbb{N}_{D} \; \pi_{\tau_G}(Q,\cL_G)
 \nn\\[2mm]
 &=& \sum_{D}\; \mathbb{N}_{D} \; \sum_{\tau_D}\pi_{\tau_D}(Q,\cL_G)\, ,
 \label{eq:tau-to-D}
\eea
where the second equality follows from the fact that the sum of all time orders $\tau_G$ of graph $G$ is the sum over posets, $D$, and within each poset the sum over time orders contained in $D$,
\be
\sum_{\tau_G}=\sum_{D} \sum_{\t_D} \; .
\label{eq:summation-identity}
\ee
Our interest is in obtaining  an expression for $\p_{D}(Q,\cL_G)$ that is free of pseudo-physical cuts. We first notice that if a poset $D$ has a single minimum $i$ (where the external momentum flows in), and a single maximum $o$ (where the external momentum flows out), every time order we induce on the poset exclusively carries physical cuts. To see this, we argue that in an arbitrary time order, and on any cut $C$ of that time order, we may start at vertex $v$ that lies to the left of $C$, and follow a sequence of vertices that are less than $v$ (in the poset ordering) to reach the unique minimum $i$. Therefore every vertex on the left of the cut $C$ is ordered with respect to $i$. A similar argument shows that every vertex to the right of such a cut is also ordered with respect to the unique maximum $o$.
There may or may not be more than one time order for such a poset, depending on the incidence matrix, but every cut of such a time-ordered diagram will be physical.

If, on the other hand, there is more than one extremum in a poset, the poset is guaranteed to include time orders that differ by exchanging the relative positions of the minima or maxima among themselves.
Our strategy in handling posets without a unique minimum or a unique maximum element is to reduce the problem to one where there exists a unique extremum.   This will involve combining time orders that exchange extrema and other unordered pairs of vertices.

\begin{figure}[h]
\begin{center}
  \includegraphics[width=5cm]{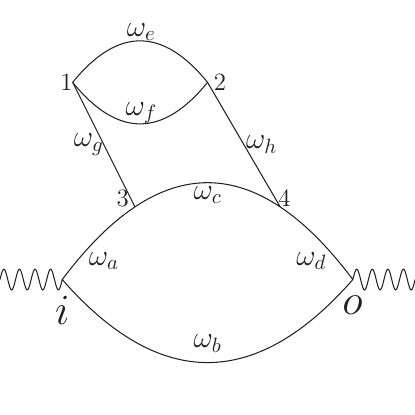}
  \caption{A three loop example of a graph with two minima. Vertices $1$ and $i$ are both minima of this graph, and our procedure eliminates the minimum $1$, through the process of successive integration. This graph is to be understood as a poset graph rather than a time-ordered graph.}
  \label{example1}
    \end{center}
\end{figure}

As a warm-up consider an example with a unique maximum but two minima, the three loop graph in Fig.\ \ref{example1}.   Although the figure shows a particular time order of its six vertices, $i$, $o$, and $b_1 \dots b_4$, it represents only the poset structure.   We recall that in TOPT every line carries energy forward in time (to the right in a TOPT diagram).   This information provides us with an incidence matrix, and hence a poset.   In this case, the relevant binary relations are
\bea
&\ & \hspace{10mm}  i<b_3<b_4<o\, ,
\nn\\[2mm]
&\ & \hspace{10mm} b_1 <b_ 2 < b_4\, ,
\nn\\[2mm]
&\ & \hspace{10mm} b_1 <b_3\, ,
\nn\\[2mm]
&\ & \hspace{10mm} b_1 \sim i\, ,
\nn\\
&\ & \hspace{10mm} b_2 \sim i,b_3\, .
\label{eq:poset-3-loops}
\eea
Within this poset, there are two minima, $b_1$ and $i$.  Vertex $b_1$ is covered by both $b_2$ and $b_3$.
There are five time orders.  Two time orders have $t_{b_2}>t_{b_3}$, which we label by permutation as
\bea
&\ & t_{b_1} < t_i \quad (1i324o)\, ,
\nn\\[2mm]
&\ & t_i < t_{b_1} \quad (i1324o)\, .
\label{eq:2>3}
\eea
The other three time orders, which have $t_{b_3}>t_{b_2}$, are given by
\bea
&\ & t_i < t_{b_1} < t_{b_2} < t_{b_3} \quad (i1234o)\, ,
\nn\\[2mm]
&\ & t_{b_1}< t_i < t_{b_2}<t_{b_3} \quad (1i234o)\, ,
\nn\\[2mm]
&\ & t_{b_1} < t_{b_2}<t_i<t_{b_3}  \quad (12i34o)\, .
\label{eq:3>2}
\eea
In TOPT, after time integration, each of these orderings gives one or more pseudo-physical cuts.   For example, the contribution of $(i1324o)$ can be written as
\bea
\pi_{(i1324o)}\ &=&\ e^{iQt_o}\; \int_{-\infty}^{t_o} dt_{b_4} e^{-i(\omega_{ch} -\omega_d+i\epsilon )t_{b_4}}  \int_{-\infty}^{t_{b_4}} dt_{b_2} e^{-i(\omega_{ef} - \omega_h+i\epsilon )t_{b_2}} \int_{-\infty}^{t_{b_2}} dt_{b_3} e^{-i(\omega_{ag} - \omega_f+i\epsilon )t_{b_3}} 
\nn\\[2mm]
&\ & \times\ \int_{-\infty}^{t_{b_3}} dt_{b_1} e^{-i( -\omega_{efg}+i\epsilon )t_{b_1}} \int_{-\infty}^{t_{b_1}} dt_i e^{-i(Q - \omega_{ab}+i\epsilon )t_i}
\nn\\[2mm]
&=&\ \frac{i}{Q-\omega_{bd}+i\epsilon}\, \frac{i}{Q-\omega_{bch}+i\epsilon}\, \frac{i}{Q-\omega_{bcef}+i\epsilon}   \frac{i}{Q-\omega_{abefg}+i\epsilon} \, \frac{i}{Q-\omega_{ab}+i\epsilon}\, ,
\label{eq:pi-topt-eg}
\nn\\
\eea
where for compactness we use the notation
\bea
\omega_{ab \dots c}\ =\ \omega_a+\omega_b+\dots + \omega_c\, .
\label{eq:omega-abc-def}
\eea
We have also set the outgoing energy $Q'$ to equal $Q$, to suppress the phase that provides the energy conserving delta function after the integral over 
$t_o$ in this case.
In this representative TOPT term, the denominator $Q-\omega_{abefg}$ corresponds to a pseudo-physical final state because the amplitude with this final state consists of two disconnected parts, one that includes vertex $i$ and provides particles $a$ and $b$ in the final state, and one in which particles $e$, $f$, and $g$ emerge from the vacuum into the final state.   We shall see, however, that such singularities are absent in an evaluation of integrals based on the poset.   In other words, pseudo-physical denominators all cancel in the sum over the time orderings that make up the poset.   This will be the case separately for the combinations of time orders in Eq.\ (\ref{eq:2>3}) and (\ref{eq:3>2}).

The case of the two orders in Eq.\ (\ref{eq:2>3}) is particularly simple.   Adding the two orders together, we find that times $t_{b_1}$ and $t_i$ integrate independently to $t_{b_3}$, and the pseudo-physical cut disappears
in TOPT, even though each of these orderings gives one or more pseudo-physical cuts,
\bea
\pi_{(i1324o)}+\pi_{(1i324o)} &=& e^{iQt_o}\; \int_{-\infty}^{t_o} dt_{b_4} e^{-i(\omega_{ch} -\omega_d+i\epsilon )t_{b_4}}  \int_{-\infty}^{t_{b_4}} dt_{b_2} e^{-i(\omega_{ef} - \omega_h+i\epsilon )t_{b_2}} 
\nn\\[2mm]
&\ & \times\ \int_{-\infty}^{t_{b_2}} dt_{b_3} e^{-i(\omega_{ag} - \omega_f+i\epsilon )t_{b_3}} 
\nn\\[2mm]
&\ & \times\ \int_{-\infty}^{t_{b_3}} dt_{b_1} e^{-i( -\omega_{efg}+i\epsilon )t_{b_1}} \int_{-\infty}^{t_{b_3}} dt_i e^{-i(Q - \omega_{ab}+i\epsilon )t_i}
\nn\\[2mm]
&\ &\hspace{-20mm} =\ \left (\frac{i}{Q-\omega_{bd}+i\epsilon}\, \frac{i}{Q-\omega_{bch}+i\epsilon}\, \frac{i}{Q-\omega_{bcef}+i\epsilon}\, \frac{i}{Q-\omega_{ab}+i\epsilon} \right )\, \frac{i}{-\omega_{efg}}\, ,
\label{eq:pi-poset-2>3}
\nn\\
\eea
where the rightmost fraction, which is negative semi-definite for massless lines, is the result of the integral $\int \limits_{-\infty}^{t_{b_3}} dt_{1}e^{-i(-\o_{efg} +i\e)t_{b_1}}$.  The remainder of the integral gives the factor in parentheses, consisting of four denominators, each of which provides a unitarity cut, corresponding to a physical final state.

\begin{figure}[H]
\begin{center}
\subfloat[\label{ex1-int-a}]{\includegraphics[width=5cm]{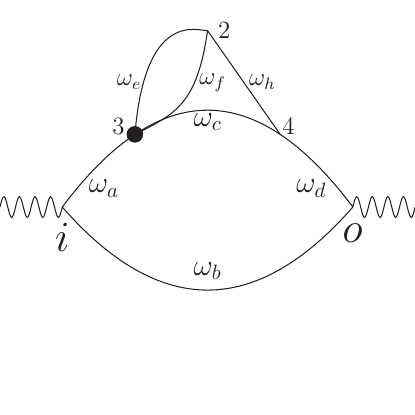}}\\
\subfloat[\label{ex1-int-b}]{\includegraphics[width=5cm]{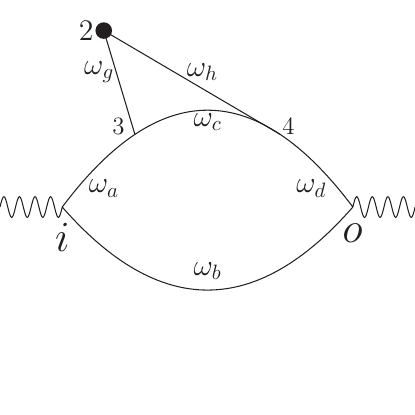}}
\subfloat[\label{ex1-int-c}]{\includegraphics[width=5cm]{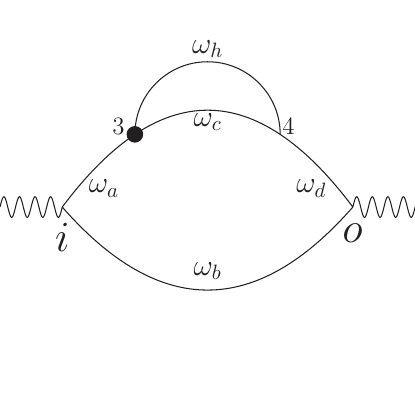}}
\end{center}
\caption{ 
    \protect\subref{ex1-int-a}
   The poset ordered graph in Fig.\ \ref{example1}. after integrating out the minimum vertex $1$, with the choice $2\geq 3$. Here the black dot represents the composite, modified vertex $3$.   
    \protect\subref{ex1-int-b}
   Integrating the time of vertex $1$, making the choice $3\geq 2$. The composite vertex $2$ is represented with a black dot. $2$ is still a minimum of the new poset.
    \protect\subref{ex1-int-c}
    The time of the composite vertex $2$, which was a new minimum in the choice $3\geq 2$, has been integrated. The new composite vertex $3$ is represented with a black dot. Here, all vertices lie between $i,o$.
    }
\label{example1-integration}
\end{figure}

A representation of the diagram after the $t_{b_1}$ integral is given in Fig.\ \ref{ex1-int-a}, which includes a modified vertex, which absorbs the fraction $\frac{i}{-\omega_{efg} }$.  This vertex, the denominator combined with coupling constants, is real and is effectively local in time (in this case at $t_{b_3}$).     At this stage,  all remaining vertices are ordered between $i$ and $o$ in the poset.   We have thus achieved what we set out to do, reduce the diagram in Fig.\ \ref{example1} to one with a unique maximum and unique minimum.  The cuts of the modified diagram in Fig.\ \ref{ex1-int-a} do not disconnect the graph into more than two connected subdiagrams, and are therefore unitarity cuts, that is, physical cuts of the forward scattering graph. 

We can now turn to the other possibility, $t_{b_3} > t_{b_2}$, the component of this poset given by the orders in Eq.\ (\ref{eq:3>2}).    
Rather than reproducing the four-dimensional time integral as in the previous case, we shall simply show the integrals over minimum vertices.
Examining the ranges of integrations possible in Eq.\ (\ref{eq:3>2}), we find that we can carry out the $t_1$ integral from $-\infty $ to $t_2$ independently of the value of $t_i$.   This gives the $t_2$-dependent factor
\be 
\int \limits_{-\infty}^{t_2} dt_{1}\; e^{-i
(-\o_{efg} + i\e)t_1}=\frac{i}{-\o_{efg}}\; e^{-i(-\o_{efg}+i\e)t_2}.
\ee
The composite graph obtained after integrating out vertex $1$ is  shown in Fig.\ \ref{ex1-int-b}.
In this case, the vertex $2$ is still a minimum of the poset, and we would like to carry out the $t_2$ integral as well.  We can do this because in the combination
of time orders in Eq.\ (\ref{eq:3>2}), the upper limit of the $t_2$ integral is $t_3$, independent of $t_i$.   We thus find,
 \be
  \int \limits_{-\infty}^{t_3} dt_{2} \; e^{-i
(-\o_{efg} + i\e)t_2} e^{-i(\o_{ef} -\o_h+i\e)t_2}=\frac{i}{-\o_{gh}} \; e^{-i(-\o_{gh} + i\e)t_3}.
 \ee
 A graphical representation of the diagram that results after this step is shown in Fig.\ \ref{ex1-int-c}. 
The complete result for the time orders in Eq.\ (\ref{eq:3>2}), analogous to Eq.\ (\ref{eq:pi-poset-2>3}) for the time orders of Eq.\ (\ref{eq:2>3}), is now easily found to be
\bea
\pi_{(i1234o)}+\pi_{(1i234o)}+\pi_{(12i34o)} &=& \left ( \frac{i}{Q-\o_{bd}+i\e}\, \frac{i}{Q-\o_{cbh}+i\e}\, \frac{i}{Q-\o_{ab}+i\e}\, \right ) \frac{i}{-\o_{gh}}\, \frac{i}{-\o_{efg}}\, .
\nn\\
\eea
 Together, the two negative semi-definite denominators combine with couplings to form the composite vertex 3 in Fig.\ \ref{ex1-int-c}.
Again, the remaining denominators are either physical (in parentheses) or negative semi-definite.
The full expression for this poset ($D$), $\pi_D$, is given by the sum of this result with Eq.\ (\ref{eq:pi-poset-2>3}).

\begin{figure}[H]
\begin{center}
  \includegraphics[width=6cm]{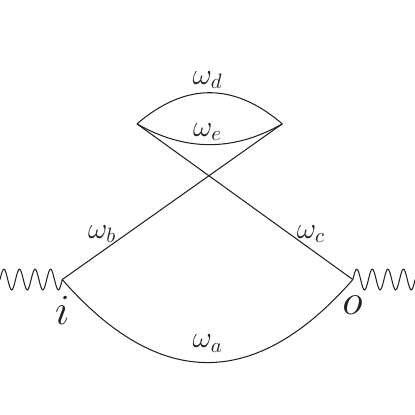}
  \caption{A two loop example of a graph with two minima and two maxima. Vertices $1$ and $i$ are both minima of this graph, and vertices $2$ and $o$ are both maxima of this graph.}
  \label{example2}
    \end{center}
\end{figure}
\begin{figure}[H]
\centering
\subfloat[\label{ex2-int-a}]{\includegraphics[width=5cm]{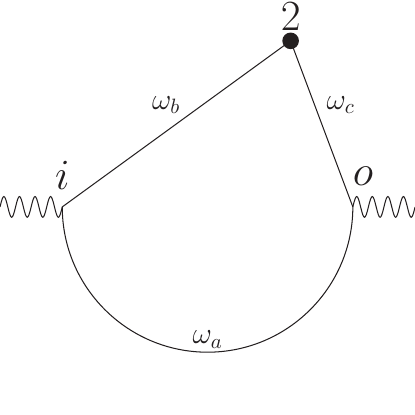}}
\subfloat[\label{ex2-int-b}]{\includegraphics[width=5cm]{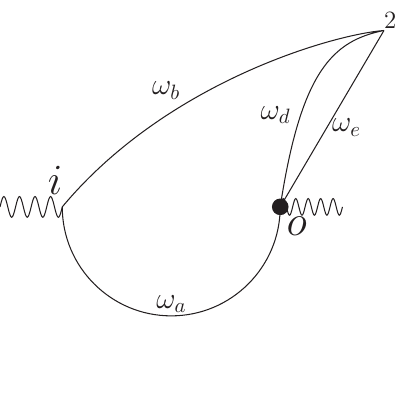}}
\subfloat[\label{ex2-int-c}]{\includegraphics[width=5cm]{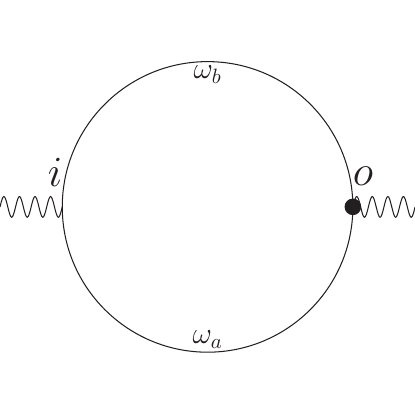}}
\caption{ 
\protect\subref{ex2-int-a}
The poset ordered graph in Fig.\ \ref{example2}. after integrating the time of the minimum vertex, $1$, with the choice $o\geq 2$. Here the black dot represents the composite, modified vertex $2$.
\protect\subref{ex2-int-b}
Integrating $t_1$, making the choice $2\geq o$. The composite vertex $o$ is represented with a black dot. Vertex $2$ is the new maximum of the poset. However, since the maximum is not $o$, we continue by integrating the time for vertex $2$.
\protect\subref{ex2-int-c}
The composite vertex $o$, after $t_2$ has been integrated. The new composite vertex $o$ is represented with a black dot.
}
\label{example2-integration}
\end{figure}

 
 For the next example, consider the two loop graph in Fig.\ \ref{example2}. In this example, at the level of the poset, there exists no unique minimum or unique maximum. 
Our process of eliminating the additional extrema starts with eliminating  the vertex we have labeled $1$. The minimum vertex, $1$ is connected to mutually un-ordered vertices $2$ and $o$. To integrate out the minimum $1$, we will order these vertices as before. The two choices in ordering are $t_o > t_2$ and $t_2 > t_o$, and as before, we must sum over both choices. Making the choice $t_o > t_2$ and integrating $t_1$, we obtain a unique minimum $i$ and a unique maximum $o$, giving
\be
 \int \limits_{-\infty}^{t_u} dt_{1} \;e^{-i(-\o_{dec}+i\e)t_1} =\frac{i}{-\o_{dec}} \; e^{-i(-\o_{dec}+i\epsilon)t_u},
\ee
 where the upper limit $t_u$ is $t_2$ if $t_o>t_2$ and  $t_o$ if $t_2>t_o$. Having made the choice $t_o>t_2$, the rest of the graph is fully ordered and we may use TOPT for the remaining denominators, 
\begin{equation}
\p_D^{(1)}= \left( \frac{i}{Q-\o_{ac}+i\epsilon}\, \frac{i}{Q-\o_{ab}+i\epsilon}\right )\, \frac{i}{-\o_{ced}}.
\end{equation}
The ordered graph that yields these denominators is shown in Fig.\ \ref{ex2-int-a}.

The choice of ordering $t_2 > t_o$, leaves us with a single maximum vertex $2$, which is not $o$. We would therefore like to  integrate the maximum time, $t_2$, from $t_o $ to $\infty$,
\be 
\int \limits^{\infty}_{t_o} dt_{2} \;e^{-i
(\o_{bed}-i\e)t_2} =\frac{i}{-\o_{bed}} \; e^{-i(\o_{bed} - i\e)t_o}.
\ee
This procedure is represented diagrammatically in Figs.\ \ref{ex2-int-b}, \ref{ex2-int-c}. What remains is the integral of $t_i$ in the range $-\infty$ to $t_o$ which is easily carried out. The contribution of the order $t_2>t_o$  to the poset denominator is  
\begin{equation}
\p_D^{(2)}= \left( \frac{i}{Q-\o_{ab}+i\epsilon}\right)\, \frac{i}{-\o_{bed}} \, \frac{i}{-\o_{ced}}\, .
\end{equation}
As before, the full expression for the integrand from this poset is the sum of the two choices in orderings included in the poset, $\p_D=\p_D^{(1)}+\p_D^{(2)}$.

\begin{figure}[H]
\begin{center}
  \includegraphics[width=6cm]{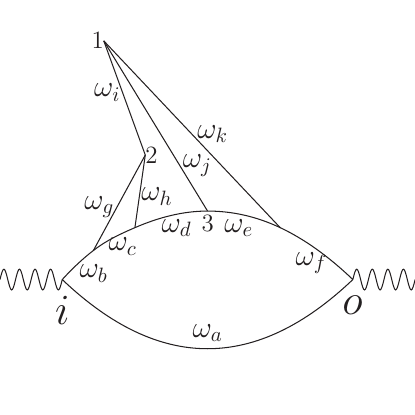}
  \caption{A four loop example of a graph with two minima and two maxima. Vertices $1$ and $i$ are both minima of this graph, and vertices $2$ and $o$ are both maxima of this graph.}
  \label{example3}
    \end{center}
\end{figure} 

\begin{figure}[H]
\begin{center}
\subfloat[\label{ex3-int-a}]{\includegraphics[width=5cm]{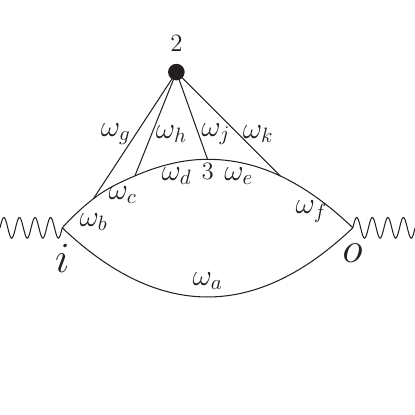}}
\subfloat[\label{ex3-int-b}]{\includegraphics[width=5cm]{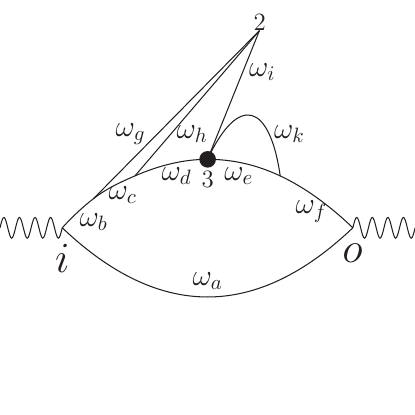}}
\subfloat[\label{ex3-int-c}]{\includegraphics[width=5cm]{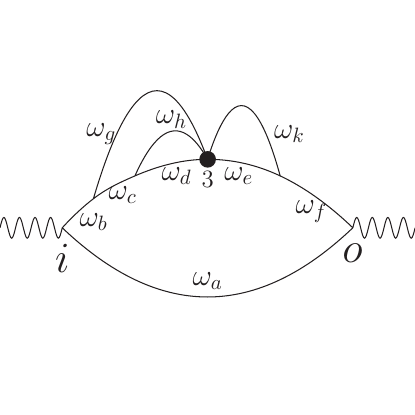}}
\end{center}
\caption{\protect\subref{ex3-int-a}
The poset ordered graph in Fig \ref{example3} after integrating  the time of the minimum vertex $1$, with the choice $3\geq 2$. Here the black dot represents the composite, modified vertex $2$.
\protect\subref{ex3-int-b}
Integrating  vertex $1$, making the choice $2\geq 3$. The composite vertex $3$ is represented with a black dot. Vertex $2$ is a  new maximum of the poset. However, since the maximum is not $o$, we continue integrating out vertex $2$.
\protect\subref{ex3-int-c}
The vertex $2$, which was the new maximum in the choice $2\geq 3$, after $t_2$ has been integrated. The new composite vertex $3$ is represented with a black dot. 
\label{example3-integration}
}
\end{figure}

As a final example, consider the poset graph in Fig.\  
\ref{example3}, which has both a minimum (vertex 1) and a maximum (vertex 2).  
This example captures our procedure at four loops.   In this poset, time $t_1$ is bounded from above by $t_2$ if $t_2<t_3$, and by $t_3$ if $t_3<t_2$.   In either case,  
\be
\int \limits_{-\infty}^{t_u} dt_{1} \;e^{-i
(-\o_{ijk} + i\epsilon)t_1} =\frac{i}{-\o_{ijk}} \; e^{-i(-\o_{ijk} + i\epsilon)t_u},
\ee
 where the upper limit $t_u$ is $t_2$ if $t_3>t_2$ and  $t_3$ if $t_2>t_3$.
Making the first choice, $t_3 > t_2$,  confines every remaining vertex to lie between $i \text{ and } o$. 
A graphical representation of this reduced graph is in Fig \ref{ex3-int-a}.  It is now fully ordered and as above, we can reconstruct the remaining integrals from the rules of TOPT, to find.
\begin{equation}
\pi_D^{(1)}=     \left (  \frac{i}{Q-\o_{af}}\,  \frac{i}{Q-\o_{kae}}\, \frac{i}{Q-\o_{adjk}}\,  \frac{i}{Q-\o_{ghad}} \,  \frac{i}{Q-\o_{gcd}}\, \frac{i}{Q-\o_{ab}} \right ) \frac{i}{-\o_{ijk}}\, ,
\end{equation}
where we have suppressed the $+i\epsilon$ terms present in each physical denominator.

Making the other choice, $t_2> t_3$, and integrating the minimum time, $t_1$ from $-\infty $ to $ t_3$ leaves $2$ as a maximum element.  The reduced graph we obtain at this stage is shown in Fig.\ \ref{ex3-int-b}. We can now integrate over $t_2$, 
\be 
\int \limits^{\infty}_{t_3} dt_{2} \;e^{-i
(\o_{igh} - i\epsilon )t_2} =\frac{i}{-\o_{igh}} \; e^{-i(\o_{igh}-i\epsilon)t_3}.
\ee
 The reduced graph we obtain is shown in Fig.\ \ref{ex3-int-c}. 
This is a fully ordered diagram, for which every vertex lies between $i$ and $o$. We can proceed using TOPT rules for the new diagram, giving the final result 
\begin{equation}
\pi_D^{(2)}=  \left( \frac{i}{Q-\o_{af}}\,  \frac{i}{Q-\o_{kae}}\, \frac{i}{Q-\o_{ghad}}\, \frac{i}{Q-\o_{gcd}}\, \frac{i}{Q-\o_{ab}} \right )\, \frac{i}{-\o_{igh}}\, \frac{i}{-\o_{ijk}} \, ,
\end{equation}
where again we suppress $+i\epsilon$ terms in all the denominators.
 The full poset integrand is once again $\p_D=\pi_D^{(1)}+\pi_D^{(2)}$.
 
 Let us summarize the process of integrating out of extrema that we have seen so far in these examples. We first identify a minimum vertex, $v$ ( $v\neq i$), and order the vertices that cover $v$. We then integrate out the minimum, at the cost of adding the energy carried by $v$ into the next vertex, say $v+1$. If there were more than one minimum, we will sequentially repeat these steps for all minima. We continue this process until each vertex that remains is greater than $i$.  We use an identical procedure with maxima, integrating them to generate negative definite denominators. After the elimination of both minima and maxima, the remaining vertices are all ordered relative to $i$ and $o$ and  therefore carry only physical cuts of the graph. 
 
 We have seen in these examples that the resulting ordered diagrams include composite vertices with multiple lines emanating from or flowing into them.  At each stage, the graph with composite vertices yields a new time-ordered graph, with fewer pseudo-physical cuts but the same physical cuts as in the original graph. The new graph thus obtained has fewer vertices and a modified incidence matrix. The composite vertex denotes a factorized expression for long time processes that were initiated in the vacuum and do not affect the cuts of the short distance process. The composite vertices may be interpreted intuitively as effective vertices in the short distance function, where all the long distance behavior has been absorbed into factorized, negative semi-definite, vacuum denominators.    We next turn to a general implementation of these methods.

\section{PTOPT:  TOPT without unphysical singularities} 
\label{Formula}

In this section, we describe a generalization to arbitrary order of the examples of the previous section, based on graphs' poset structures.  We will term this approach, ``partially time-ordered perturbation theory" (PTOPT).
 We will construct an algorithm that enables us to re-express the sum over time-ordered diagrams of any graph into a sum with fewer terms, all of whose denominators are either negative semi-definite, or physical.  In this construction, all pseudo-physical cuts will be eliminated.
We will present our discussion for an arbitrary amplitude with any numbers of incoming and outgoing momenta.  
Our result, given in Eq.\ (\ref{eq:final-general-amp})  below, is essentially equivalent to that given Ref.\ \cite{Capatti:2022mly}.   In the following section, we show how PTOPT can be applied to cross sections written as sums over cuts,
as in the weighted cross sections of Sec.\ \ref{sec:TOPT}, to provide expressions that involve only physical singularities, and which separate universal and process-dependent dynamics.

In this construction, it will be useful to turn our attention to integral representation Eq.\ (\ref{eq:integral-form}) for the time-ordered denominators of functions like $\pi_{\tau_G}$ in Eq.\ (\ref{eq:poset-denominator}), written as sums over posets.  
Compared to Eq.\ (\ref{eq:integral-form}), however, we associated with {\it every} vertex $\alpha$ an external energy $E_\alpha$, defined to flow into that vertex.  In a multiloop diagram, with a fixed number of external lines, most $E_\alpha$ are zero.  For an outgoing physical momentum flowing out of vertex $\alpha$, $E_\alpha$ is negative.
The resulting functions are then not limited to vacuum polarizations and may represent any scattering configuration, with arbitrary numbers of incoming and outgoing lines.   To 
emphasize the generality of these considerations, we denote our functions as $F_G(\{E_\a\},\cL_G)$ for graph $G$, and  $F_D(\{E_\a\},\cL_G)$ for poset $D$.   When we return to the analysis of vacuum polarizations, we revert to the notation of $\pi_G$ or $\pi_D$.

In these terms, our expression for the integral associated with an arbitrary graph $G$ with $n$ vertices at fixed spatial loop momenta is
\bea
  2\pi \delta \left ( \sum_\alpha E_\alpha \right )\, F_G \left (\{E_\alpha\},\cL_G \right ) &=&\sum_{\t} \mathbb{N}_\t\, \prod_{\alpha=1}^{n} \int_{-\infty}^{t_{\alpha+1}}dt_{\alpha} e^{-i(E_\alpha + \eta^{(\tau_\alpha)}_j (\omega_j-i\epsilon))t_{\alpha}}\, \nn \\
  &=&  2\pi \left ( \sum_\alpha E_\alpha \right )\, \sum_\t\  \mathbb{N}_\tau F_\t \left (\{E_\alpha\},\cL_G \right ) \label{eq:integral-form-2}\\
  &=& 2\pi \delta \left ( \sum_\alpha E_\alpha \right )\, \sum_D \sum_{\t_D} \mathbb{N}_D\, F_{\t_D} \left (\{E_\alpha\},\cL_G \right )
  \nn \\ &=& 2\pi \delta \left ( \sum_\alpha E_\alpha \right )\, \sum_D  \mathbb{N}_D\, F_D \left (\{E_\alpha\},\cL_G \right )\, ,
  \nn
\eea
where we define, $t_{n+1}=\infty$ in the first equality.   The final, $t_n$, integral gives the delta function that enforces the energy conservation in $G$.
Here, $\eta^{(\t_\alpha)}_{j}$ is the incidence matrix of the vertex $b_{\t_\alpha}$ in the time order labeled by $\t$.  The second equality defines $F_\t$ as the coefficient of the energy-conserving delta function from the full integral associated with time order $\tau$.   In the third, we reorganize the sum over time orders into a sum over posets, $D$, as in Eq.\ (\ref{eq:tau-to-D}), recalling that every time order is associated with a single poset.  Finally, the fourth equality defines $F_D$ as the complete contribution to $F_G$ from all the time orders within poset $D$.   The functions $F_D$ will be the subject of the following discussion.

\subsection{Time integrals of extremal vertices; the covering set}
\label{sec:integration}

We consider the time integrals allowed within a given partial order for an arbitrary graph. Again, in this section, we will not restrict to forward scattering graphs that one encounters in leptonic annihilation.    Rather, we treat graphs with a generic flow of external momenta.   We will label this general partially-ordered diagram, or poset diagram, as $D$, defined on a set of vertices $V$ with 
ordering $\ge$.   In the following, we will use $D$ to refer both to the poset and to the ordered diagram.  Note that we will only encounter vertices that are local in time, whose time integrals we will carry out.  After our first time integral, however, the composite vertices we encounter generally will not be local in space.

The object of interest is the function $F_D$ for a fixed poset $D=(V,\geq)$, which as we have noted before, is fixed by an incidence matrix $\eta^{(\alpha)}_j$.  (For compactness of notation, we do not label $\eta^{(\a)}_j$ by $D$.)   We represent schematically the multidimensional region of vertex times restricted to the orders associated with a specific poset $D$ as
\begin{equation}
  2\pi \delta \left (\sum_{\alpha=1}^n E_\alpha \right)\, F_D\left(\{E_\a\},\cL_G \right) =\prod_{\alpha=1}^{|V|} \int_{X_D}dt_{\alpha} e^{-i(E_\alpha + \eta^{(\alpha)}_j (\omega_j-i\epsilon))t_{\alpha}}\, ,
\label{eq:integral-form-poset}
\end{equation}
where we have \textit{formally} carried out the sum over time orders consistent with poset $D$, corresponding to the integral of times over a  region labeled by $X_D$. We let $|V|$ represent the number of vertices.
The region $X_D$ is defined by
\begin{equation}
X_D=\{(t_1,t_2\ldots t_{|V|})| b_k \geq b_j \implies t_k \geq t_j\}\, .
\label{eq:region}
\end{equation}
The union of all possible regions $X_D$ is the full set of time integrals in Eq.\ (\ref{eq:integral-form-2}).

For a given poset  $D$, we can identify a unique set $M^1_D$ of {\it embedded extrema} (minima and maxima) of $D$.   We define an embedded vertex $v\in M^1_D$, as one whose removal leaves behind a connected diagram consisting of vertices $V/v$.   Every connected diagram has a non-empty set of embedded vertices (see Appendix \ref{app:index}) .   The time integrals of minima in $X_D$ extend independently to negative infinity, and those of maxima to positive infinity.  We can always begin the integration process for $X_D$ by  picking a subset of embedded extremal vertices that form an anti-chain (mutually incomparable elements).   We will do so for a subset of the extremal embedded vertices, chosen to have the largest number of such vertices  that form an anti-chain.
Such a subset may include both maxima and minima, because the minima may not be ordered with respect to all maxima, and vice-versa.  

 Let us call this set the first {\it extremal antichain}, $A^1$, defined to satisfy,
\bea
A^1\ &=&\   \{x\in M^1| x,y \in A^1 \implies x \sim y \}\nn \\ &=& \left\{ b_{1,1}, \dots, b_{1,r_1}\, ;\,  d_{1,1}, \dots, d_{1,s_1} \right \} \ \equiv\ \left\{ e_{1,i} \right \}\, ,
\label{eq:antichain-def-1}
\eea
where we denote by $r_1$ the number of minimal elements, $b_{1,i}$, in $A^1$ and by $s_1$ the number of maximal elements, $d_{1,i}$, and where $e_{i,j}$ labels these vertices collectively, with $j=1\dots r_1+s_1$. $A^1$ represents a set of vertices whose times can be integrated simultaneously and independently of each other.\footnote{This choice for $A^1$ is not unique, and we need not adhere to $A^1$ being the largest possible antichain in any one step.}   Clearly, each of these integrals must extend from a fixed upper (lower) limit to infinity (negative infinity).   We now turn to the determination of these limits.

Having identified the set $A^1$, we identify a companion set, $C^1$, of vertices that cover the minimal vertices in 
$A^1$ or are covered by the maximal vertices in $A^1$, that is, vertices that are connected to one or more elements of $A^1$ by single line(s) with no intervening vertices,
\bea
C^1&=&\{x\in V|\exists v \in A^1: (v \prec x) \, {\rm or}\, (x\prec v)\}\nn \\
&=& \bigcup\limits_{i=1}^{r_1} C_{b_{1,i}}\bigcup\limits_{j=1}^{s_1} C_{d_{1,j}},
\label{eq:C1-def}
\eea
where we have defined the covering set for each element in $A^1$ by
\bea
C_{b_{1,i}}&=&\{ x\in V|b_{1,i} \prec x\},\nn \\
C_{d_{1,i}}&=&\{ x\in V|x \prec d_{1,i} \}.
\label{eq:C_b1-def}
\eea
By construction, vertices in $C^1$ are connected to extrema by lines that either emerge directly from minima or flow directly into maxima, as denoted by the covering relation $\prec$, defined in Sec.\ \ref{sec:def}.   If there are several such vertices for any given extremal vertex, they form an anti-chain, that is, they are not mutually ordered within the poset $D$.  We note as well that a single element of $C^1$ can cover more than a single extremum.  We will refer to the elements of $C^1$ as the first {\it covering set} for the first extremal antichain, $A^1$.   We will use the covering set $C^1$ to organize the time integrals for the extremal vertices in $A^1$.  This will enable us to begin a recursive analysis over  induced poset diagrams and to do all subsequent time integrals in the same way as the first.

In the time integral over $X_D$, corresponding to the partially ordered diagram $D$, Eq.\ (\ref{eq:integral-form-poset}), we must include all choices within poset $D$ of the earliest covering vertices for all minimum elements of $A^1$ and of the latest covered vertex for all maximum elements.  Region $X_D$, and thus our integral, is given by a sum over these choices, which we will denote as $\gamma^1 \subset C^1$.  It is worth pointing out that it is possible to construct the full list of compatible choices $\gamma^1$ of next-to-extremal covering vertices systematically, by summing over available choices of covering (covered) vertices
for each minimum (maximum)  in turn.   Because next-to-extremal vertices may cover more than one extremum, the number of resulting sets is not a simple product of the number of covering vertices for the extrema.  For a concrete algorithm to generate all consistent $\g^1$, see Appendix \ref{app:algorithm}.

Given a choice of next-to-extremal vertices, the integration range of each extremal vertex is -$\infty$ to the time of its earliest covering vertex for a minimum, and from the time of its latest covered vertex to $ \infty$ for each maximum.   
We shall describe these choices collectively as sets of next-to-extremal covering vertices, and label them among the elements of $C^1$ as $x^{(e_{1,i})}_a$, where index $e_{1,i}$ identifies the extremal vertex connected to the vertex $x_a^{(e_{1,i})}$, while index $a$ identifies the choice of covering (covered) vertex in set  $C_{e_{1,i}}$ that covers (is covered by) the minimal (maximal) vertex $e_{1,i}$.  

In terms of the next-to-extremal covering elements, $\gamma^1$ is given by
\bea
\gamma^1\ =\  \left \{ \cup_{e_i=1\dots r_1+s_1}\;  x_a^{(e_{1,i})} \right \}\, .
\label{eq:gamma-1-def}
\eea
Whenever two extrema, say $e_{1,i}$ and $e_{1,i'}$, have the same next-to-extremal covering vertex, we have $x^{(e_{1,i})}_a=x^{(e_{1,i'})}_{a'}$.  The number of elements in $\gamma^1$ is therefore less than or equal to the number of elements in $C^1$.
In this notation, once we choose $\gamma^1$, the integrals over the times of all extremal vertices in set $A^1$, minima and maxima, can be done explicitly and independently.   

To relate the integration region for each $\gamma^1$ to the original integration region $X_D$ for poset $D$ in Eq.\ (\ref{eq:region}), we proceed as follows.
As noted above, the next-to-extremal vertices for any $e_{1,i}$ form an antichain (mutually incomparable vertices) and therefore need additional ordering at the poset level to specify the ranges of the $e_{1,i}$ time integrals.  This additional ordering we impose is fixed by our choice in $x_a^{(e_{1,i})}$.      For each choice of $\gamma^1$, we construct a partial order on the antichains in $C^1$,
by defining a new binary relation, 
denoted $\geq_{\gamma^1}$ over the set $V$ by
\bea
&& \hspace{15mm} {\rm If}\ a \geq  b,\ {\rm then}\ a \geq_{\gamma^1}b\, ,
\nn\\
&& \hspace{15mm}  \forall y \in {C_{b_{1,i}}}:  y\geq_{\g^1} x_a^{(b_{1,i})} >_{\g^1} b_{1,i}\, ,
\nn\\
&& \hspace{15mm}  \forall y \in {C_{d_{1,i}}}:  d_{1,i} >_{\g^1} x_a^{(d_{1,i})} \geq_{\g^1} y\, .
\label{eq:ge-gamma1-def}
\eea
Here, the subscript on $\geq_{\gamma^1}$  reflects the ordering that remains after we identify each extremal vertex $e_{1,i}$ with its corresponding vertex $x_a^{(e_{1,i})} \in C_{e_{1,i}}$ in the covering set $\gamma^1$.   

The relations, Eq.\ (\ref{eq:ge-gamma1-def}) that define the ordering $\ge_{\gamma^1}$ identify new posets, whose mutually-disjoint integration regions allow us to integrate the times $t_{e_{1,i}}$ explicitly.  
Again, although we formally have constructed $\geq_{\g^1}$ as a ``stronger" binary relationship than 
the original ordering $\ge$ of the poset, it is  the natural ordering inherited from $\ge$  when every extremum $e_{1,i}$ is identified with the specific covering element $x_a^{(e_{1,i})}$, that is, contracting the vertex $e_{1,i}$ onto the corresponding $x_a^{(e_{1,i})}$ in $\gamma^1$. We, therefore, think of a new, lower-order graph  with all $e_{1,i}$ removed from the set of vertices.  Every line that previously emanated from a minimum $b_{1,i}$, and was not absorbed at vertex  $x^{(b_{1,i})}_a$ now emanates from $x^{(b_{1,i})}_a$, and therefore $x^{(b_{1,i})}_a$ is ``less than" every other vertex of $C^1$ that was connected to $b_{1,i}$, since after the contraction of $b_{1,i}$ onto $x^{(b_{1,i})}_a$ there is  at least one line starting from $x^{(b_{1i})}_a$ and ending on any such vertex.  Maxima are
contracted on their nearest covered  vertices in an exactly analogous fashion.

The combination of the new, contracted vertex and the induced order defines a new poset, 
\bea
D\left [\gamma^1 \right ]\ =\ \left \{V\left[ \gamma^1 \right ] \, ,\, \geq_{\gamma^{1}} \right \}\, , \quad \quad V \left[ \gamma^1 \right ] \ =\  V \setminus  \cup_i e_{1,i} \, ,
\label{eq:D-gamma-1-def}
\eea
with $\ge_{\gamma^1}$ defined by Eq.\ (\ref{eq:ge-gamma1-def}).  The integration region for post $D[\gamma^1]$, analogous to the original region, Eq.\ (\ref{eq:region}), is given by
\begin{equation}
X_{D\left [ \gamma^1 \right ]}=\{(t_1,t_2\ldots \hat{t}_{e_{1,1}} \ldots \hat{t}_{e_{1,r_1+s_1}} \dots t_{|V|})| b_k \geq_{\gamma^1} b_j \implies t_k \geq t_j\} \, ,
\end{equation}
where $\hat{t}_{e_{1,i}}$ means that $t_{e_{1,i}}$ is excluded from the list.   Again, the notation $X_{D\left [ \gamma^1 \right ]}$ identifies the subregion of $X_D$ where the vertices in $\gamma^1$ are the full
set of next-to-extremal covering vertices.   

It is clear that any time order in region $X_D$ has a unique set $\gamma^1$, and that distinct choices of $\gamma^1$ correspond to different time orders.   Therefore, when we sum over all choices, of $\gamma^1$, we exhaust all  time orders within $X_D$.  Explicitly, we may represent the full poset integration region, $X_D$ as
\bea
X_{D}\ \ &=&\ \bigcup\limits_{\gamma^1} X_{D\left [ \gamma^1 \right ]}\times \bigcup_{ \substack{t_{b_{i,1}} \\ i=1}}^{r_1} (-\infty, t_{ x_a^{(b_{1,i}) } }) \times \bigcup_{ \substack{t_{d_{i,1}} \\ i=1}}^{s_1} (t_{ x_a^{(d_{1,i}) }} , \infty)\, ,
\label{eq:regions-union-1}
\eea
where again, $\gamma^1$ is specified, as in Eq.\ (\ref{eq:gamma-1-def}), as the choice of  next-to-extremal elements $x_a^{(e_{1,i})}$.   

In summary, for a given choice of $\gamma^1$, we can carry out the $t_{e_{1,i}}$ integrals up to (or down to) the times $t_{x_a^{(e_{1,i})}} \in \gamma^1$.   
 We can thus rewrite the full integral for $F_D$, Eq.\ (\ref{eq:integral-form-poset}) using Eq.\ (\ref{eq:regions-union-1}) as
\bea
  2\pi \delta \left (\sum_{\alpha=1}^n E_\alpha \right)\, F_D\left(\{E_\a\}, \cL_G \right)\ &=& \    \sum_{\gamma^1}\; 
  \prod_{\alpha=1}^{|V[\gamma^1]|} \int_{X_D[\gamma^1]}dt_{\alpha} e^{-i(E_\alpha + \eta^{(\alpha)}_j (\omega_j-i\epsilon))t_{\alpha}}
  \nn\\[2mm]
  &\ & \hspace{-40mm} \times\   \prod_{i=1}^{r_1}  \int_{-\infty}^{t [ x^{(b_{1,i})}_a ]} dt_{1,i}\, e^{ -i \left ( E_i + \sum_j \eta^{(b_{1,i})}_j (\omega_j -i\epsilon )\right ) t_{1,i} }
  \nn\\[2mm]
  &\ & \hspace{-40mm} \times \   \prod_{i=1}^{s_1} \int^{\infty}_{t[x^{(d_{1,i})}_a]} dt_{1,i}\, e^{ -i \left ( E_i +\sum_j \eta^{(d_{1,i})}_j (\omega_j -i\epsilon )\right ) t_{1,i} }\, .
\label{eq:integral-form-poset-gamma-1}
\eea
Here we have labeled the times of the extremal vertices, over which we are integrating, with the subscripts of the corresponding vertices $e_{1,i}$ themselves, and the times of the next-to-extremal covering vertices $x^{(e_{1,i})}_a$ in a hopefully obvious notation. 
We note that each of the limits in the integrals over extremal vertices are times that appear in the remaining integration measure of $X_{D[\gamma^1]}$.

In Eq.\ (\ref{eq:integral-form-poset-gamma-1}), the  time integrals for minima all take the form
\bea
\int_{-\infty}^{t [ x^{(b_{1,i})}_a ]} dt_{1,i}\, e^{ -i \left ( E_{b_{1i}} + \sum_j \eta^{(b_{1,i})}_j (\omega_j -i\epsilon )\right ) t_{1,i}  } \ =\ \ \frac{i}{ \Delta_{b_{1,i}} [\gamma^1]  }\, e^{ -i \left ( E_{b_{1i}} + \sum_j \eta^{(b_{1,i})}_j (\omega_j -i\epsilon)\right ) t [ x^{(b_{1,i})}_a ] }
\nn\\
\label{eq:min-integrals}
\eea
where we define
\bea
 \Delta_{b_{1,i}} [\gamma^1] \ =\ E_{b_{1i}} - \sum_j \left| \eta^{(b_{1,i})}_j\right| \omega_j + i\epsilon\, \, .
\label{eq:min-denom}
\eea
We note that every nonzero $\eta_j^{(b_{1,i})}=-1$, since they all correspond to minimum vertices, which only emit particles.   Such a denominator has the standard form of an energy deficit in the channel of vertex $b_{1,i}$, where external energy $E_{b_{1,i}}$ flows in.   We will refer to the set of lines, $j$ for which $\eta^{(b_{1,i})}_j=-1$, all of whose on-shell energies flow out of vertex $b_{1,i}$, as the ``cut" of diagram $D$ that separates it into two diagrams, one consisting of vertex $b_{1,i}$ and one with the remaining vertices, $V \setminus b_{1,i}$.   Since all $b_{1,i}$ are embedded vertices, all the diagrams $V\setminus b_{1,i}$ are connected.

Similarly, for all the time integrals for maxima in Eq.\ (\ref{eq:integral-form-poset-gamma-1}) we have
\bea
\int^{\infty}_{t[x^{(d_{1,i})}_a]} dt_{1,i}\, e^{ -i \left ( E_{d_{1i}} + \sum_j \eta^{(d_{1,i})}_j (\omega_j -i\epsilon )\right ) t_{1,i}  } &= & \frac{i}{ \Delta_{d_{1,i}} [\gamma^1] }\, e^{ -i \left ( E_{d_{1i}} + \sum_j \eta^{(d_{1,i})}_j (\omega_j -i\epsilon)\right ) t[x^{(d_{1,i})}_a]  }\, .
\nn\\
\label{eq:max-integrals}
\eea
In this case we define 
\bea
\Delta_{d_{1,i}} [\gamma^1]  &=&\ - \left (E_{d_{1i}} + \sum_j \eta^{(d_{1,i})}_j \omega_j - i\epsilon \right )
\nn\\[2mm]
&=&\  (-E_{d_{1i}}) - \sum_j  \eta^{(d_{1,i})}_j \omega_j + i\epsilon \, ,
\label{eq:max-denom}
\eea
where in the second equality we use that $\eta^{(d_{1,i})}$ is either zero or $1$ for maximum vertices.  When $-E_{d_{1,i}}$ is a positive energy flowing out of vertex $d_{1,i}$, we again have
a standard energy deficit form in the channel of vertex $d_{1,i}$.   As for minimal vertices, will refer to the set of lines, $j$ for which $\eta^{(d_{1,i})}_j=1$,  as the ``cut" of diagram $D$ that separates vertex $d_{1,j}$ from a connected diagram with vertices $V \setminus d_{1,j}$.

 Substituting the integrals in Eqs.\ (\ref{eq:min-integrals}) to (\ref{eq:max-denom}) into the expression for $F_D$, Eq.\ (\ref{eq:integral-form-poset-gamma-1}), we now have
 \bea
  2\pi \delta \left (\sum_{\alpha=1}^n E_\alpha \right)\, F_D\left(\{E_\a\}, \cL_G\right )  &=&  \sum_{\gamma^1}\;  \prod_{i=1}^{r_1} \frac{i}{ \Delta_{b_{1,i}} [\gamma^1]  } \; \prod_{i=1}^{s_1}  \frac{i}{ \Delta_{d_{1,i}} [\gamma^1]  }
  \nn\\[2mm]
  &\ & \hspace{-10mm} \times \ \prod_{\alpha=1}^{|V[\gamma^1]|} \int_{X_D[\gamma^1]}dt_{\alpha} e^{-i(E_\alpha + \eta^{(\alpha)}_j (\omega_j-i\epsilon))t_{\alpha}}
  \nn\\[2mm]
  &\ & \hspace{-30mm} \times\  \prod_{i=1}^{r_1}  e^{ -i \left ( E_{b_{1,i}} + \sum_j \eta^{(b_{1,i})}_j (\omega_j -i\epsilon)\right ) t [ x^{(b_{1,i})}_a ] }
   \prod_{i=1}^{s_1} e^{ -i \left ( E_{d_{1,i}} + \sum_j \eta^{(d_{1,i})}_j (\omega_j -i\epsilon)\right ) t[x^{(d_{1,i}) }_a ] }\, .
   \nn\\
   \label{eq:integral-min-max-gamma-1}
  \eea
  In this expression,  every next-to-extremal vertex $x^{(e_{1,i})}_a$ appears in the product over the vertices ($\alpha$) of diagram $D[\gamma^1]$.   The integral may thus be written in 
  a more compact form by combining these contributions in the phases, using the definitions
\bea
E[\gamma^1]_\alpha\ &=& \ E_\alpha + \sum_{i=1}^{r_1} E_{b_{1,i}} \, \delta_{\alpha,x^{(b_{1,i})}_a} +  \sum_{i=1}^{s_1} E_{d_{1,i}} \, \delta_{\alpha,x^{(d_{1,i})}_a}
\nn\\[2mm]
 \eta[\gamma^1]^{(\alpha)}_a\ &=& \  \eta_j^{(\alpha)} + \sum_{i=1}^{r_1} \eta^{(b_{1,i})}_j \delta_{\alpha,x^{(b_{1,i})}_a}    +   \sum_{i=1}^{s_1}  \eta^{(d_{1,i})}_j  \delta_{\alpha,x^{(d_{1,i})}_a}\, .
 \label{eq:E-eta-gamma-1}
 \eea
An important feature of the new incidence matrix $\eta[\gamma^1]_j^{(\a)}$ is that every line $j$ for which it takes a non-zero value is connected to vertex $\alpha$ in $D[\gamma^1]$, either directly or through an embedded extremal vertex whose time has been integrated up to the time of vertex $\alpha$.  
 
We can now write the full poset integral as 
 \bea
   2\pi \delta \left (\sum_{\alpha=1}^n E_\alpha \right)\, F_D\left(\{E_\a\}, \cL_G \right )    &=& \ \sum_{\gamma^1}\;  \prod_{i=1}^{r_1} \frac{i}{ \Delta_{b_{1,i}} [\gamma^1]  } \; \prod_{j=1}^{s_1}  \frac{i}{ \Delta_{d_{1,j}} [\gamma^1]  }
   \nn\\
   &\ & \hspace{-20mm} \times \ \prod_{\alpha=1}^{|V[\gamma^1]|} \int_{X_D[\gamma^1]}dt_{\alpha} e^{-i(E[\gamma^1]_\alpha + \eta[\gamma^1]^{(\alpha)}_j (\omega_j-i\epsilon))t_{\alpha}}\, .
   \label{eq:integral-first-step}
   \eea
In this form, the original integral for poset $D$ is expressed as a sum over $\gamma^1$ (which is constructed explicitly in App.\ \ref{app:algorithm}) of products of energy deficit denominators multiplied by an integral of the original form, Eq.\ (\ref{eq:integral-form-poset}), this
time for the poset diagram $D[\gamma^1]$, with a reduced number of time integrals remaining.  The energy deficit denominators correspond to the cuts of $D$ associated with each of the embedded extremal vertices in the set $A^1$.   Because these vertices make up an antichain, these cuts have no lines in common.  Finally, we recall from Eq.\ (\ref{eq:D-gamma-1-def}) that the original list of vertices, $V$ is given by
the union of $V[\gamma^1]$ with the extremal vertices,
\bea
V[\gamma^1]  \prod_{i=1}^{r_1} \cup\, b_{1,i}\,  \prod_{j=1}^{s_1} \cup\, d_{1,j}\ =\ V\, .
\label{eq:v-union-1}
\eea
We will encounter  generalizations of all of these features below.

Equation (\ref{eq:integral-first-step}) already exhibits the basic structure of our results.    Let us summarize its  essential features.    The new, {\it induced}, poset diagram $D[\gamma^1]$, defined by Eqs.\ (\ref{eq:ge-gamma1-def}) and (\ref{eq:D-gamma-1-def}), is a connected diagram with a set of vertices $V[\gamma^1]$, all of which are local in time.   The number of vertices has decreased by identifying pairs of embedded extremal and next-to-extremal vertices, with the resulting ``composite" vertices inheriting the next-to-extremal times $t[x_j^{(e_{1,i})}]$.   The extremal vertices of $D[\gamma^1]$ can be labeled $\{b_{2,i},d_{2,i}\}$.

For these induced, embedded minimal vertices, the new incidence matrix defined in Eq.\ (\ref{eq:E-eta-gamma-1}),  $\eta[\gamma^1]^{(b_{2,i})}_j$ takes only the values $0,-1$, with only outgoing lines, although now these lines may have been emitted originally by the first round of vertices, $\{e_{1,i}\}$, over whose times we have already integrated.   These lines thus emerge from a subdiagram of the original poset $D$ that is connected.   Similar considerations apply to maximal extrema in $D[\gamma^1]$, for which $\eta[\gamma^1]^{(d_{2,i})}_j=0,+1$.
In the new diagram, by analogy to Eqs.\ (\ref{eq:C1-def}) and (\ref{eq:C_b1-def}) we can  identify the set of next-to-extremal vertices, $C^2=\{ C_{e_{2,i}} \}$, which will be labeled $x^{(e_{2,i})}_a$.

Examples of this process are given above in the integrations from Fig.\ \ref{example1} to Fig.\ \ref{ex1-int-a} and to Fig.\ \ref{ex1-int-b}, which correspond to different choices of $\gamma^1$ in this case.    The explicit denominators of Eq.\ (\ref{eq:integral-first-step}) provide the results of the time integrals of the chosen antichain $A^1$ of embedded extremal vertices.   By construction, they correspond to non-overlapping cuts, each separating the original diagram into two connected parts.  As such, they are all ``physical" denominators, in the sense we have identified above.   The poset $D[\gamma^1]$ and the integral over the region $X_{D[\gamma^1]}$ in Eq.\ (\ref{eq:integral-first-step}) have all the properties of the original poset $D$ and integral over region $X_D$ of Eq.\ (\ref{eq:integral-form-poset}) that we used in the forgoing analysis.  In the following subsection, we use this recursive structure to derive a general expression for $F_D$ in which all unphysical denominators are eliminated. 

\subsection{Integrals of induced extrema}

The relation, Eq.\ (\ref{eq:integral-first-step}) clearly carries fewer time integrals, and the remaining time integrals are in the same form that we encountered in Eq.\ (\ref{eq:integral-form-poset}). It is therefore straightforward to extend the reasoning recursively in order to carry out all the time integrals. 

Let us assume that we are given the result after $k-1$ iterations of the procedure outlined above.   Each step consists of integration over the induced embedded extrema of an induced poset.    We assume that the result takes the form 
\bea
   2\pi \delta \left (\sum_{\alpha=1}^n E_\alpha \right)\, F_D\left(\{E_\a\}, \cL_G \right )    &=& \ \sum_{\gamma^{k-1}}\; \prod_{l=1}^{k-1} \prod_{i=1}^{r_l} \frac{i}{ \Delta_{b_{l,i}} [\gamma^{l}]  } \; \prod_{j=1}^{s_l}  \frac{i}{ \Delta_{d_{l,j}} [\gamma^{l}]  }
   \nn\\
   &\ & \hspace{-30mm} \times \ \prod_{\alpha=1}^{|V[\gamma^{k-1}]|} \int_{X_D[\gamma^{k-1}]}dt_{\alpha} e^{-i(E[\gamma^{k-1}]_\alpha + \eta[\gamma^{k-1}]^{(\alpha)}_j (\omega_j-i\epsilon))t_{\alpha}}\, .
   \label{eq:integral-k-1-steps}
   \eea
This integral has a set of properties that hold in the initial case, Eq.\ (\ref{eq:integral-first-step}), corresponding to $k=2$, and which will remain true recursively.
   
   (1) {\it Poset structure.}  $D[\gamma^{k-1}]$ is a poset with integration region $X_{D[\gamma^{k-1}]}$, where $\gamma^{k-1}$ is specified by a set of vertices from the original diagram $D$, and all vertices in $V_{D[\gamma^{k-1}]}$ are local in time. That is, each remaining vertex $\alpha$ is associated with a time integral over $t_\alpha$.   As for any such diagram, among the vertices of $D[\gamma^{k-1}]$, there is a non-empty set $M^k$ of embedded extremal vertices, $\{b_{k,i},d_{k,i}\}$, each with a corresponding covering set, $C^k=\cup_i C^k_{e_{k,i}}$, of potential next-to-extremal vertices, $C^k_{e_{k,i}} = \{x^{(e_{k,i})}_a\}$.  
   
   (2) {\it Inductive functional dependence.}   In Eq.\ (\ref{eq:integral-k-1-steps}), the explicit denominators and phases are defined by
\bea
 \Delta_{b_{l,i}} [\gamma^l] \ =\ E_{b_{l,i}}[\gamma^{l-1}] - \sum_j \left| \eta[\gamma^{l-1}]^{(b_{l,i})}_j\right| \omega_j + i\epsilon\, \, ,
\label{eq:min-denom-k}
\eea
and
\bea
\Delta_{d_{l,i}} [\gamma^l] 
&=&\  \left(-E_{d_{l,i}}[\gamma^{l-1}]\right) - \sum_j  \eta[\gamma^{l-1}]^{(d_{l,i})}_j \omega_j + i\epsilon \, ,
\label{eq:max-denom-k}
\eea
in terms of the inductive relations
\bea
E_\alpha[\gamma^l] &=& \ E_\alpha[\gamma^{l-1}] + \sum_{i=1}^{r_l} E_{b_{l,i}}[\gamma^{l-1}] \, \delta_{\alpha,x^{(b_{l,i})}_a} +  \sum_{i=1}^{s_l} E_{d_{k,i}}[\gamma^{l-1}] \, \delta_{\alpha,x^{(d_{l,i})}_a}\, , \nn
\\[2mm]
 \eta[\gamma^l]^{(\alpha)}_j\ &=& \  \eta[\gamma^{l-1}]_j^{(\alpha)} + \sum_{i=1}^{r_l} \eta[\gamma^{l-1}]^{(b_{l,i})}_j \delta_{\alpha,x^{(b_{l,i})}_a}    +   \sum_{i=1}^{s_l}  \eta[\gamma^{l-1}]^{(d_{l,i})}_j  \delta_{\alpha,x^{(d_{l,i})}_a}. \nn \\
 \label{eq:E-eta-gamma-k}
 \eea
The expressions, Eqs.\ (\ref{eq:min-denom-k}), (\ref{eq:max-denom-k}) and (\ref{eq:E-eta-gamma-k}) are direct generalizations of Eqs.\ (\ref{eq:min-integrals}), (\ref{eq:max-integrals}) and (\ref{eq:E-eta-gamma-1}), with $\gamma^1$ replaced by $\gamma^l$, and $e_{1,i}$ by $e_{l,i}$.  
      
   (3) {\it Denominators, cuts, and merged sets.}  The explicit denominators, $\Delta_{e_{l,i}}$ ($e=b,d$) in Eq.\ (\ref{eq:integral-k-1-steps}), with $l\le k-1$ all correspond to cuts of the original diagram, $D$, each of which separates $D$ into exactly two connected components. One of these components is associated with an embedded extremal vertex, $e_{l,i} \in A^{l-1}$, of $D[\gamma^{l-1}]$, and with a connected set of vertices, $\lambda[e_{l,i}]$ of the original diagram. We will refer to $\lambda[e_{l,i}]$ as the ``merged set" of vertices for $e_{l,i}$, corresponding to one or more series of time integrals  that terminate at $t_{e_{l,i}}$.    
It   may be a minimum or maximum.   In either case, its merged set is given by 
      \bea
      \lambda[e_{1,i}]\ &=& e_{1,i}\, ,
      \nn\\[2mm]
  \lambda[e_{l,i}]\ &=&\ \left \{ e_{l,i}\,  \prod_{e_{l',j}}\cup\, \lambda[e_{l',j}]\, \bigg | \, x^{(e_{l',j})}_{a'} = e_{l,i}\, , l' \le l-1 \right  \}\, ,
  \label{eq:lambda-def}
  \eea
  where the range of index $i$ depends on  the set $\gamma^{l-1}$.   That is, the merged set of extremal vertex $e_{l,i}$ is found by merging all the merged sets of extremal vertices $e_{l',i}$, $l'\le l-1$ for which the vertex $e_{l,i}$ is the nearest vertex in its covering set.
    For a minimal extremal vertex in $D[\gamma^{l-1}]$, all lines that cross the cut $\Delta_{b_{l,i}}[\gamma^{l}]$ emerge from the minimum $b_{l,i}$ and are absorbed in $D[\gamma^{l-1}] \setminus b_{l,i}$.       For a maximal extremal vertex, lines that cross the cut $\Delta_{d_{l,i}}[\gamma^{l}]$ emerge from $D[\gamma^{l-1}] \setminus d_{l,i}$ and are absorbed at $d_{l,i}$.   In both cases, in terms of the original poset diagram $D$, the lines of each cut connect $\lambda[e_{l,i}]$ with $D\setminus \lambda[e_{l,i}]$, both of which are connected.
    
   (4)  {\it Partial nesting.}  Any set of vertices picked from different  $\lambda[b_{k,i}]$ form an antichain, and similarly for maximal sets $\lambda[d_{k,i}]$.       Together, they satisfy the following properties, which may be described as partial nesting  \cite{Capatti:2022mly},
   \bea
    \lambda[e_{l_1,i}] \subset \lambda[e_{l_2,i'}] \quad {\rm or}\quad  \lambda[e_{l_1,i}] \cap \lambda[e_{l_2,i'}] \ &=&\ 0\, ,  l_1 \ < l_2\, ,
      \nn\\[2mm]
      \lambda[e_{l,i_1}] \cap \lambda[e_{l,i_2}] \ &=&\ 0\, ,  i_1 \ \ne i_2\, ,
      \nn\\[2mm]     V[\gamma^{k}]  \prod_i\, \cup\,  \lambda[b_{k,i}]\, \prod_l\, \cup\,  \lambda[d_{k,l}]\ &=&\ V\, .
      \label{eq:v-union-k-1}
\eea
Taken together, these conditions imply that  the sets of vertices $\lambda[e_{l,i}]$ are either nested, according to index $l$, or disjoint, and that, together with $V[\gamma^{k}]$, they include all the vertices of the original diagram $D$.   This implies that the cuts represented by the denominators $\Delta_{e_{l,i}}$ do not cross, since this would lead to a non-nested relationship between at least two successive sets $\lambda[e_{l,i}]$ and $\lambda[e_{l+1,j}]$.
Comparing to Eq.\ (\ref{eq:v-union-1}) above for the first set of integrals, we observe that indeed
$\lambda[b_{1,i}]\ =\ b_{1,i}$
and
$\lambda[d_{1,i}]\ =\ d_{1,i}$,
as in Eq.\ (\ref{eq:lambda-def}).

The specific manner in which the posets and the sets $\gamma^{k-1}$ and $\lambda[e_{k,i}]$ appear will emerge from the following analysis,
where we describe how to carry out the $k$th iteration,  the next set of time integrals in Eq.\ (\ref{eq:integral-k-1-steps}).  We will see that all of the features (1)\, --\ (4) of $D[\gamma^{k-1}]$ are  inherited by the resulting expression in terms of a poset $D[\gamma^k]$.

We begin by characterizing the extremal time integrals in Eq.\ (\ref{eq:integral-k-1-steps}).   In fact, we need only repeat the steps in Eqs.\ (\ref{eq:antichain-def-1}) to (\ref{eq:D-gamma-1-def}) that we applied to the original integral form for $F_D(\{E_\alpha\}, \cL_G )$ in Eq.\ (\ref{eq:integral-form-poset}).   We can do so because these steps depend only on the poset structure of the diagram $D[\gamma^{k-1}]$.

To organize the $k$th set of integrals, given the poset $D[\g^{k-1}]$, we identify the set of its embedded extrema $M^k$ (See App.\ \ref{app:index}). Next, we pick a maximal antichain (a set of extrema that are not related pairwise), which we label $A^{k}$. Generically, it contains both embedded minima and embedded maxima. We denote the number of minima in $A^k$ by $r_k$ and the number of maxima by $s_k$.  Analogous to Eq.\ (\ref{eq:antichain-def-1}), we can write
\bea
A^k\ &=&\   \{x\in M^k| x,y \in A^k \implies x \sim y \}\nn \\ &=& \left\{ b_{k,1}, \dots, b_{k,r_k}\, ;\,  d_{k,1}, \dots, d_{k,s_k} \right \} \ \equiv\ \left\{ e_{k,i} \right \}\, ,
\label{eq:antichain-def-k}
\eea
where, as before, in Eq. (\ref{eq:antichain-def-k}) the $b_{k,i}$ represent the minima in $A^k$ and the $d_{k,i}$ represent the maxima in $A^{k}$. We use the symbol $e_{k,i}$ to represent elements of the combined set.

Next, we identify a companion set $C^k$, which is the set of next-to-extremal vertices.  Analogous to Eqs.\ (\ref{eq:C1-def}) and (\ref{eq:C_b1-def}) we define

\bea
C^k&=&\left \lbrace x\in V[\g^{k-1}]\bigg|\; \exists v \in A^k: (v \prec_{
\g^{k-1}} x)\, {\rm or}\, (x\prec_{\g^{k-1}} v)\right \rbrace \nn \\
&=& \bigcup\limits_{i=1}^{r_k} C_{b_{k,i}}\bigcup\limits_{j=1}^{s_k} C_{d_{k,j}},
\label{eq:Ck-def}
\eea
where we have defined the covering (or covered) set for each element in $A^k$ by
\bea
C^k_{b_{k,i}}&=&\left \lbrace x\in V[\g^{k-1}] \bigg|\; b_{k,i} \prec_{\g^{k-1}} x\right \rbrace,\nn \\
C^k_{d_{k,i}}&=&\left \lbrace x\in V[\g^{k-1}] \bigg|\; x \prec_{\g^{k-1}} d_{k,i} \right \rbrace.
\label{eq:C_bk-def}
\eea
As before, we identify every consistent set of next-to extremal elements. We label the next-to-extremal element of $e_{k,i}$ by $x^{(e_{k,i})}_a$. As above, the index $a$ labels different  choices in the next-to-extremal element $x$, given the choice in next-to-extremal elements for all $e_{l,m}, \; l \leq k-1, m \leq r_l+s_l$, as well as $e_{k,m}, \; m \leq i-1$ (See App. \ref{app:algorithm} for more details). We can now define the object $\g^{k}$ representing one of the consistent choices for next-to-extremal vertices, up to the stage $k$,
\bea
\gamma^k\ =\  \left \{ \g^{k-1} \bigcup_{i=1}^{r_k+s_k}\;  x_a^{(e_{k,i})} \right \}\, .
\label{eq:gamma-k-def}
\eea
We now observe that a choice in $\g^k$ naturally induces a poset structure on the set of vertices $V\left[ \g^k \right]$, which is defined by analogy to Eq.\ (\ref{eq:D-gamma-1-def}) as
\be
V\left[ \g^k \right] =V\left[ \g^{k-1} \right] \bigg\backslash \bigcup_{i=1}^{r_k+s_k}e_{k,i}.
\ee 
The corresponding binary relationship for the set of vertices $V[\gamma^k]$ is uniquely defined by requiring that the chosen covering vertex of a given minimum, $b_{k,i}$, say $x^{(b_{k,i})}_{j}$, is itself a minimum in the set $C_{b_{k,i}}$ and the chosen covered vertex of a given maximum, $d_{k,i}$, say $x^{(d_{k,i})}_{j}$ is itself a maximum in the set $C_{d_{k,i}}$. We may therefore conclude that the new binary relation, constructed by analogy to Eq.\ (\ref{eq:ge-gamma1-def}), and labeled $\geq_{\g^k}$, is given by
\bea
&& \hspace{10mm} {\rm for}\  a,\, b \in V[\gamma^k],\ \mathrm{and}\ a \geq_{\g^{k-1}}  b,\ {\rm then}\ a \geq_{\gamma^k}b
\nn\\
&& \hspace{10mm}   \forall y \in {C_{b_{k,i}}}:  y\geq_{\g^k} x_j^{(b_{k,i})} >_{\g^k} b_{k,i} \, , 
\nn\\
&& \hspace{10mm}  \forall y \in {C_{d_{k,i}}}:  d_{k,i}>_{\g^k} x_j^{(d_{k,i})} \geq_{\g^k} y\, .
\label{eq:ge-gamma-k-def}
\eea
Together with the set of vertices, $V^k$, the binary relation $\ge_k$ defines a new poset $D[\gamma^k]$.
 
With the new poset structure in place, we can carry out the time integrals corresponding to the extrema $e_{k,i}$. To do so, we rewrite Eq.\ (\ref{eq:integral-k-1-steps}), separating out the integrals over extremal elements in $A^k$, in analogy with Eq.\ (\ref{eq:integral-form-poset-gamma-1}). As before such a decomposition will enable us to carry out the extremal integrals, here in
\bea
   2\pi \delta \left (\sum_{\alpha=1}^n E_\alpha \right)\, F_D\left(\{E_\a\}, \cL_G \right )    &=& \ \sum_{\gamma^{k}}\; \prod_{l=1}^{k-1} \prod_{i=1}^{r_l} \frac{i}{ \Delta_{b_{l,i}} [\gamma^{l}]  } \; \prod_{j=1}^{s_l}  \frac{i}{ \Delta_{d_{l,j}} [\gamma^{l}]  }
   \nn\\
   &\ & \hspace{-20mm} \times \prod_{\alpha=1}^{|V[\gamma^{k}]|} \int_{X_D[\gamma^{k}]}dt_{\alpha} e^{-i(E[\gamma^{k-1}]_\alpha + \eta[\gamma^{k-1}]^{(\alpha)}_j (\omega_j-i\epsilon))t_{\alpha}}\nn \\[2mm]
  &\ & \hspace{-20mm} \times\   \prod_{i=1}^{r_k}  \int_{-\infty}^{t [ x^{(b_{k,i})}_a ]} dt_{k,i}\, e^{ -i \left ( E_i[\gamma^{k-1}]+ \sum_j \eta[\gamma^{k-1}]^{(b_{k,i})}_j (\omega_j -i\epsilon )\right ) t_{k,i} }
  \nn\\[2mm]
  &\ & \hspace{-20mm} \times \   \prod_{j=1}^{s_k} \int^{\infty}_{t[x^{(d_{k,i})}_a]} dt_{k,i}\, e^{ -i \left ( E_i[\gamma^{k-1}] +\sum_j \eta[\gamma^{k-1}]^{(d_{k,i})}_j (\omega_j -i\epsilon )\right ) t_{k,i} }\, .
  \nn\\
\label{eq:integral-form-poset-gamma-k}
   \eea
 We can repeat the steps from Eqs.\ (\ref{eq:min-integrals}) to (\ref{eq:max-denom}) to perform extremal time integrals and define the denominators that arise at this stage of the integration procedure.   The results are of exactly the same form
 as for the case $k=2$, and reproduce the inductive forms of momentum dependence in denominators and phases in Eqs.\ (\ref{eq:min-denom-k}) to (\ref{eq:E-eta-gamma-k}), 
 \bea
   2\pi \delta \left (\sum_{\alpha=1}^n E_\alpha \right)\, F_D\left(\{E_\a\}, \cL_G\right )    &=& \ \sum_{\gamma^{k}}\; \prod_{l=1}^{k} \prod_{i=1}^{r_l} \frac{i}{ \Delta_{b_{l,i}} [\gamma^{l}]  } \; \prod_{j=1}^{s_l}  \frac{i}{ \Delta_{d_{l,j}} [\gamma^{l}]  }
   \nn\\
   &\ & \hspace{-20mm} \times \ \prod_{\alpha=1}^{|V[\gamma^{k}]|} \int_{X_D[\gamma^{k}]}dt_{\alpha} e^{-i(E[\gamma^{k}]_\alpha + \eta[\gamma^{k}]^{(\alpha)}_j (\omega_j-i\epsilon))t_{\alpha}}\, .
   \label{eq:integral-k-steps}
   \eea
   Comparing the expression in Eq.\ (\ref{eq:integral-k-steps}) with the form of the integral after 
   $k-1$ time integrals in Eq.\ (\ref{eq:integral-k-1-steps}), and the four conditions that follow that expression, we can check that we have completed an inductive construction of Eq.\ (\ref{eq:integral-k-steps}) with the quantities $\Delta_{b_{k,i}} [\gamma^k],\Delta_{d_{k,i}} [\gamma^k]$ and $ E_\alpha\ [\gamma^k], \eta[\gamma^k]^{(\alpha)}_j$ defined recursively through Eqs. (\ref{eq:min-denom-k}), (\ref{eq:max-denom-k}) and (\ref{eq:E-eta-gamma-k}) respectively.  We also see that the four conditions that follow Eq.\ (\ref{eq:integral-k-1-steps}) remain true after the $k$th integral.
   
    (1) {\it Poset structure.} By construction $D[\gamma^k]$ is a new poset, which again represents a diagram with all vertices local in time, with a non-empty set of embedded extremal vertices $M^{k+1}$.
    
    (2) {\it Functional dependence.}  The time integrals in 
    Eq.\ (\ref{eq:integral-form-poset-gamma-k}) leading to Eq.\ (\ref{eq:integral-k-steps}) give precisely the results for denominators and phases specified by Eqs.\ (\ref{eq:min-denom-k}) - (\ref{eq:E-eta-gamma-k}), with $l=k$.
    
    (3) {\it Denominators, cuts, and merged sets.}   The new denominators $\Delta_{b_{k,i}}$ and $\Delta_{d_{k,i}}$ arise from the time integrals of embedded extremal vertices  in the chosen set $A^k$ of $D[\gamma^{k-1}]$.  Their explicit expressions in Eq.\ (\ref{eq:min-denom-k}) show that they have the standard forms of energy deficits.    We can verify that they correspond to cuts of the original diagram, $D$ that separate $D$ into two connected components as follows.  Consider first the minimal vertices, $b_{k,i}$ of $D[\gamma^{k-1}]$.   Each emits lines from the set of vertices in the merged set, $\lambda[b_{k,i}]$, all of whose elements are unordered with respect to the vertices in any other merged set $\lambda[b_{k,j}]$, $j\ne i$. As a result, all vertices in $\lambda[b_{k,i}]$ are connected to the remainder of the original diagram $D$ only through lines emitted by vertex $b_{k,i}$, which may connect to any non-minimal vertices in $D[\g^{k-1}]$.   
    Then, cutting all lines emitted by $b_{k,i}$ separates all vertices in $\lambda[b_{k,i}]$ from the remainder of the diagram $D$.  But $\lambda[b_{k,i}]$ is connected by construction, and so is $V[\g^{k-1}] \setminus b_{k,i}$ since $b_{k,i}$ is by construction an embedded minimal vertex.  Thus, cutting the set of lines emerging from any 
    $b_{k,i}$ cuts the diagram into two connected components. Identical considerations apply to the embedded maximal vertices, $d_{k,i}$ of $D[\gamma^{k-1}]$.
    Lines emerging from (for minimal) or absorbed into (for maximal) extremal vertices of the next poset $D[\gamma^k]$, labeled $e_{k+1,i}$,  are emitted or absorbed by sets of merged vertices, which we label $\lambda[e_{k+1,i}]$,   of the original diagram, $D$.   These sets are defined. as in Eq.\ (\ref{eq:lambda-def}), by
  \bea
  \lambda[e_{k+1,i}]\ =\ \left \{ e_{k+1,i} \, \prod_{e_{k',j}}\, \cup\, \lambda[e_{k',j}]\, \bigg | \, x^{(e_{k',j})}_a = e_{k+1,i}\ ,\, k' \le k-1 \right  \}\, ,
  \label{eq:lambda-e-k+1-def}
  \eea
  where index $i$ varies over all choices of the set $\gamma^k$. 
     
  (4)  {\it Partial nesting.}   Finally, the sets $\lambda[e_{k+1,i}]$, given by Eq.\ (\ref{eq:lambda-e-k+1-def}), associated with posets $D[\gamma^{k}]$, have the same properties as the $\lambda[e_{k,i}]$ associated with the posets $D[\gamma^{k-1}]$, as described in Eq.\ (\ref{eq:v-union-k-1}).  The nesting  features of Eq.\ (\ref{eq:v-union-k-1}) are inherited by the sets defined by Eq.\ (\ref{eq:lambda-e-k+1-def}) precisely because the $\lambda[e_{k+1,i}]$ are disjoint unions of smaller sets that satisfy (\ref{eq:v-union-k-1}).    A consequence is that elements from different $\lambda[b_{k+1,i}]$s are unordered (form an antichain), and similarly for elements in different $\lambda[d_{k+1,i}]$.  In addition, the $k+1$st layer of merged sets satisfy the same relation as the $k$th, Eq.\ (\ref{eq:v-union-k-1}),
  \bea
   V[\gamma^{k}]  \prod_i\, \cup\,  \lambda[b_{k+1,i}]\, \prod_l\, \cup\,  \lambda[d_{k+1,l}]\ &=&\ V\, ,
   \label{eq:v-union-k}
  \eea
because the difference between the set $V[\gamma^k]$ and the corresponding quantity $V[\gamma^{k-1}]$ in Eq.\ (\ref{eq:v-union-k-1}) is precisely the chosen set of embedded extremal vertices of $D[\g^{k-1}]$, $A^{k-1}$, which are identified with the new vertices in the sets $\lambda[e_{k+1,i}]$.  Thus, any vertices that are in $V[\g^{k-1}]$ but not in $V[\g^k]$ are absorbed into the union of the $\lambda[e_{k+1,i}]$, along with all of the $\lambda[e_{k,i}]$. 
   
   In summary, all of the recursive features of the integrals of extrema have been confirmed.  In making the steps from the $(k-1)$st to $k$th sets of time integrals, we used only the partial ordering and the existence of at least one embedded extremal vertex at each step.
   Starting with any finite-order diagram, however, it is clear that eventually, the process terminates at $k=\kappa$, when the poset $D[\gamma^\kappa]$ consists of a single vertex, $e_{\kappa+1}$.
   Such a vertex will connect to no internal lines (none remain) but will connect to {\it all} external energies, $E_\a$.  Its time integral yields the momentum conserving 
   delta function in Eq.\ (\ref{eq:integral-form-2}).
   
 The general form for an arbitrary poset $D$ is thus the energy conserving delta function in Eq.\ (\ref{eq:integral-form-poset}), times the amplitude function,
  \bea
   F_D\left(\{E_\a\}, \cL_G \right )    &=& \ \sum_{\gamma^\kappa}\; \prod_{l=1}^{\k} \prod_{i=1}^{r_l} \frac{i}{ E_{b_{l,i}}[\gamma^{l-1}] - \sum_n \left| \eta[\gamma^{l-1}]^{(b_{k,i})}_n\right| \omega_j + i\epsilon\  } 
   \nn\\
   &\ & \times  \prod_{j=1}^{s_l}  \frac{i}{  \left(-E_{d_{l,i}}[\gamma^{l-1}]\right) - \sum_n  \eta[\gamma^{l-1}]^{(d_{l,i})}_n \omega_j + i\epsilon  },
  \label{eq:final-general-amp}
   \eea
   where we have used the explicit forms of denominators in Eqs.\ (\ref{eq:min-denom-k}) and (\ref{eq:max-denom-k}).
   Here all denominators correspond to  cuts that divide the graph into exactly two connected components, and the sum over $\gamma^\kappa$ represents all complete, recursive choices of next-to-extremal sets, starting with those of the original embedded extremal vertices of $D$.   Each denominator is of the form of an energy deficit, either with respect to a sum of energies flowing out of or into the diagram.   In contrast to the normal TOPT form, however, singularities associated with any diagram are all physical, in that they divide the diagram into two connected subdiagrams.   
   
   In practical cases,  most external energies are zero to begin with (corresponding to collections of internal vertices).   Such denominators are negative semi-definite even for massless theories.   An intriguing feature of Eq.\ (\ref{eq:final-general-amp}) is that denominators for which $E_i[\gamma^k]=0$ are universal, in the sense that they are independent of the underlying process.   Such denominators describe a series of states that emerge from the vacuum, and which couple diagrammatically to the process described by amplitude $F$ at composite vertices of the sort illustrated in Figs.\ \ref{example1-integration} and \ref{example3-integration} above.   We shall not pursue this concept of universality here, and leave it as a subject for future work. 
   
     As noted above, an equivalent expression for a general amplitude has been derived in Ref.\ \cite{Capatti:2022mly} by a somewhat different graphical analysis.   In the next section, we apply the method developed above to study weighted cross sections in
     leptonic annihilation.
   
   We close this section by remarking briefly on the application of these methods to light-cone ordered perturbation theory (LCOPT) \cite{Chang:1968bh,Kogut:1969xa,Brodsky:1997de}.  In LCOPT, vacuum orderings are absent altogether.   The method here may be applied, however, whenever there are more than two external momenta, which leads, as in time ordering, to pseudo-physical cuts.   The result is just of the form of Eq.\ (\ref{eq:final-general-amp}), with the $E_i$ and $\omega_j$ replaced by external and on-shell minus momenta, respectively (in $x^+$ ordering).   In this way, pseudo-physical cuts are again eliminated. 

\section{Vacuum polarization graphs, leptonic annihilation, and weighted cross sections}
\label{sec:weighted-cs}

In this section, we will adapt the general result for poset contributions to amplitudes, Eq.\ (\ref{eq:final-general-amp}), to lepton annihilation processes, and their weighted cross sections.    The essential feature of this analysis is that it eliminates the pseudo-physical denominators altogether.   A specific diagram in the form of Eq.\ (\ref{eq:final-general-amp}), however, does not have a manifest ordering between potential physical cuts.   To provide such an order, we will adapt our procedure to the process at hand, in this case, leptonic annihilation.

Given a vacuum polarization graph $G$, we single out two vertices $i$ and $o$, where the current carries positive and negative energy into the diagram, respectively.  
As in the vacuum polarization diagrams of Sec.\ \ref{sec:TOPT}, the external energies of all vertices $\alpha$ vanish, with the exception of $i$ and $o$,
\bea
E_\alpha\ =\ Q\delta_{\alpha i} - Q'\delta_{\a,o}\, ,
\label{eq:E-flow-vp}
\eea
where energy conservation will require $Q=Q'$.
Next, we  define the first set of embedded extrema $\hat M^1$, using the  extremal set $\hat{S}=\left( \text{Min}(D) \cup \text{Max}(D)\right) \setminus \{i,o\}$. We then continue with our process of integrating out the largest antichain in $\hat{M}^1$. The main idea is to remove the two extrema $i,o$ at each stage in the procedure until all other extrema, whose time integrations result in vacuum denominators (and soft pinches) only. This separates out, at the level of the integrand, long time processes that only carry soft singularities in the IR from short time processes that have the interpretation of taking place at the ``hard scale".  
As in Sec.\ \ref{Formula}, we define a set of next-to-extremal vertices $\g^1$, and carry out the first set of integrals.   In general, $\hat M^1$ contains $r_1$ embedded minima and $s_1$ embedded maxima, and we find
 \bea
  2\pi \delta \left (Q-Q'\right)\, \p_D\left(Q,\cL_G \right )  &=&  \sum_{\gamma^1}\;  \prod_{i=1}^{r_1} \frac{i}{ \tilde{\Delta}_{b_{1,i}} [\gamma^1]  } \; \prod_{i=1}^{s_1}  \frac{i}{ \tilde{\Delta}_{d_{1,i}} [\gamma^1]  }
  \nn\\[2mm]
  &\ & \hspace{-10mm} \times \ \prod_{\alpha=1}^{|V[\gamma^1]|} \int_{X_D[\gamma^1]}dt_{\alpha} e^{-i(E_\alpha + \eta^{(\alpha)}_j[\g^1] (\omega_j-i\epsilon))t_{\alpha}},
   \label{eq:integral-min-max-1-ep}
  \eea
where we have used notation identical to that introduced in Sec.\ \ref{Formula}. The difference from Eq.\ (\ref{eq:integral-min-max-gamma-1}), however is that $E_\a [\g^1]=E_\a$, given in Eq.\ (\ref{eq:E-flow-vp}), by construction. This follows from our choice to integrate out a set of vertices that do not carry external energy. As a consequence, the denominators $\tilde{\Delta}_{b_{1,i}}$ and $\tilde{\Delta}_{d_{1,i}}$ are of the ``vacuum" form,
\bea
 \tilde{\Delta}_{b_{1,i}} [\gamma^1] \ =\ - \sum_j \left| \eta^{(b_{1,i})}_j\right| \omega_j \, \, .
\label{eq:min-denom-ep}
\eea
for minima and
\bea
\tilde{\Delta}_{d_{1,i}} [\gamma^1]  
&=&\  - \sum_j  \eta^{(d_{1,i})}_j \omega_j  \, ,
\label{eq:max-denom-ep}
\eea
for maxima,
where once again we observe that the difference from Eqs.\ (\ref{eq:min-denom})  and (\ref{eq:max-denom}) is simply the absence of external energy. These denominators therefore do not vanish anywhere except when all lines in the state carry zero momenta (which is a genuine pinch surface in massless theories). In Eqs.\ (\ref{eq:min-denom-ep})  and (\ref{eq:max-denom-ep}) we have explicitly dropped the $i\epsilon$'s to emphasize the reality of these state denominators. 

We now continue the process that we have described in Sec.\ \ref{Formula}, but continue to leave out vertices $i,o$ at each stage, until these are the only extrema remaining. We emphasize that at each stage we only generate denominators of the kind we encountered in Eqs.\ (\ref{eq:min-denom-ep}), (\ref{eq:max-denom-ep}). The process terminates after $\kappa_1$ steps, when the set of embedded extrema is empty, $\hat{M}^{\kappa_1+1}=\emptyset$.   This does not mean that all vertices have been exhausted since we have explicitly excluded $i$ and $o$ from the set of extrema at each stage and non-embedded extremal vertices may also remain. 
We thus can write for the poset integrand $\pi_D$,
 \bea
   2\pi \delta \left (Q-Q'\right)\, \p_D\left(Q,\cL_
   G \right ) &=& \ \sum_{\gamma^{\kappa_1}}\; \prod_{l=1}^{\kappa_1} \prod_{i=1}^{r_l} \frac{i}{ - \sum_k \left| \eta \; [\gamma^{l}]^{(b_{l,i})}_k \right| \omega_k   } \; \prod_{j=1}^{s_l}  \frac{i}{ - \sum_k  \eta \; [\gamma^{l}]^{(d_{l,j})}_k \omega_k    }
   \nn\\
   &\ & \hspace{-20mm} \times \ \prod_{\alpha=1}^{|V[\gamma^{\kappa_1}]|} \int_{X_D[\gamma^{\kappa_1}]}dt_{\alpha} e^{-i(E_\alpha + \eta[\gamma^{\kappa_1}]^{(\alpha)}_j (\omega_j-i\epsilon))t_{\alpha}}\, .
   \label{eq:integral-Pi1-steps-ep}
   \eea
 Having separated the vacuum denominators we would like to order the remaining integrand of the vacuum polarization poset diagram.   The poset $X_D[\g^{\kappa_1}]$ has no tadpole subdiagrams, lacking external momenta and connected to the remaining diagram by a single vertex only.   This is because, as shown in App.\ \ref{app:index}, any nontrivial tadpole has an embedded extremal vertex.   But the times of all embedded extrema except $i$ and $o$ have been integrated over in arriving at Eq.\ (\ref{eq:integral-Pi1-steps-ep}).   
 
 By construction, the remaining diagram has at most two embedded extremal vertices, $i$ and $o$.   Any additional extremal vertices must be non-embedded.  Each non-embedded vertex, whether extremal or not, connects two internally connected subdiagrams, one of which must include $i$ and the other $o$. This is because there are no tadpole subdiagrams.

\begin{figure}[htb]
\begin{center}
\subfloat[\label{ileqo}]{\includegraphics[width=5cm]{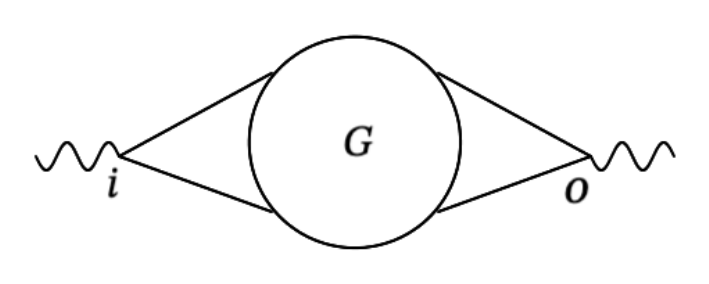}}
\subfloat[\label{oleqi}]{\includegraphics[width=5cm]{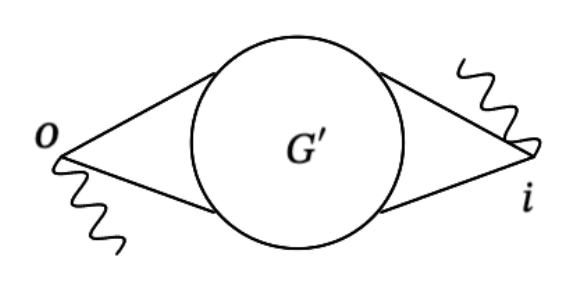}}\\
\subfloat[\label{osimi}]{\includegraphics[width=5cm]{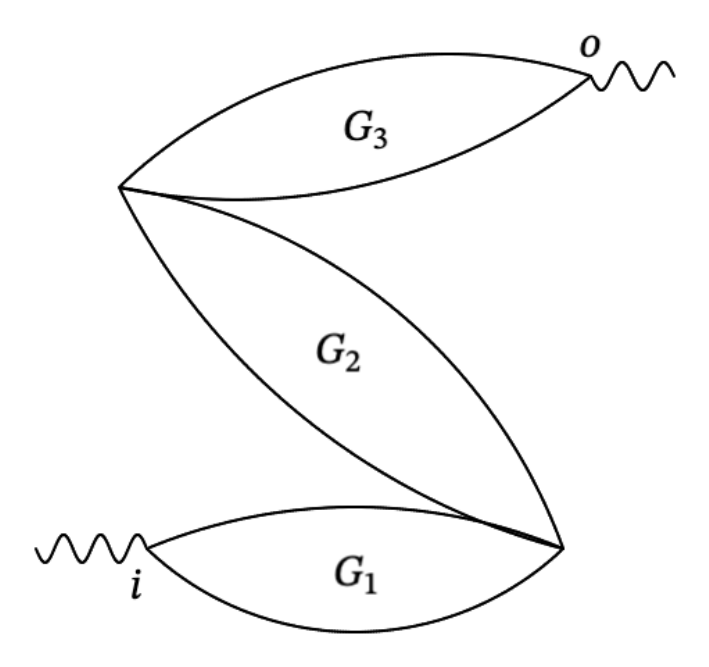}}
\subfloat[\label{osimiex}]{\includegraphics[width=5cm]{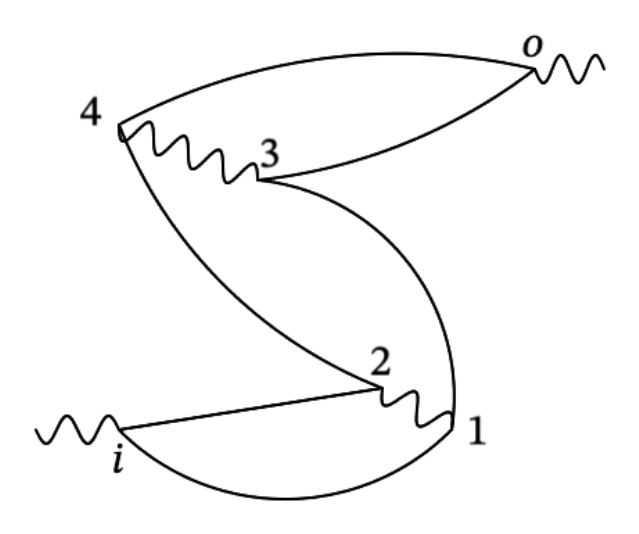}}
\end{center}
\caption{ 
    \protect\subref{ileqo}
   A poset ordered graph after $\k_1$ steps representing the situation when $i<o$. Here, $G$ is any graph containing no extrema. \\
    \protect\subref{oleqi}
  A poset ordered graph after $\k_1$ steps representing the situation when $o<i$. Here, $G'$ is any graph containing no extrema.\\
    \protect\subref{osimi}
    A poset ordered graph with $i \sim o$. Here $i$ is a minimum and $o$ is a maximum, but $i \nleq o$ and $G_1,G_2$ and $G_3$ are subgraphs containing no embedded extrema other than possibly $i$ or $o$. This represents a situation with two non-embedded extrema. Generically, $i,o$ are either minima or maxima and there may be an arbitrary number of intervening non-embedded extrema.  \\
    \protect\subref{osimiex}
    An example of a poset ordered graph with $i \sim o$ in QED at three loops. Integrating the times of vertices $4,1$ yields a poset of the type in Fig. \ref{osimi}.
    }
\label{fig:vacuum-pol}
\end{figure}

We have three possibilities,
\begin{enumerate}
\item $i \leq_{\g^{\kappa_1}}o$,
\item $o \leq_{\g^{\kappa_1}}i$,
\item $o \sim_{\g^{\kappa_1}}i$. 
\end{enumerate}
These three situations have been represented diagrammatically in Fig.\ \ref{fig:vacuum-pol}.

Let us analyze cases (1), (2) first.  Since we have integrated out all embedded extrema excluding $i,o$, one of $i,o$ is an embedded extremum. Suppose $i\leq_{\g^{\kappa_1}} o$, and $i$ were not a minimum. This would imply there exists a minimal vertex $x<_{\g^{\kappa_1}} i$ which is not embedded. By definition the cut succeeding  $x$, $\Delta_x$ would split the graph into two or more disconnected subdiagrams, say $V_1 \dots V_k$, $k\ge 2$.  Because $x$ is a minimum, vertices in different $V_i
$ are all mutually unordered.   Then $i,o$ would be in the same disconnected component, say $V_1$, since $i\leq_{\g^{\kappa_1}} o$. This would therefore imply that all the other components $V_2 \dots V_k$ carry no external momenta, that is, that there are tadpole subdiagrams, which we have shown cannot be present in $D[\g^{\kappa_1}]$.  The same reasoning shows that $i$ is itself an embedded minimum and that there are no other minima in the poset $D[\g^{\kappa_1}]$. 
A similar argument reveals that $o$ is the unique maximum in $D[\g^{\kappa_1}]$ for case (1). 

One can use similar reasoning to argue that if $o\leq_{\g^{\kappa_1}} i$, $i$ is the unique maximum and $o$ is the unique minimum  in $D[\g^{\kappa_1}]$ .  In cases $(1),(2)$ we can either impose any total order (time order)  or continue along the lines of the arguments made in Sec.\ \ref{Formula}. The remaining integrals in Eq.\ (\ref{eq:integral-Pi1-steps-ep}) can therefore be carried out either by constructing new posets or through the imposition of a sum of time orders.  Every time-ordered graph in cases (1) and (2) has a unique minimum and a unique maximum.  Carrying out the procedure outlined in Sec.\ \ref{Formula} is exactly equivalent to generating all possible time orders.  

Turning to case (3),  we treat the possibility that $o \sim_{\g^{\kappa_1}}i$. It is easy to see using the arguments made for the first two cases that both $i,o$ are embedded extrema of the graph, using the absence of tadpole subgraphs.  
 Suppose $o$ was an embedded extremum and $i$ was not an extremal element. 
 This means there is an $x\in (\text{Min}D[\g^{\k_1}]) $ and $ y\in \text{Max}(D[\g^{\k_1}])$ 
 such that $x\leq_{\g^{\k_1}} i \leq_{\g^{\k_1}} y$. 
 By assumption, both $x,y$ are not embedded, and therefore, in the absence of tadpole subdiagrams, $x$ divides the graph into precisely two additional components, $V_{1}$ and $V_2$, with  $i$ and $y$ in the same component, say $V_1$, and $o \in V_2$. However, this means $y$ cannot divide the graph into two components without a tadpole subgraph.  Thus, $y$ must have been an embedded extremum, contrary to our assumption.   Thus, $i$ must be an embedded extremum.

In summary, we learn that in all three cases,  $i,o$ are both unique embedded extrema.   In cases (1) and (2), vertices $i$ and $o$ are in a single, connected diagram, in which they are the only extremal vertices.   That is, there are no non-embedded extrema. 
In these cases, any other vertex $y$ satisfies 
\bea
i<_{\g^{\kappa_1}} &y& <_{\g^{\kappa_1}} o\, , {\rm case}\ (1)\, ,
\nn\\[2mm]
o<_{\g^{\kappa_1}} &y& <_{\g^{\kappa_1}} i\, , {\rm case}\ (2)\, .
\label{eq:case-1-2}   
\eea
Suppose we integrate $t_i$ first, (up or down) to time $t_{x_{i'}^{(\kappa_1+1)}}$.   In the resulting poset, $D[\g^{\kappa_1+1}]$, any remaining vertices $y$ continue to satisfy the same inequalities.   That is,
\bea
x_{i'}^{(\kappa_1+1)} <_{\g^{\kappa_1+1}} &y& <_{\g^{\kappa_1+1}} o\, , \;\;\,{\rm case}\ (1)\, ,
\nn\\[2mm]
o\ <_{\g^{\kappa_1+1}} &y& <_{\g^{\kappa_1+1}} x_{i'}^{(\kappa_1+1)}\, , {\rm case}\ (2)\, .
\label{eq:case-1-2-integrated}   
\eea
Therefore, vertex $x_{i'}^{(\kappa_1+1)}$ is now an embedded extremum.
  The argument that $x_{i'}^{(\kappa_1+1)}$ is also an embedded extremum  for  case (3) in which $i$ and $o$ are in different subdiagrams, follows the same pattern, using the fact that $i$ and $o$ are unique embedded extrema.

In order to now be able to  
order the resulting expressions diagrammatically,
 we will only integrate the embedded extremum $i$ (and not $o$).   
Following the procedure outlined in Sec.\ \ref{Formula}, we integrate over a single-element $M^{\k+1}=\{i\}$.  After integrating the time $t_i$, we will find a new extremal $i'$ with external energy $Q$ flowing into it.
Because there are no additional embedded extremal vertices other than vertices $i$ and $o$, this will also  be an embedded extremal vertex, although it may be a minimum or a maximum.
  If we again choose $\k$ to represent the total number of sequences in the process, the procedure terminates after $\k -\kappa_1+1$ steps (including the final integral integral that gives the energy delta function).   At each integration step $\kappa_1+1 \le l  \le \k$, there
  is a single time integral only, corresponding to the sequence that begins with vertex $i$ at step $\kappa_1+1$.
 This results in the formula,
\bea
  \p_D\left(Q,\cL_G \right ) &=& \ \sum_{\gamma^{\kappa_1}}\; \prod_{l=1}^{\kappa_1} \prod_{i=1}^{r_l} \frac{i}{ - \sum_n \left| \eta \; [\gamma^{l}]^{(b_{l,i})}_n \right| \omega_n   } \; \prod_{j=1}^{s_l}  \frac{i}{ - \sum_n  \eta \; [\gamma^{l}]^{(d_{l,j})}_n \omega_n    }
   \nn\\
   &\ & \hspace{-20mm} \times \ \sum_{\gamma^{\k}| \g^{\k_1}}\; \prod_{l=\k_1+1}^{\k} \frac{i}{ \tilde{\Delta}_{e_{l}} [\gamma^{l}] +i\e } \; ,
   \label{eq:integral-Pi-steps-ep}
   \eea
   where the sum over $\gamma^\kappa$ is restricted to sets of next-to-extremal vertices consistent with the choice $\gamma^{\kappa_1}$ (see App.\ \ref{app:algorithm}).
Here $e_l$ is defined to be the vertex that satisfies $E_{e_l}[\g^{l}]=Q > 0$ in the notation of Eq.\ (\ref{eq:E-eta-gamma-k}), and the denominator is defined by 
\begin{equation}
\tilde{\Delta}_{e_{l}} [\gamma^{l}]= \lambda_{e_l}Q\ -\ \sum_k \left |\eta[\g^l]^{(e_l)}_{k} \right| \omega_k ,
\label{eq:denominators-cross-section}
\end{equation} 
where $\lambda_{e_l}=+1$ when vertex $e_l$ is a minimum, and $-1$ when it is a maximum.  In the latter case, the state denominators represented by $\tilde \Delta_{e_l}$ are negative definite.   We note that as
$l$ increases, the denominators change sign when vertex $e_l$ reaches a previously non-embedded extremal vertex (in case (3) above), which then becomes embedded.   We also notice that the denominators $\tilde\Delta$ in Eq.\ (\ref{eq:denominators-cross-section}) are defined slightly differently from the definition of $\Delta$  in Eqs.\ (\ref{eq:min-denom}) and (\ref{eq:max-denom}) in order to explicitly display the $i\e$.

Having obtained the expression for the forward scattering locally in phase space, we  immediately write down the contribution of a poset $D$ to  the $e^+e^-$  total cross section's integrand, in terms of a sum over ordered final states, $C$,
\bea
  \text{Im}\, \left[ \p_D (Q,\cL_G) \right ] &=& \ \sum_{\gamma^{\kappa_1}}\; \prod_{l=1}^{\kappa_1} \prod_{i=1}^{r_l} \frac{i}{ - \sum_n \left| \eta \; [\gamma^{l}]^{(b_{l,i})}_n \right| \omega_n   } \; \prod_{j=1}^{s_m}  \frac{i}{ - \sum_n  \eta \; [\gamma^{l}]^{(d_{l,j})}_n \omega_n    }
   \nn\\
   &\ & \hspace{-20mm} \times \ \sum_{\gamma^{\k}| \g^{\k_1}}\sum_{C=\kappa_1+1}^{\k} \;  \prod_{l=C+1}^{\k} \frac{i}{ \tilde{\Delta}_{e_{l}} [\gamma^{l}]-i\e }     
    2\pi \delta(\tilde{\Delta}_C) \; \prod_{l=\k_1+1}^{C-1} \frac{i}{ \tilde{\Delta}_{e_{l}} [\gamma^{l}] +i\e } \; .
   \label{eq:ep-cross-section-D}
   \eea
    This result is equivalent to the TOPT expression in Eq.\ (\ref{eq:forwardscattering-cuts}), but has been reorganized through partial ordering to eliminate all pseudo-physical cuts. 
   
    We recall that to get the total cross section as defined in Eq.\ (\ref{eq:optical}), we multiply the expression in Eq.\ (\ref{eq:ep-cross-section-D}) by the poset numerator factor, sum over posets and finally over graphs, giving the integrand
   \bea
 \sigma(Q,\cL_G) &=&  
 \ \sum_G \sum_{D}\mathbb{N}_D\, \;\prod_{i=1}^{N_G}\frac{1}{2\omega_i}\;  \sum_{\gamma^{\kappa_1}}\; \prod_{l=1}^{\kappa_1} \prod_{i=1}^{r_l} \frac{i}{ - \sum_n \left| \eta \; [\gamma^{l}]^{(b_{l,i})}_n \right| \omega_k   } 
  \nn \\ &&
   \times \prod_{j=1}^{s_l}  \frac{i}{ - \sum_n \eta \; [\gamma^{l}]^{(d_{l,j})}_n \omega_n    }   \nn \\
  &\ & \hspace{-20mm} \times  \ \sum_{\gamma^{\k}| \g^{\k_1}}\sum_{C=\kappa_1+1}^{\k} \; 
 \prod_{l=C+1}^{\k} \frac{i}{ \tilde{\Delta}_{e_{l}}[\gamma^{l}] -i\e }  \
  2\pi \delta(\tilde{\Delta}_C) \;
   \prod_{l=\k_1+1}^{C-1} \frac{i}{ \tilde{\Delta}_{e_{l}} [\gamma^{l}] +i\e }\, .
   \label{eq:ep-cross-section}
   \eea
   This is the analog of the TOPT expression for the integrand of the cross section, again without pseudo-physical cuts.  
   
     It is now straightforward to obtain a new expression for the integrand of the weighted cross section defined in Eq.\ (\ref{eq:denominator-fs}).   The relation (\ref{eq:ep-cross-section}) is fully local in loop momenta $\cL_G$, and hence specifies all contributions to the squared amplitude from each final state, $C$.   We thus have 
     \bea
 \Sigma[f,Q]  &=& \ \sum_G \int d\cL_G \;\prod_{i=1}^{N_G}\frac{1}{2\omega_i}\;  \sum_{D}\mathbb{N}_D\sum_{\gamma^{\kappa_1}}\; \prod_{l=1}^{\kappa_1} \prod_{i=1}^{r_l} \frac{i}{ - \sum_n \left| \eta \; [\gamma^{l}]^{(b_{l,i})}_n \right | \omega_n} \; \nn \\
  &\ & \hspace{-20mm} \times \prod_{j=1}^{s_l}  \frac{i}{ - \sum_n  \eta \; [\gamma^{l}]^{(d_{l,j})}_n \omega_n    }   \ \sum_{\gamma^{\k}| \g^{\k_1}} \sum_{C=\kappa_1+1}^{\k}\;
  \prod_{l=C+1}^{\k} \frac{i}{ \tilde{\Delta}_{e_{l}}[\gamma^{l}] -i\e }
  \; 2\pi \delta(\tilde{\Delta}_C) \nn \\
  &\ & \hspace{-20mm} \times f_{C}  (\vec{q}_1 \dots \vec{q}_{k_C}) \ \left( \frac{1+ \lambda_{e_C}}{2}\right) 
   \;  \prod_{l=\k_1+1}^{C-1} \frac{i}{ \tilde{\Delta}_{e_{l}} [\gamma^{l}] +i\e } \, .
   \label{eq:ep-cross-section-fs}
   \eea
In Eq.\ (\ref{eq:ep-cross-section-fs}) we have inserted a factor of $\left( \frac{1+\lambda_{e_C}}{2}\right)$ to set the weight of cuts with negative external energy, and hence $\l_{e_C}=-1$, to zero.  We can also keep them, but they will always give zero because the argument of the energy conservation delta function is negative definite for such states.    In the sum over posets $D$, only those posets with $i<_{\g^{\kappa_1}} o$ or $i\sim_{\g^{\kappa_1}} o$ can have $\lambda_{eC}=1$. 

Using Eq.\ (\ref{eq:ep-cross-section-fs}) and the $\delta$ function identity in Eq.\ (\ref{eq:delta-identity}) we can collect terms with the same denominator structure to write the analog of Eq.\ (\ref{eq:reor})
\bea
\Sigma[f,Q]  &=& \ \sum_G \int d\cL_G \;\prod_{i=1}^{N_G}\frac{1}{2\omega_i}\;  \sum_{D}\mathbb{N}_D\sum_{\gamma^{\kappa_1}}\; \prod_{l=1}^{\kappa_1} \prod_{i=1}^{r_l} \frac{i}{ - \sum_n \left| \eta \; [\gamma^{l}]^{(b_{l,i})}_n \right| \omega_n   } \label{eq:ep-cross-sec-fs-reor} \;  \\
  &\ & \hspace{-20mm} \times \prod_{j=1}^{s_l}  \frac{i}{ - \sum_n  \eta \; [\gamma^{l}]^{(d_{l,j})}_n \omega_n    }   \ \sum_{\gamma^{\k}| \g^{\k_1}} \Bigg( \sum_{C=\kappa_1+1}^{\k-1 }    \prod_{j=C+1}^{\k-1} \frac{i}{ \tilde{\Delta}_{e_{j}} [\gamma^{j}] -i\e } 
   \nn \\
 &\ & \hspace{-20mm} \times \left( f_{C}  (\vec{q}_1 \dots \vec{q}_{k_C})
    \left( \frac{1+\lambda_{e_C}}{2} \right) - f_{C+1}  (\vec{q}_1 \dots \vec{q}_{k_{C+1}}) \ \left( \frac{1+\lambda_{e_{C+1}}}{2}\right)  \right)  \; 
    \prod_{i=\kappa_1+1}^C\; \frac{i}{ \tilde{\Delta}_{e_{i}} [\gamma^{i}]  +i\e}
    \nn \\
  &\ & \hspace{-20mm} - \prod_{j=\kappa_1+1}^{\k} \frac{i}{ \tilde{\Delta}_{e_{j}} [\gamma^{j}] -i\e } f_{1}  (\vec{q}_1 \dots \vec{q}_{k_1})
    \left( \frac{1+\lambda_{e_1}}{2}\right)+\prod_{i=\kappa_1+1}^{\k} \frac{i}{ \tilde{\Delta}_{e_{i}} [\gamma^{i}]+i\e  } f_{\k}  (\vec{q}_1 \dots \vec{q}_{k_\k})
    \left( \frac{1+\lambda_{e_\k}}{2}\right)\Bigg)\, . \nn
\eea
This represents our final result for leptonic annihilation cross sections. It carries no unphysical singularities and is power counting finite everywhere in the region of integration.   It is a sum of terms corresponding to all cuts of a vacuum polarization diagram with, in general, elementary vertices and composite vertices that include vacuum denominators.   Examples can be found in Figs.\ \ref{ex3-int-a} and \ref{ex3-int-c} of Sec.\ \ref{Poset and caus}.  As in the original form of Eq.\ (\ref{eq:reor-f-to-D}), the final two terms in this expression only require  contour deformations to manifest finiteness. 

\section{Summary}

 In this work, we have used TOPT to re-express infrared safe cross sections in electroweak annihilation in a form that is manifestly power-counnting finite in four dimensions.  We observed that contributions to cross sections, in which the
 amplitude or complex conjugate are disconnected, vanish after a sum over appropriate time orders.   Generalizing to arbitrary amplitudes, we reorganized TOPT expression using their poset structure, to derive  results similar to Ref.\ \cite{Capatti:2022mly}.  We then applied this formalism to electroweak annihilation, to derive a locally finite diagrammatic expression in which all pseudo-physical cuts are eliminated.  
 
  We anticipate that it will be possible to use an expression like Eq.\ (\ref{eq:ep-cross-sec-fs-reor}) as the starting point for the numerical evaluation of weighted cross sections in leptonic annihilation, complementing the loop-tree duality treatments of Refs.\ \cite{Capatti:2019edf}-\cite{Capatti:2020ytd}.   This, of course, will require a systematic method to use contour deformation or related methods to avoid or cancel non-pinched singularities \cite{Kermanschah:2021wbk}.   Beyond leptonic annihilation, the methods described here may complement those of Refs.\ \cite{Anastasiou:2020sdt,Anastasiou:2022eym} to control infrared singularities locally in hadronic cross sections.  

 \subsection*{Acknowledgements}  We thank Charalampos Anastasiou for many helpful discussions on these and related topics.   We thank Zeno Capatti and Hoffie Hannesdottir, Valentin Hirschi, and Dario Kermanschah for very useful conversations.
 This work was supported in part by National Science Foundation Awards PHY-1915093 and PHY-2210533.
\appendix

\section{Embedded Extrema} 
\label{app:index}

We would like to show that every connected, finite-order poset diagram  has at least one embedded extremal vertex.  

First, we recall that embedded vertex, $v$ in poset diagram $D$ is one whose elimination from $D$ leaves the remaining diagram, $D\setminus v$, connected.  If, on the contrary, $u$ is a ``non-embedded" vertex, we must have  $D\setminus u=\cup_{i=1}^{p_u} D^{[u]}_i$, with $p_u$ the number of subdiagrams of $D$ that share vertex $u$ and are otherwise disconnected.   This is the case whether $u$ is an extremum or not.

We begin our argument by picking an arbitrary extremum, $e_0$, which for definiteness we may choose to be a minimum.  Say $e_0$ is not embedded.   Then, starting at $e_0$, we choose arbitrarily a path $x_1, x_2 \dots$, that follows the relation $x_{j+1}>x_i$.   We refer to this as an ``increasing" path.   Because $D$ is finite-order, this path must terminate at some maximum vertex, $e_1$.  Either $e_1$ is embedded, or not.  If not, we know that $D\setminus e_1=\cup_{i=1}^{p_{e_1}} D^{[e_1]}_i$.  We repeat the process for vertex $e_1$ in subdiagram, $D^{[e_1]}_i \subset D$, chosen arbitrarily, where $e_0 \not\in D^{[e_1]}_i$.   If $e_1$ is not a maximum in $D^{[e_1]}_i$, we can again follow an increasing path in the $D^{[e_1]}$ to a second maximum.   If $e_1$ happens to be a maximum in $D^{[e_1]}_i$, we can follow a decreasing path until we reach a minimum.  In either case, the extremum, $e_2$ at which we arrive may or may not be embedded.   If it is not embedded we repeat the process by choosing any of the subdiagrams $D^{[e_2]}_j \subset D_i^{[e_1]}$,  $e_1 \not\in D^{[e_2]}_j$.   Because the complete poset $D$ is of finite order, 
\bea
0\, < \, |D^{[e_2]}_j|\, < \, |D^{[e_1]}_i|\, , \label{eq:reducing-seq}
\eea
and this sequence of steps must terminate at an extremal vertex, $e_n$ in some subdiagram $D_k^{[e_{n-1}]}$, whose removal does not disconnect $D_k^{[e_{n-1}]}$.  This is because  Eq.\ (\ref{eq:reducing-seq}) represents the first two terms in a sequence of strictly reducing positive integers. Such a sequence must terminate at $0$.  The extremal  vertex at which the sequence terminates is embedded.\footnote{Notice that extremum $e_{n-1}$ is considered part of $D_k^{[e_{n-1}]}$.}   

 We conclude by observing that the arguments above apply to diagrams that are of the tadpole topology: no external momenta and connected to the remainder of the diagram by only a single vertex, which may or may not be extremal.   Thus, as observed in Sec.\ \ref{sec:weighted-cs}, every tadpole has an embedded extremal vertex, by definition without an external momentum.  In the intermediate expression of Eq.\ (\ref{eq:integral-Pi1-steps-ep}), all such time integrals have been carried out, and in the process all tadpoles subdiagrams have been eliminated in poset $D[\g^{\kappa_1}]$. 

\section{An explicit algorithm to identify $\g^{k+1}$}
\label{app:algorithm}
In this Appendix, we would like to present one explicit algorithm to identify all the possible consistent choices in $\g^{k+1}$,  as defined for instance in Eq.\ (\ref{eq:gamma-1-def}), given a poset $D\left[\g^{k}\right]=\{V\left[ \g^{k}\right],\geq_{\g^{k}}\}$ and a choice in maximal embedded anti-chain $A^{k+1}=\{b_{k+1,1}\dots b_{k+1,r_{k+1}}, d_{k+1,1} \dots d_{k+1,s_{k+1}}\} =\{e_{k+1,j}\}$.  We emphasize that the ordering of extremal elements in $A^{k+1}$ is arbitrary. As in Sec.\ \ref{sec:integration}  we define next-to-extremal sets,  
\bea
C_{b_{k+1,j}}=\{ x\in V\left[ \g^{k}\right]\bigg|\; b_{k+1,j} \prec_{\g^k} x\},\nn \\
C_{d_{k+1,j}}=\{ x\in V\left[ \g^{k}\right]\bigg|\; x \prec_{\g^k} d_{k+1,j}\},\nn\\
C^{k+1}= \bigcup\limits_{j=1}^{r_{k+1}} C_{b_{k+1,j}}\bigcup\limits_{l=1}^{s_{k+1}} C_{d_{k+1,l}}.
\label{eq:next-to-extremal}
\eea
We also define a set of next-to-extremal vertices,
\bea
\g^{k}_{i} \ \equiv \ \bigcup\limits_{q\in [1,k]} \bigcup\limits_{a \in [1,\hat{r}_q(k,i)]} x_{{j_{q,a}}}^{(e_{q,a})}\, ,
\label{eq:gamma-ki-def}
\eea
where
\bea
 \hat{r}_q(k,i) &=&r_q+s_q \text{ if } q\in[1,k-1] \, , \nn \\ \hat{r}_q(k,i) &=& i \text{ if } q=k \, .
\label{eq:union_notation}
\eea
In the notation of Eq.\ (\ref{eq:union_notation}), $D\left[\g^{k}_0\right] = D\left[\g^{k}\right]$, with $D[\g^k]$ defined in terms of the bilinear relation after the $k$th set of time integrals, Eq.\ (\ref{eq:ge-gamma-k-def}).  We start with $D[\gamma^k]$ and construct the full explicit sum over $D[\g^{k+1}]$, one extremal vertex at a time.   We will iteratively define the sum over $D\left[\g^{k+1}\right]$ in terms of a sum over intermediate posets leading to $D\left[\g^{k}_{r_{k+1}+s_{k+1}}\right]$. 

Suppose we are given the poset $D\left[\g^{k}_i\right]$, which is, as usual, a set of vertices $D\left[\g^k_i\right]$, and a binary relation $\ge_{\g^k_i}$, defined by direct analogy to Eq.\ (\ref{eq:ge-gamma-k-def}). To identify the poset $D\left[\g^{k}_{i+1}\right]$, we would like the minimum (maximum) $e_{k+1,i+1}$ to be covered by (to cover) a unique element $x_{j}^{(e_{k+1,i+1})}$. To do this, we find the set
\bea
\tilde{C}_{e_{k+1,i+1}}&=&\{ x\in V\left[\g^{k}_{i+1}\right]\bigg|\; e_{k+1,i+1} \prec_{\g^k_i} x\}\;\; \text{if } \;\; i+1 \leq r_{k+1}\nn\\ 
\tilde{C}_{e_{k+1,i+1}}&=&\{ x\in V\left[\g^{k}_{i+1}\right]\bigg|\; x \prec_{\g^k_i} e_{k+1,i+1}\} \;\; \text{if } \;\; i+1 > r_{k+1}.\nn\\
&=&\{x^{(e_{k+1},i+1)}_{1},x^{(e_{k+1},i+1)}_{2} \dots x^{(e_{k+1},i+1)}_{c_{k+1,i+1}} \}.
\label{eq:covering-vertices-onestep}
\eea
We notice that generically, $\tilde{C}_e$ is distinct from $C_e$ because we use the binary $\geq_{\g^k}$ in Eq.\ (\ref{eq:next-to-extremal}) while we use the binary $\geq_{\g^k_i}$ in Eq.\ (\ref{eq:covering-vertices-onestep}). In fact, one generally finds  $\tilde{C}_e \subseteq C_e$ since we use the more restrictive binary $\geq_{\g^k_i}$. We understand this to follow from the observation that having identified unique covering (covered) vertices for the extrema $e_{k+1,j}\;\; j \in [1,i]$, there are fewer consistent covering (covered) vertices for $e_{k+1,i+1}$. We now define a new binary $\g^k_{i+1}$ that satisfies
\medskip
\begin{itemize}
\item If $a \geq_{\g^{k}_i}  b$, then $a \geq_{\g^{k}_{i+1}}  b$
\item 
If $i+1 \leq r_{k+1}$, we choose the earliest element $ x_{j_{k,i+1}}^{(e_{k,i+1})} \in \tilde{C}_{e_{k+1,i+1}}: y \geq_{\g^{k}_{i+1}}  x_{j_{k,i+1}}^{(e_{k,i+1})}, \forall y \in   \tilde{C}_{e_{k+1,i+1}}$,
\item 
If $i+1 > r_{k+1}$, we choose the latest element $ x_{j_{k,i+1}}^{(e_{k,i+1})} \in \tilde{C}_{e_{k+1,i+1}}: y \leq_{\g^{k}_{i+1}}  x_{j_{k,i+1}}^{(e_{k,i+1})}, \forall y \in   \tilde{C}_{e_{k+1,i+1}}$.
\end{itemize}

\medskip
With this definition of the binary relationship $\g^{k}_{i+1}$, we have identified a unique element that provides  a limit of integration for the extremum element $e_{k,i+1}$. We may now define an intermediate poset $D\left[\g^{k}_{i+1}\right]$,
\be
D\left[\g^{k}_{i+1}\right] = \left\lbrace V\left[ \g^{k}\right]\setminus  \left(\bigcup\limits_{j=1}^{i+1}e_{k,j} \right),\geq_{\g^{k}_{i+1}}\right\rbrace.
\ee 
Using this definition, we can start from $\gamma^k_0$ and recursively find $\gamma^{k+1}=\gamma^{k}_{r_{k+1}+s_{k+1}}$ .
 It is clear from these definitions, that every consistent set of next  to extremal vertices is generated. An inductive proof of this follows from assuming the result for $\g^k_i$, and demonstrating that it follows for $\g^k_{i+1}$.
 
 It is now possible to define a recursive notation that makes explicit the many sums over sets $\gamma^k$ in the main text,
 \be
 \sum\limits_{\g^{k}} F\left(\left\lbrace\bigcup \limits_{m=1}^{k}\bigcup \limits_{l=1}^{r_{k}+s_{k}} x_{j_{m,l}}^{(e_{m,l})}\right \rbrace\right)=\sum \limits_{j_{k,r_{k}+s_{k}}=1}^{c_{k,r_{k}+s_{k}}}\dots\sum \limits_{j_{k,2}=1}^{c_{k,2}} \sum \limits_{j_{k,1}=1}^{c_{k,1}}\sum\limits_{\g^{k-1}} F\left(\left\lbrace\bigcup \limits_{m=1}^{k}\bigcup \limits_{l=1}^{r_{k}+s_{k}} x_{j_{m,l}}^{(e_{m,l})}\right \rbrace\right).\
 \label{eq:summation-notation}
 \ee
In Eq.\ (\ref{eq:summation-notation}) $F$ is any function of the next-to-extremal vertices and $c_{k,i}$ is the number of elements in the companion set $\tilde{C}_{k,i}$ as in Eq.\ (\ref{eq:covering-vertices-onestep}). Notice that the sum in Eq.\ (\ref{eq:summation-notation}) is a nested sum where the outer summations depend on the inner summations. The summation over $\g^k$ depends also on our choice in anti-chain $A^k$.


\begin{thebibliography}{99}

\bibitem{Heinrich:2020ybq}
G.~Heinrich,
``Collider Physics at the Precision Frontier,''
Phys. Rept. \textbf{922}, 1-69 (2021)
doi:10.1016/j.physrep.2021.03.006
[arXiv:2009.00516 [hep-ph]].

\bibitem{Caola:2022ayt}
F.~Caola, W.~Chen, C.~Duhr, X.~Liu, B.~Mistlberger, F.~Petriello, G.~Vita and S.~Weinzierl,
``The Path forward to N$^3$LO,''
[arXiv:2203.06730 [hep-ph]].

\bibitem{Boughezal:2022cbl}
R.~Boughezal, Z.~Ligeti, W.~Altmannshofer, S.~Das Bakshi, F.~Caola, M.~Chala, A.~Diaz-Carmona, W.~Chen, N.~Darvishi and B.~Henning, \textit{et al.}
``Theory Techniques for Precision Physics -- Snowmass 2021 TF06 Topical Group Report,''
[arXiv:2209.10639 [hep-ph]].

\bibitem{TorresBobadilla:2020ekr}
W.~J.~Torres Bobadilla, G.~F.~R.~Sborlini, P.~Banerjee, S.~Catani, A.~L.~Cherchiglia, L.~Cieri, P.~K.~Dhani, F.~Driencourt-Mangin, T.~Engel and G.~Ferrera, \textit{et al.}
``May the four be with you: Novel IR-subtraction methods to tackle NNLO calculations,''
Eur. Phys. J. C \textbf{81}, no.3, 250 (2021)
doi:10.1140/epjc/s10052-021-08996-y
[arXiv:2012.02567 [hep-ph]].

\bibitem{Anastasiou:2020sdt}
C.~Anastasiou, R.~Haindl, G.~Sterman, Z.~Yang and M.~Zeng,
``Locally finite two-loop amplitudes for off-shell multi-photon production in electron-positron annihilation,''
JHEP \textbf{04}, 222 (2021)
doi:10.1007/JHEP04(2021)222
[arXiv:2008.12293 [hep-ph]].

\bibitem{Anastasiou:2022eym}
C.~Anastasiou and G.~Sterman,
`Locally finite two-loop QCD amplitudes from IR universality for electroweak production,''
JHEP \textbf{05}, 242 (2023)
doi:10.1007/JHEP05(2023)242
[arXiv:2212.12162 [hep-ph]].

\bibitem{Hernandez-Pinto:2015ysa}
R.~J.~Hernandez-Pinto, G.~F.~R.~Sborlini and G.~Rodrigo,
``Towards gauge theories in four dimensions,''
JHEP \textbf{02}, 044 (2016)
doi:10.1007/JHEP02(2016)044
[arXiv:1506.04617 [hep-ph]].

\bibitem{Sborlini:2016gbr}
G.~F.~R.~Sborlini, F.~Driencourt-Mangin, R.~Hernandez-Pinto and G.~Rodrigo,
``Four-dimensional unsubtraction from the loop-tree duality,''
JHEP \textbf{08}, 160 (2016)
doi:10.1007/JHEP08(2016)160
[arXiv:1604.06699 [hep-ph]].

\bibitem{Sborlini:2016hat}
G.~F.~R.~Sborlini, F.~Driencourt-Mangin and G.~Rodrigo,
``Four-dimensional unsubtraction with massive particles,''
JHEP \textbf{10}, 162 (2016)
doi:10.1007/JHEP10(2016)162
[arXiv:1608.01584 [hep-ph]].

\bibitem{Aguilera-Verdugo:2019kbz}
J.~J.~Aguilera-Verdugo, F.~Driencourt-Mangin, J.~Plenter, S.~Ram\'{i}rez-Uribe, G.~Rodrigo, G.~F.~R.~Sborlini, W.~J.~Torres Bobadilla and S.~Tracz,
``Causality, unitarity thresholds, anomalous thresholds and infrared singularities from the loop-tree duality at higher orders,''
JHEP \textbf{12}, 163 (2019)
doi:10.1007/JHEP12(2019)163
[arXiv:1904.08389 [hep-ph]].

\bibitem{Capatti:2019ypt}
Z.~Capatti, V.~Hirschi, D.~Kermanschah and B.~Ruijl,
``Loop-Tree Duality for Multiloop Numerical Integration,''
Phys. Rev. Lett. \textbf{123}, no.15, 151602 (2019)
doi:10.1103/PhysRevLett.123.151602
[arXiv:1906.06138 [hep-ph]].

\bibitem{Capatti:2019edf}
Z.~Capatti, V.~Hirschi, D.~Kermanschah, A.~Pelloni and B.~Ruijl,
``Numerical Loop-Tree Duality: contour deformation and subtraction,''
JHEP \textbf{04}, 096 (2020)
doi:10.1007/JHEP04(2020)096
[arXiv:1912.09291 [hep-ph]].

\bibitem{Aguilera-Verdugo:2020set}
J.~J.~Aguilera-Verdugo, F.~Driencourt-Mangin, R.~J.~Hern\'andez-Pinto, J.~Plenter, S.~Ramirez-Uribe, A.~E.~Renteria Olivo, G.~Rodrigo, G.~F.~R.~Sborlini, W.~J.~Torres Bobadilla and S.~Tracz,
``Open Loop Amplitudes and Causality to All Orders and Powers from the Loop-Tree Duality,''
Phys. Rev. Lett. \textbf{124}, no.21, 211602 (2020)
doi:10.1103/PhysRevLett.124.211602
[arXiv:2001.03564 [hep-ph]].

\bibitem{Capatti:2020ytd}
Z.~Capatti, V.~Hirschi, D.~Kermanschah, A.~Pelloni and B.~Ruijl,
``Manifestly Causal Loop-Tree Duality,''
[arXiv:2009.05509 [hep-ph]].

\bibitem{Sterman:1978bj}
G.~F.~Sterman,
``Mass Divergences in Annihilation Processes. 2. Cancellation of Divergences in Cut Vacuum Polarization Diagrams,''
Phys. Rev. D \textbf{17}, 2789 (1978)
doi:10.1103/PhysRevD.17.2789

\bibitem{Sterman:1979uw}
G.~F.~Sterman,
``Zero Mass Limit for a Class of Jet Related Cross-sections,''
Phys. Rev. D \textbf{19}, 3135 (1979)
doi:10.1103/PhysRevD.19.3135

\bibitem{Bodwin:1984hc}
G.~T.~Bodwin,
``Factorization of the Drell-Yan Cross-Section in Perturbation Theory,''
Phys. Rev. D \textbf{31}, 2616 (1985)
[erratum: Phys. Rev. D \textbf{34}, 3932 (1986)]
doi:10.1103/PhysRevD.34.3932

\bibitem{Collins:1985ue}
J.~C.~Collins, D.~E.~Soper and G.~F.~Sterman,
``Factorization for Short Distance Hadron - Hadron Scattering,''
Nucl. Phys. B \textbf{261}, 104-142 (1985)
doi:10.1016/0550-3213(85)90565-6

\bibitem{Sterman:1993hfp}
G.~F.~Sterman,
``An Introduction to quantum field theory,''
Cambridge University Press, 1993,
ISBN 978-0-521-31132-8

\bibitem{Sterman:1995fz}
G.~F.~Sterman,
``Partons, factorization and resummation, TASI 95,''
[arXiv:hep-ph/9606312 [hep-ph]].

\bibitem{Catani:2008xa}
S.~Catani, T.~Gleisberg, F.~Krauss, G.~Rodrigo and J.~C.~Winter,
``From loops to trees by-passing Feynman's theorem,''
JHEP \textbf{09}, 065 (2008)
doi:10.1088/1126-6708/2008/09/065
[arXiv:0804.3170 [hep-ph]].

\bibitem{TorresBobadilla:2021ivx}
W.~J.~Torres Bobadilla,
``Loop-tree duality from vertices and edges,''
JHEP \textbf{04}, 183 (2021)
doi:10.1007/JHEP04(2021)183
[arXiv:2102.05048 [hep-ph]].

\bibitem{Sborlini:2021owe}
G.~F.~R.~Sborlini,
``Geometrical approach to causality in multiloop amplitudes,''
Phys. Rev. D \textbf{104}, no.3, 036014 (2021)
doi:10.1103/PhysRevD.104.036014
[arXiv:2102.05062 [hep-ph]].

\bibitem{Borinsky:2022msp}
M.~Borinsky, Z.~Capatti, E.~Laenen and A.~Salas-Bern\'ardez,
JHEP \textbf{01}, 172 (2023)
doi:10.1007/JHEP01(2023)172
[arXiv:2210.05532 [hep-th]].

\bibitem{Capatti:2022mly}
Z.~Capatti,
``Exposing the threshold structure of loop integrals,''
Phys. Rev. D \textbf{107}, no.5, L051902 (2023)
doi:10.1103/PhysRevD.107.L051902
[arXiv:2211.09653 [hep-th]].

\bibitem{Komiske:2020qhg}
P.~T.~Komiske, E.~M.~Metodiev and J.~Thaler,
``The Hidden Geometry of Particle Collisions,''
JHEP \textbf{07}, 006 (2020)
doi:10.1007/JHEP07(2020)006
[arXiv:2004.04159 [hep-ph]].

\bibitem{Coleman:1965xm}
S.~Coleman and R.~E.~Norton,
``Singularities in the physical region,''
Nuovo Cim. \textbf{38}, 438-442 (1965)
doi:10.1007/BF02750472

\bibitem{Collins:2020euz}
J.~Collins,
``A new and complete proof of the Landau condition for pinch singularities of Feynman graphs and other integrals,''
[arXiv:2007.04085 [hep-ph]].

\bibitem{Sterman:1978bi}
G.~F.~Sterman,
``Mass Divergences in Annihilation Processes. 1. Origin and Nature of Divergences in Cut Vacuum Polarization Diagrams,''
Phys. Rev. D \textbf{17}, 2773 (1978)
doi:10.1103/PhysRevD.17.2773

\bibitem{Collins:2011zzd}
J.~Collins,
``Foundations of perturbative QCD,''
Camb. Monogr. Part. Phys. Nucl. Phys. Cosmol. \textbf{32}, 1-624 (2011)
Cambridge University Press, 2013,
ISBN 978-1-00-940184-5
doi:10.1017/9781009401845

\bibitem{Sterman:1977wj}
G.~F.~Sterman and S.~Weinberg,
``Jets from Quantum Chromodynamics,''
Phys. Rev. Lett. \textbf{39}, 1436 (1977)
doi:10.1103/PhysRevLett.39.1436

\bibitem{Stevenson:1978td}
P.~M.~Stevenson,
``Comments on Sterman-Weinberg jet formula
,''
Phys. Lett. B \textbf{78}, 451-454 (1978)
doi:10.1016/0370-2693(78)90483-5

\bibitem{Dukes:2013gea}
M.~Dukes, E.~Gardi, H.~McAslan, D.~J.~Scott and C.~D.~White,
``Webs and Posets,''
JHEP \textbf{01}, 024 (2014)
doi:10.1007/JHEP01(2014)024
[arXiv:1310.3127 [hep-th]].

\bibitem{Erdogan:2017gyf}
O.~Erdogan and G.~Sterman,
``Path description of coordinate-space amplitudes,''
Phys. Rev. D \textbf{95}, no.11, 116015 (2017)
doi:10.1103/PhysRevD.95.116015
[arXiv:1705.04539 [hep-th]].

\bibitem{Chang:1968bh}
S.~J.~Chang and S.~K.~Ma,
``Feynman rules and quantum electrodynamics at infinite momentum,''
Phys. Rev. \textbf{180}, 1506-1513 (1969)
doi:10.1103/PhysRev.180.1506

\bibitem{Kogut:1969xa}
J.~B.~Kogut and D.~E.~Soper,
``Quantum Electrodynamics in the Infinite Momentum Frame,''
Phys. Rev. D \textbf{1}, 2901-2913 (1970)
doi:10.1103/PhysRevD.1.2901

\bibitem{Brodsky:1997de}
S.~J.~Brodsky, H.~C.~Pauli and S.~S.~Pinsky,
``Quantum chromodynamics and other field theories on the light cone,''
Phys. Rept. \textbf{301}, 299-486 (1998)
doi:10.1016/S0370-1573(97)00089-6
[arXiv:hep-ph/9705477 [hep-ph]].

\bibitem{Kermanschah:2021wbk}
D.~Kermanschah,
``Numerical integration of loop integrals through local cancellation of threshold singularities,''
JHEP \textbf{01}, 151 (2022)
doi:10.1007/JHEP01(2022)151
[arXiv:2110.06869 [hep-ph]].


\end{thebibliography}
\end{document}